\documentclass[11pt,aps,showpacs,notitlepage]{revtex4-1}
\usepackage{amssymb,amsmath}
\usepackage{bm}
\usepackage[pdftex]{graphicx}
\usepackage{subfig}
\usepackage{color}

\newcommand{\beq}{\begin{equation}}
\newcommand{\beqn}{\begin{eqnarray}}
\newcommand{\eeq}{\end{equation}}
\newcommand{\eeqn}{\end{eqnarray}}

\begin{document}

\title{Preheating after Multifield Inflation with Nonminimal Couplings, I: \\ Covariant Formalism and Attractor Behavior}
\author{Matthew P. DeCross$^1$\footnote{Now at the Department of Physics and Astronomy, University of Pennsylvania}, David I. Kaiser$^1$, Anirudh Prabhu$^1$\footnote{Now at the Department of Physics, Stanford University}, C. Prescod-Weinstein$^2$, and Evangelos I. Sfakianakis$^3$\footnote{Now at NIKHEF and Leiden University} }
\email{Email addresses: mdecross@sas.upenn.edu ; dikaiser@mit.edu ; aniprabhu@stanford.edu ; cprescod@uw.edu ;\\ evans@nikhef.nl}
\affiliation{$^1$Department of Physics, 
Massachusetts Institute of Technology, Cambridge, Massachusetts 02139 USA\\
$^2$ Department of Physics, University of Washington, Seattle, Washington 98195-1560
\\
$^3$Department of Physics, University of Illinois at Urbana-Champaign, Urbana, Illinois 61801}
\date{\today}
\begin{abstract} This is the first of a three-part series of papers, in which we study the preheating phase for multifield models of inflation involving nonminimal couplings. In this paper, we study the single-field attractor behavior that these models exhibit during inflation and quantify its strength and parameter dependence. We further demonstrate that the strong single-field attractor behavior persists after the end of inflation. Preheating in such models therefore generically avoids the ``de-phasing" that typically affects multifield models with minimally coupled fields, allowing efficient transfer of energy from the oscillating inflaton condensate(s) to coupled perturbations across large portions of parameter space. We develop a doubly-covariant formalism for studying the preheating phase in such models and identify several features specific to multifield models with nonminimal couplings, including effects that arise from the nontrivial field-space manifold. In papers II and III, we apply this formalism to study how the amplification of adiabatic and isocurvature perturbations varies with parameters, highlighting several distinct regimes depending on the magnitude of the nonminimal couplings $\xi_I$.
\end{abstract}
\pacs{98.80.Cq ; 95.30.Cq.  Preprint MIT-CTP/4716.}
\maketitle

\section{Introduction} 

This is the first paper in a three-part series that examines the early stages of post-inflation reheating in models that involve multiple scalar fields, each nonminimally coupled to gravity. (The companion papers are Refs.~\cite{MultiPreheat2,MultiPreheat3}.)

Post-inflation reheating is a critical phase in the history of the cosmos, necessary to connect early-universe inflation to the usual successes of the standard hot big bang scenario. Reheating falls between two regimes that are well constrained by observations, and which match the latest observations remarkably well: production of a spatially flat universe seeded with nearly scale-invariant primordial curvature perturbations during inflation \cite{GuthKaiser,BTW,LythLiddle,Baumann,MartinRingeval,GKN,LindePlanck,MartinRev}, and production of specific abundances of light nuclei during big-bang nucleosynthesis \cite{Steigman,FieldsBBN,Cyburt}. Though it remains difficult to relate the reheating phase directly to specific, testable predictions for observations, the process of reheating remains critical in order to compare predictions from the inflationary era with present-day observations, since relating comoving scales at different cosmological epochs requires knowledge of the intervening expansion history of the universe \cite{AdsheadEasther,Dai,Creminelli,MartinReheat,GongLeungPi,CaiGuoWang,Cook,Heisig}. See \cite{BTW,Allahverdi,Frolov,AHKK,AminBaumann} for recent reviews of reheating.

The post-inflation reheating phase not only must bring the early universe to thermal equilibrium in a radiation-dominated phase at an appropriately high temperature; reheating should also populate the universe with matter like the kind we see around us today. During inflation, the energy density of the universe was presumably dominated by one or more scalar ``inflaton" fields. After reheating, the energy density should include contributions from multiple species of matter, including the Standard Model particles or (at least) types of matter that decay into Standard Model particles prior to big-bang nucleosynthesis. Such interactions could address other long-standing challenges in cosmological theory, such as generating the observed baryon - antibaryon asymmetry \cite{MPHBaryogenesis,Mustafa,Adshead:2015jza,Adshead:2015kza}. Reheating therefore must be a multifield phenomenon.

Arguably, inflation itself should be treated as a process involving multiple fields. Realistic models of high-energy particle physics typically include many distinct scalar fields at high energies \cite{LythRiotto,WandsReview,MazumdarRocher,VenninWands,GongReview}. Hence we consider multiple scalar fields to be a central ingredient of realistic models of inflation. Nonminimal couplings between the scalar fields and the Ricci spacetime curvature scalar are also a generic feature of realistic models of the early universe. Many theoretical motivations for nonminimal couplings derive from high-energy model-building, including dilatons and moduli fields, but a more basic motivation comes from renormalization: as has long been known, models with self-interacting scalar fields in curved spacetime require nonminimal couplings as counterterms in order to remain self-consistent at high energies. Nonminimal couplings are induced by quantum corrections even in the absence of bare couplings; they are a generic feature of scalar fields in curved spacetime \cite{Callan,Bunch,BirrellDavies,Buchbinder,ParkerToms,Odintsov,Markkanen}. Moreover, such couplings arise even in a classical background spacetime. Thus their effects can be important at energy scales relevant to inflationary or post-inflationary dynamics, even for models in which quantum-gravitational corrections to the Einstein-Hilbert action --- which would presumably be quadratic or higher order in the spacetime curvature --- may remain subdominant at those energy scales \cite{Rsquared}.

In recent work \cite{KMS,GKS,KS,SSK} we have studied the dynamics during inflation from multifield models with nonminimal couplings, including generalizations of ``Higgs inflation" \cite{BezrukovShaposhnikov}. These papers have demonstrated that such models generically predict observable quantities (related to the spectrum of primordial curvature perturbations) squarely in the most-favored region of the latest observations. Moreover, such models exhibit a strong attractor behavior: across broad regions of parameter space and phase space the fields relax to an effectively single-field trajectory early in inflation. Hence the predictions for observable quantities from these models show little dependence on coupling constants or initial conditions \cite{KS}. Such attractor behavior is a generic feature of multifield models with nonminimal couplings, including the so-called ``$\alpha$ attractors" \cite{Lindealpha}.

In this paper we focus on the dynamics of such models immediately after inflation, during the ``preheating" phase. During preheating, the scalar-field condensate(s) that drove inflation decay resonantly into higher-momentum quanta. We develop a doubly-covariant formalism that incorporates metric perturbations and field fluctuations self-consistently (to first order), and which also respects the reparameterization freedom of the nontrivial field-space manifold. We restrict attention to the early stages of preheating, for which an approximation linear in the fields' fluctuations remains reliable, and only consider decays into scalar fields rather than fermions or gauge fields. Our approach complements previous studies that have examined reheating in models with nonminimally coupled fields \cite{BassettGeometric,ShinjiPapers,ShinjiBassett,WatanabeWhite,Racioppi,vanBruck,Ema}, including Higgs inflation \cite{HiggsReheat1,HiggsReheat2,HiggsReheat3,HiggsReheat4}, as well as with noncanonical kinetic terms or other string-inspired features of the action~\cite{NonCanKin1,NonCanKin2,NonCanKin3,NonCanKin4,Mflation}. In our companion papers \cite{MultiPreheat2,MultiPreheat3}, we analyze the amplification of perturbations in this family of models semi-analytically and numerically across wide regions of parameter space. 

We find three principal distinctions from the well-studied cases of preheating with minimally coupled fields. First, the conformal stretching of the scalar fields' potential in the Einstein frame affects the oscillations of the background fields, compared to the case of minimal couplings. In particular, for strong nonminimal couplings $\xi_I \gg 1$, the background fields' oscillations interpolate between the behavior of minimally coupled models with quadratic and quartic self-couplings. Second, the single-field attractor behavior during inflation typically leads to greater efficiency during preheating than in corresponding multifield models with minimal couplings, in which de-phasing of the background fields' oscillations usually damps resonances \cite{AHKK,Barnaby,Battefeld}. Third, the nontrivial field-space manifold contributes differently to the effective masses for fluctuations in the adiabatic and isocurvature directions, leading to distinct behavior depending on whether the nonminimal couplings are small ($\xi_I <{\cal O} (1)$), intermediate ($\xi_I \sim {\cal O} (1 - 10)$), or large ($\xi_I \geq {\cal O} (100)$).

In Section \ref{CovariantFormalism} we review the doubly-covariant formalism with which we study the dynamics of background fields and fluctuations. In Section \ref{BackgroundDynamics} we examine the background dynamics for a two-field model during and after inflation, highlighting distinctions between oscillations during preheating with and without nonminimal couplings. The behavior of the background fields during the oscillating phase is critical for understanding the resonant production of particles during preheating. In Section \ref{EvolutionFluctuations} we introduce a covariant mode expansion for the fluctuations and derive multifield generalizations of the ``adiabatic parameter" with which to characterize the resonant, nonperturbative growth of fluctuations. Concluding remarks follow in Section \ref{Conclusions}.

\section{Doubly-Covariant Formalism}
\label{CovariantFormalism}

When studying multifield models with nonminimal couplings, one must consider two types of gauge transformations: the usual spacetime coordinate transformations, $x^\mu \rightarrow x^{\mu \prime}$, as well as transformations of the field-space coordinates, $\phi^I \rightarrow \phi^{I \prime}$. To address the first type of transformation, we adopt the usual (spacetime) gauge-invariant perturbation formalism \cite{Bardeen,Sasaki,Mukhanov}; see Refs.~\cite{Brandenberger,BTW,MalikWands} for reviews. To address the multifield aspects, we build on the methods of Refs.~\cite{SasakiStewart,GWBR,Nibbelink,WandsBartolo,SeeryLidsey,LeeSasaki,WandsReview,Yokoyama,Langlois,Tegmark,GongTanaka,Seery}. Together, these yield a doubly-covariant formalism for studying fluctuations in these multifield models \cite{KMS}.

We follow closely the notation and parameterization of \cite{KMS,GKS,KS,SSK}. We work in $(3+1)$ spacetime dimensions and adopt the spacetime metric signature $(-,+,+,+)$. We consider models with $N$ real-valued scalar fields, each of which is coupled to the Ricci spacetime curvature scalar. In the Jordan frame, the action takes the form
\beq
S = \int d^4 x \sqrt{-\tilde{g} } \left[ f (\phi^I ) \tilde{R} - \frac{1}{2} \delta_{IJ} \tilde{g}^{\mu\nu} \partial_\mu \phi^I \partial_\nu \phi^J - \tilde{V} (\phi^I ) \right] ,
\label{SJ}
\eeq
where upper-case Latin letters label field-space indices, $I, J = 1, 2, ... , N$, Greek letters label spacetime indices, $\mu, \nu = 0, 1, 2, 3$, and tildes denote Jordan-frame quantities. We will use lower-case Latin letters for spatial indices, $i, j = 1, 2, 3$. 

We may perform a conformal transformation to bring the gravitational portion of the action into canonical Einstein-Hilbert form, by rescaling $\tilde{g}_{\mu\nu} (x) \rightarrow g_{\mu\nu} (x) = \Omega^2 (x)\> \tilde{g}_{\mu \nu} (x)$. The conformal factor $\Omega^2 (x)$ is related to the nonminimal-coupling function,
\beq
g_{\mu\nu} (x) = \frac{2}{M_{\rm pl}^2} f (\phi^I (x)) \> \tilde{g}_{\mu\nu} (x),
\label{gtildeg}
\eeq
where $M_{\rm pl} \equiv 1 / \sqrt{8 \pi G} = 2.43 \times 10^{18}$ GeV is the reduced Planck mass. The action may then be rewritten \cite{DKconf}
\beq
S = \int d^4 x \sqrt{-g} \left[ \frac{ M_{\rm pl}^2}{2} R - \frac{1}{2} {\cal G}_{IJ} (\phi^K ) g^{\mu\nu} \partial_\mu \phi^I \partial_\nu \phi^J - V (\phi^I ) \right] .
\label{SE}
\eeq
(See also Ref.~\cite{Abedi}.) The potential in the Einstein frame is stretched by the conformal factor,
\beq
V (\phi^I) = \frac{ M_{\rm pl}^4 }{4 f^2 (\phi^I) } \tilde{V} (\phi^I ) .
\label{VE}
\eeq
In addition, the nonminimal couplings induce a curved field-space manifold in the Einstein frame, with associated field-space metric ${\cal G}_{IJ} (\phi^K)$. Because the induced field-space manifold is not conformal to flat for $N \geq 2$, no combination of rescalings of $g_{\mu\nu}$ and $\phi^I$ can retain the Einstein-Hilbert form for the gravitational portion of the action while also bringing the fields' kinetic terms into canonical form \cite{DKconf}. The components of ${\cal G}_{IJ}$ take the form 
\beq
{\cal G}_{IJ} (\phi^K ) = \frac{ M_{\rm pl}^2 }{2 f (\phi^K) } \left[ \delta_{IJ} + \frac{3}{ f (\phi^K) } f_{, I} f_{, J} \right] ,
\label{GIJ}
\eeq
where $f_{, I} = \partial f / \partial \phi^I$. The field-space metric satisfies ${\cal G}^{IJ} {\cal G}_{JK} = \delta^I_{\>K}$, and field-space indices are raised and lowered with ${\cal G}_{IJ}$. 

Varying the action of Eq. (\ref{SE}) with respect to $g_{\mu \nu}$ yields the field equations 
\beq
R_{\mu\nu} - \frac{1}{2} g_{\mu\nu} R = \frac{1}{M_{\rm pl}^2} T_{\mu\nu} ,
\label{EFE}
\eeq
with the energy-momentum tensor given by \cite{KMS}
\beq
T_{\mu\nu} = {\cal G}_{IJ} \partial_\mu \phi^I \partial_\nu \phi^J - g_{\mu\nu} \left[ \frac{1}{2} {\cal G}_{IJ} g^{\alpha \beta} \partial_\alpha \phi^I \partial_\beta \phi^J + V (\phi^I ) \right] .
\label{Tmn}
\eeq
Varying Eq. (\ref{SE}) with respect to $\phi^I$ yields the equation of motion
\beq
\Box \phi^I + g^{\mu\nu} \Gamma^I_{\> JK} \partial_\mu \phi^J \partial_\nu \phi^K - {\cal G}^{IJ} V_{, J} = 0 ,
\label{eomphi}
\eeq
where $\Box \phi^I \equiv g^{\mu\nu} \phi^I_{\>\> ; \mu  \nu}$ and $\Gamma^I_{\> JK} (\phi^L)$ is the Christoffel symbol constructed from the field-space metric ${\cal G}_{IJ}$.

We expand the scalar fields and the spacetime metric to first order in perturbations. We are interested in the behavior of the fields at the end of inflation, so we consider scalar metric perturbations around a spatially flat Friedmann-Lema\^{i}tre-Robertson-Walker (FLRW) line element,
\beq
\begin{split}
ds^2 &= g_{\mu\nu} (x) \> dx^\mu dx^\nu \\
&= - (1 + 2 A) dt^2 + 2 a \left( \partial_i B \right) dx^i dt + a^2 \left[ \left( 1 - 2 \psi \right) \delta_{ij} + 2 \partial_i \partial_j E \right] dx^i dx^j ,
\end{split}
\label{ds}
\eeq
where $a(t)$ is the scale factor. We also expand the fields,
\beq
\phi^I (x^\mu) = \varphi^I (t) + \delta \phi^I (x^\mu) .
\label{phivarphi}
\eeq
The fluctuations $\delta \phi^I$ represent finite displacements from the fields' classical trajectory through field space; the fluctuations $\delta \phi^I$ are gauge dependent with respect to both $x^\mu \rightarrow x^{\mu \prime}$ and $\varphi^I \rightarrow \varphi^{I \prime}$. We therefore proceed in two steps. First, following Ref.~\cite{GongTanaka}, we introduce a vector ${\cal Q}^I$ to represent the field fluctuations covariantly with respect to the field-space metric, ${\cal G}_{IJ}$. (See also Ref.~\cite{Seery}.) The field-space vectors $\phi^I (x^\mu)$ and $\varphi^I (t)$ may be connected by a geodesic along the field-space manifold with some affine parameter $\lambda$. We take $\phi^I (\lambda = 0) = \varphi^I$ and $\phi^I (\lambda = \epsilon) = \varphi^I + \delta \phi^I$. (We may take $\epsilon = 1$ at the end.) These boundary conditions uniquely specify the vector ${\cal Q}^I$ that connects $\phi^I$ and $\varphi^I$, such that $\phi^I \vert_{\lambda = 0} = \varphi^I $ and ${\cal D}_\lambda \phi^I \vert_{\lambda = 0} = ( d \phi^I / d\lambda ) \vert_{\lambda = 0} = {\cal Q}^I$, where ${\cal D}_\lambda$ is a covariant derivative with respect to the affine parameter. Then \cite{GongTanaka}
\beq
\delta \phi^I = {\cal Q}^I - \frac{1}{2!} \Gamma^I_{\> JK} {\cal Q}^J {\cal Q}^K + \frac{1}{3!} \left[ \Gamma^I_{\> LM} \Gamma^M_{\> JK} - \Gamma^I_{\> JK ; L} \right] {\cal Q}^J {\cal Q}^K {\cal Q}^L + ...
\label{deltaphiQ}
\eeq
where the $\Gamma^I_{\> JK}$ are evaluated at background order, as functions of $\varphi^I$. Note that $\delta \phi^I \rightarrow {\cal Q}^I$ to first order in fluctuations, but one must take care to distinguish the two when working to higher order, as we will do in Section \ref{ActionQ} when we expand the action to second order in ${\cal Q}^I$. Next, we follow Ref.~\cite{KMS} and define a linear combination of ${\cal Q}^I$ and the metric perturbation $\psi$ to form a generalization of the gauge-invariant Mukhanov-Sasaki variable: 
\beq
Q^I \equiv {\cal Q}^I + \frac{ \dot{\varphi}^I }{H} \psi .
\label{Qdef}
\eeq
The vector $Q^I$ is doubly covariant, with respect to spacetime gauge transformations (to first order in metric perturbations) as well as transformations of the field-space coordinates $\varphi^I$. To first order in perturbations, $Q^I \rightarrow {\cal Q}^I \rightarrow \delta \phi^I$ in the spatially flat gauge.

For an arbitrary vector in the field space, $A^I$, we may define the usual covariant derivative with respect to the field-space metric,
\beq
{\cal D}_J A^I = \partial_J A^I + \Gamma^I_{\> JK} A^K ,
\label{covDJ}
\eeq
and a (covariant) directional derivative with respect to the affine parameter, cosmic time, $t$,
\beq
{\cal D}_t A^I \equiv \dot{\varphi}^J {\cal D}_J A^I = \dot{A}^I + \Gamma^I_{\> JK} \dot{\varphi}^J A^K ,
\label{covDt}
\eeq
where overdots denote partial derivatives with respect to $t$. To background order, we may then write the equation of motion for the fields $\varphi^I$ from Eq. (\ref{eomphi}),
\beq
{\cal D}_t \dot{\varphi}^I + 3 H \dot{\varphi}^I + {\cal G}^{IJ} V_{, J} = 0 ,
\label{eomvarphi}
\eeq
while Eqs. (\ref{EFE})-(\ref{Tmn}) yield the usual dynamical equations at background order,
\beq
\begin{split}
H^2 &= \frac{1}{3 M_{\rm pl}^2} \left[ \frac{1}{2} {\cal G}_{IJ} \dot{\varphi}^I \dot{\varphi}^J + V (\varphi^I ) \right] , \\
\dot{H} &= - \frac{1}{2 M_{\rm pl}^2} {\cal G}_{IJ} \dot{\varphi}^I \dot{\varphi}^J .
\end{split}
\label{Friedmann}
\eeq
In Eqs. (\ref{eomvarphi})-(\ref{Friedmann}), $H \equiv \dot{a} / a$ is the Hubble parameter, and the field-space metric is evaluated at background order, ${\cal G}_{IJ} (\varphi^K)$.

To first order in $Q^I$, Eqs. (\ref{EFE})-(\ref{eomphi}) may be combined to yield the equation of motion for the gauge-invariant perturbations \cite{Langlois,KMS,SebastianRP}
\beq
{\cal D}_t^2 Q^I + 3 H {\cal D}_t Q^I + \left[ \frac{k^2}{a^2} \delta^I_{\> J} + {\cal M}^I_{\> J}  \right] Q^J = 0 ,
\label{eomQ}
\eeq
where the mass-squared tensor takes the form
\beq
{\cal M}^I_{\> J} \equiv {\cal G}^{IK} \left( {\cal D}_J {\cal D}_K V \right) - {\cal R}^I_{\> LMJ} \dot{\varphi}^L \dot{\varphi}^M - \frac{1}{ M_{\rm pl}^2 a^3} {\cal D}_t \left( \frac{ a^3}{H} \dot{\varphi}^I \dot{\varphi}_J \right)
\label{MIJ}
\eeq
and ${\cal R}^I_{\> LMJ}$ is the Riemann tensor for the field-space manifold. All expressions in Eqs. (\ref{eomQ}) and (\ref{MIJ}) involving ${\cal G}_{IJ}$, $V$, and their derivatives are evaluated at background order in the fields, $\varphi^I$. The term in Eq. (\ref{MIJ}) that is proportional to $1/M_{\rm pl}^2$ arises from the coupled metric perturbations.

\section{Couplings and Background Dynamics}
\label{BackgroundDynamics}

Renormalization of models with self-coupled scalar fields in curved spacetime requires counter-terms of the form $\xi \phi^2 R$ for each nonminimally coupled field \cite{Callan,Bunch,BirrellDavies,Buchbinder,ParkerToms,Odintsov,Markkanen}. Here we consider a two-field model, $\phi^I = \{\phi, \chi\}^T$, and take $f (\phi^I)$ to be of the form
\beq
f (\phi, \chi ) = \frac{1}{2} \left[ M_{\rm pl}^2 + \xi_\phi \phi^2 + \xi_\chi \chi^2 \right] .
\label{f2field}
\eeq
Each scalar field $\phi^I$ couples to the Ricci scalar with its own nonminimal-coupling constant, $\xi_I$; conformal coupling corresponds to $\xi_I = - 1/6$. The field-space metric, ${\cal G}_{IJ} (\varphi^K)$, is determined by the form of $f (\phi^I)$ and its derivatives, as in Eq. (\ref{GIJ}). Explicit expressions for ${\cal G}_{IJ}$ and related quantities for this model may be found in Appendix A.

We consider a simple, renormalizable form for the potential in the Jordan frame,
\beq
\tilde{V} (\phi, \chi) = \frac{\lambda_\phi}{4} \phi^4 + \frac{g}{2}  \phi^2 \chi^2 + \frac{ \lambda_\chi}{4} \chi^4 .
\label{VJphichi}
\eeq
We take $\lambda_I  > 0$ and neglect bare masses $m_I^2$, in order to focus on effects from the quartic self-couplings and direct interaction terms within a parameter space of manageable size. The effects from nonzero $m_I^2$ may be incorporated using the methods developed here.

Several types of considerations may be used to bound the range of $\xi_I$ of interest. Perhaps most fundamentally, vacuum stability (under renormalization-group flow) requires $\xi_I \geq - 0.03$ \cite{MossXi}. Meanwhile, earlier studies of single-field models had found that $\vert \xi \vert \leq 10^{-3}$ for $\xi < 0$ in order to yield sufficient inflation \cite{Futamase,Salopek,Fakir,Makino,DKnGET}. These constraints leave a very narrow window of parameter space for $\xi_I < 0$ that could still be viable. Moreover, as we will see below, the behavior of such models with $\vert \xi_I \vert \ll 1$ tends to show only modest departures from the well-studied minimally coupled case, whereas qualitatively new behavior arises for $\vert \xi_I \vert \gg 1$. Hence we restrict attention here to positive couplings, $\xi_I > 0$.

Next we may consider observational constraints, such as the present bound on the primordial tensor-to-scalar ratio, $r \leq 0.1$ \cite{PlanckBICEP}, which corresponds to the bound $H_* \leq 3.4 \times 10^{-5} \> M_{\rm pl}$. (Asterisks indicate values of quantities at the time during inflation when observationally relevant perturbations first crossed outside the Hubble radius.) Models in our class predict \cite{KMS,GKS,KS,SSK} 
\beq
r = \frac{16 \epsilon}{1 + T_{\cal RS}^2 } ,
\label{rgeneral}
\eeq
where $\epsilon$ is the usual slow-roll parameter,
\beq
\epsilon \equiv - \frac{ \dot{H} }{H^2} ,
\label{epsilon}
\eeq
and $T_{\cal RS}^2$ is the transfer function for long-wavelength modes between the adiabatic $({\cal R})$ and isocurvature $({\cal S})$ directions. As analyzed in Refs.~\cite{KMS,GKS,KS,SSK} and discussed further in the next subsection, models in this class generically display strong single-field attractor behavior. Within an attractor the background fields' trajectory does not turn, and hence $T_{\cal RS}^2 \rightarrow 0$. Furthermore, given our covariant framework, we may consider the case in which the fields move along the direction $\chi = 0$ during inflation without loss of generality. In the limit $\xi_\phi \gg 1$, we find to good approximation \cite{KS}
\beq
H_* \simeq \sqrt{ \frac{\lambda_\phi}{12 \xi_\phi^2 } } \> M_{\rm pl}, \>\> N_* \simeq \frac{3}{4} \delta_*^2 ,\>\>
\epsilon \simeq \frac{3}{4 N_*^2} ,
\label{Nstarbigxi}
\eeq
where
\beq
\delta^2 \equiv \frac{\xi_\phi \phi^2}{M_{\rm pl }^2 } ,
\label{delta}
\eeq
and $N_*$ is the number of efolds before the end of inflation when relevant scales crossed outside the Hubble radius. (See also Ref.~\cite{Bezrukov13}.) Assuming $50 \leq N_* \leq 60$, we find $r \sim {\cal O} (10^{-3})$ in the limit $\xi_\phi \gg 1$, and $H_* \leq 3.4 \times 10^{-5} \> M_{\rm pl}$ for $\lambda_\phi / \xi_\phi^2 \leq 1.4 \times 10^{-8}$. In models like Higgs inflation \cite{BezrukovShaposhnikov}, one typically finds $\lambda_\phi \sim {\cal O} ( 10^{-2} - 10^{-4} )$ at the energy scales of inflation (the range stemming from uncertainty in the value of the top-quark mass, which affects the running of $\lambda_\phi$ under renormalization-group flow) \cite{Barvinsky,BezrukovRunning,Allison}. The range of $\lambda_\phi$, in turn, requires $\xi_\phi \sim {\cal O} ( 10^2 - 10^3 )$ at high energies  --- a reasonable range, given that $\xi_\phi$ typically rises with energy scale under renormalization-group flow with no UV fixed point \cite{Buchbinder}. Even for such large values of $\xi_I$, the inflationary dynamics occur at energy scales well below any nontrivial unitarity cut-off scale. (See Ref.~\cite{SSK} and references therein for further discussion.) 

For the opposite limit, with $0 < \xi_\phi \ll 1$, a similar analysis yields \cite{DKnGET,Bezrukov13} 
\beq
H_* \simeq \sqrt{\frac{\lambda_\phi}{12 \xi_\phi^2} } \frac{ \delta_*^4}{ (1 + \delta_*^2 )^2 }  M_{\rm pl}, \>\> N_* \simeq \frac{1}{8 \xi_\phi} \delta_*^2, \>\> \epsilon \simeq \frac{1}{ N_* (1 + 8 \xi_\phi N_* )} ,
\label{Nstarsmallxi}
\eeq
where $\delta^2$ is again defined as in Eq. (\ref{delta}). In this limit, the bound $r \leq 0.1$ requires $\xi_\phi \geq 0.006$ (for $N_* = 50$) or $\xi_\phi \geq 0.004$ (for $N_* = 60$), which in turn yields a constraint on $\lambda_\phi$ typical of minimally coupled models: $\lambda_\phi \sim {\cal O} (10^{-12})$ in order to keep $H_* \leq 3.4 \times 10^{-5} \> M_{\rm pl}$ \cite{Freese,DKconstraints}. Thus in the remainder of this analysis, we focus our attention to the range $10^{-3} \leq \xi_I \leq 10^4$.

\subsection{Single-Field Attractor}
\label{AttractorBehavior}

\begin{figure}
\centering
\includegraphics[width=0.48\textwidth]{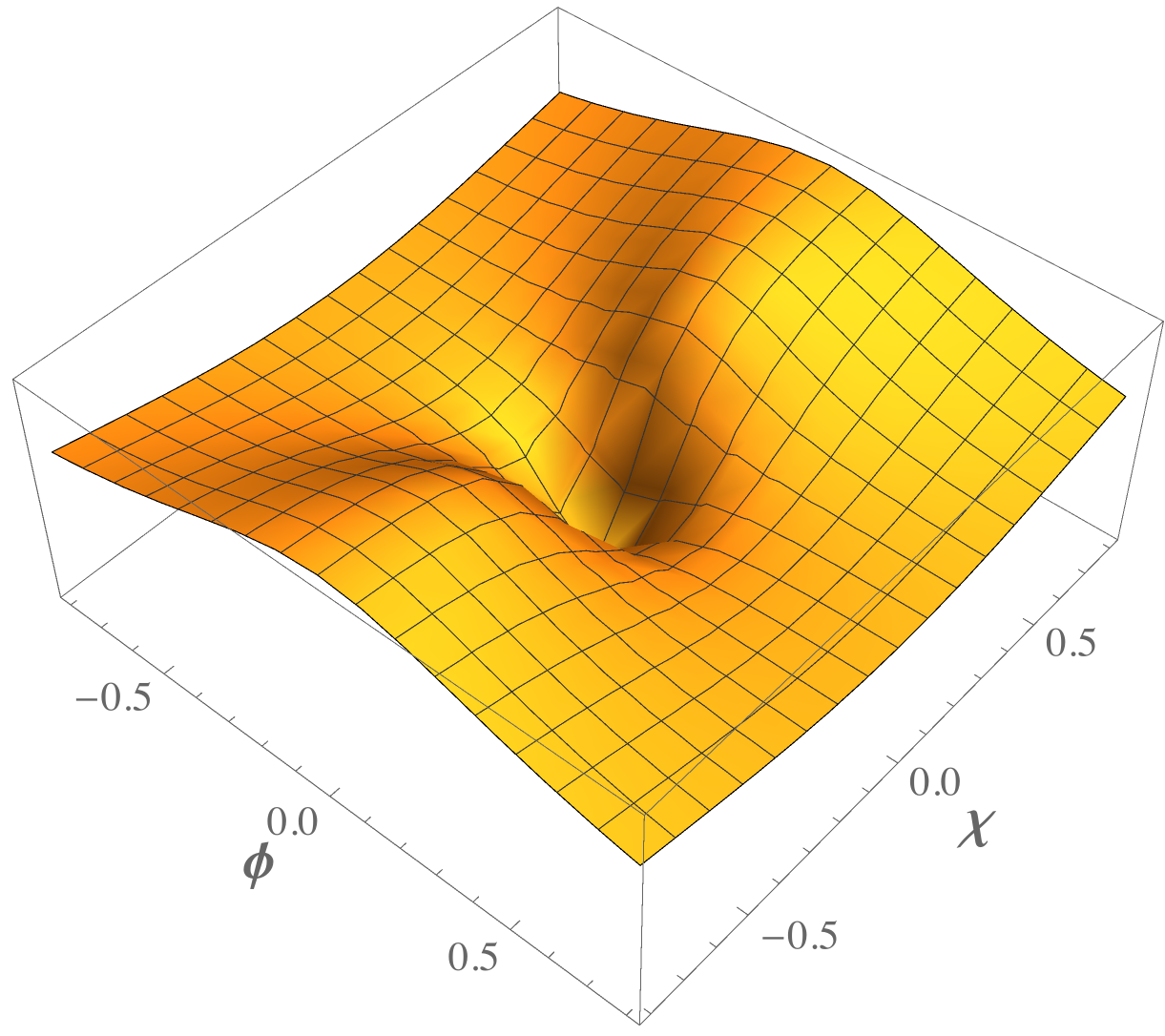} \quad \includegraphics[width=0.48\textwidth]{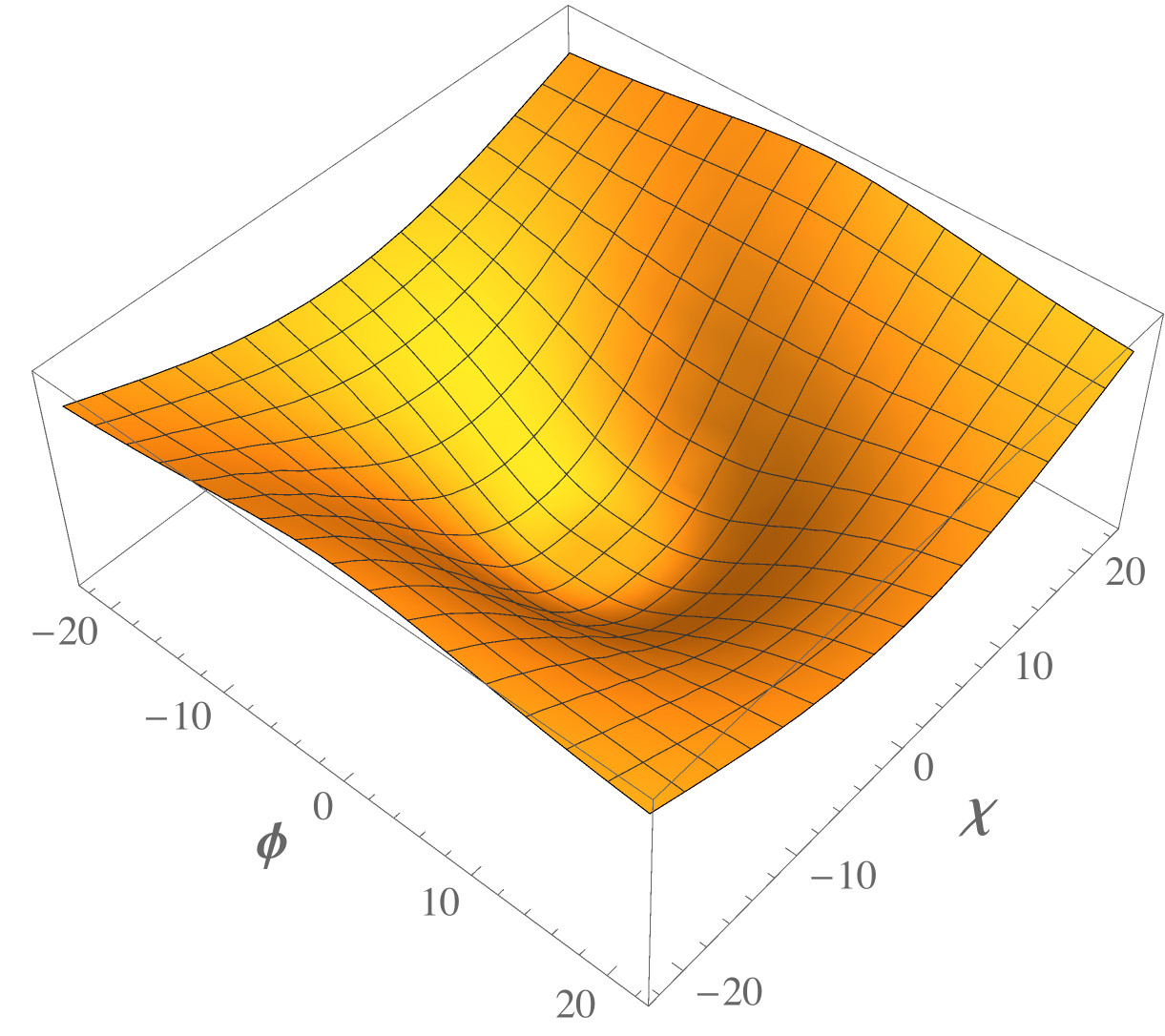}
\caption{ \small \baselineskip 11pt Potential in the Einstein frame, $V (\phi^I)$, for a two-field model with $\xi_\chi / \xi_\phi = 0.8, \lambda_\chi / \lambda_\phi = 1.25 $, and $g / \lambda_\phi  = 1$, for $\xi_\phi  =10^2$ ({\it left}) and $\xi_\phi = 10^{-2}$ ({\it right} ). Field values are in units of $M_{\rm pl}$. }
\label{VEfig}
\end{figure}

Inflation begins in a regime in which $\xi_J (\phi^J )^2 > M_{\rm pl}^2$ for at least one component, $J$. The potential in the Einstein frame becomes asymptotically flat along each direction of field space, as each field $\phi^I$ becomes arbitrarily large:
\beq
V (\phi^I) \rightarrow \frac{M_{\rm pl}^4}{4} \frac{\lambda_I}{\xi_I^2} \left[ 1 + {\cal O} \left( \frac{ M_{\rm pl}^2}{\xi_I (\phi^I)^2 } \right) \right]
\label{Vasympt}
\eeq
(no sum on $I$). Unless some explicit symmetry constrains all coupling constants in the model to be identical ($\lambda_\phi = g = \lambda_\chi$, $\xi_\phi = \xi_\chi$), then the potential in the Einstein frame will develop ridges and valleys. Both the ridges and the valleys satisfy $V > 0$, and hence the system will inflate (albeit at different rates) whether the fields evolve along a ridge or a valley toward the global minimum of the potential. As seen in Fig. \ref{VEfig}, even in the case of $\xi_I\ll1$, in which inflation can occur for field values $\phi^I$ for which the potential has not reached its asymptotically flat form, the potential still exhibits ridges and valleys, all of which are capable of supporting inflation.

\begin{figure}
\centering
\includegraphics[width=0.48\textwidth]{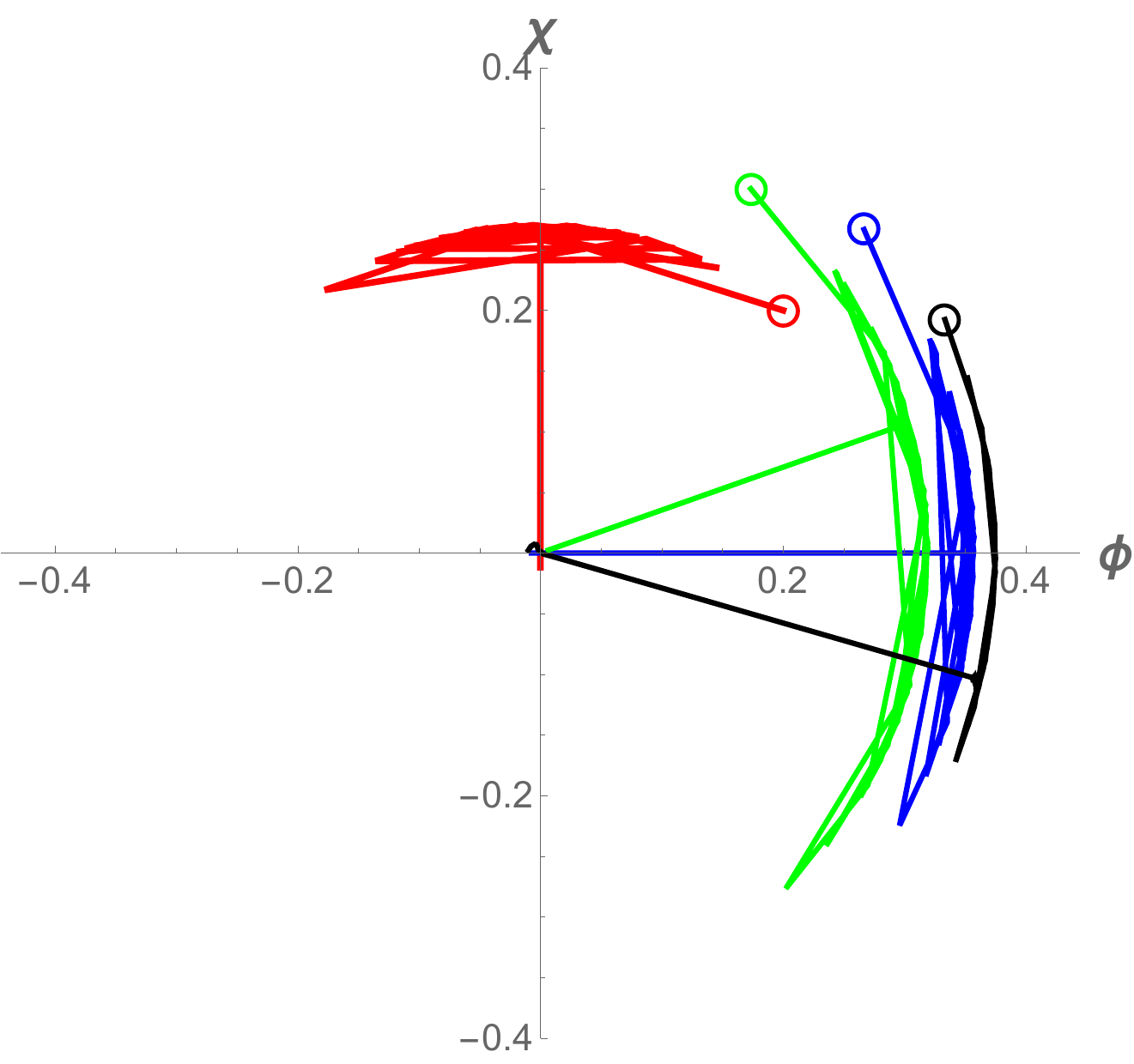} \quad \includegraphics[width=0.48 \textwidth]{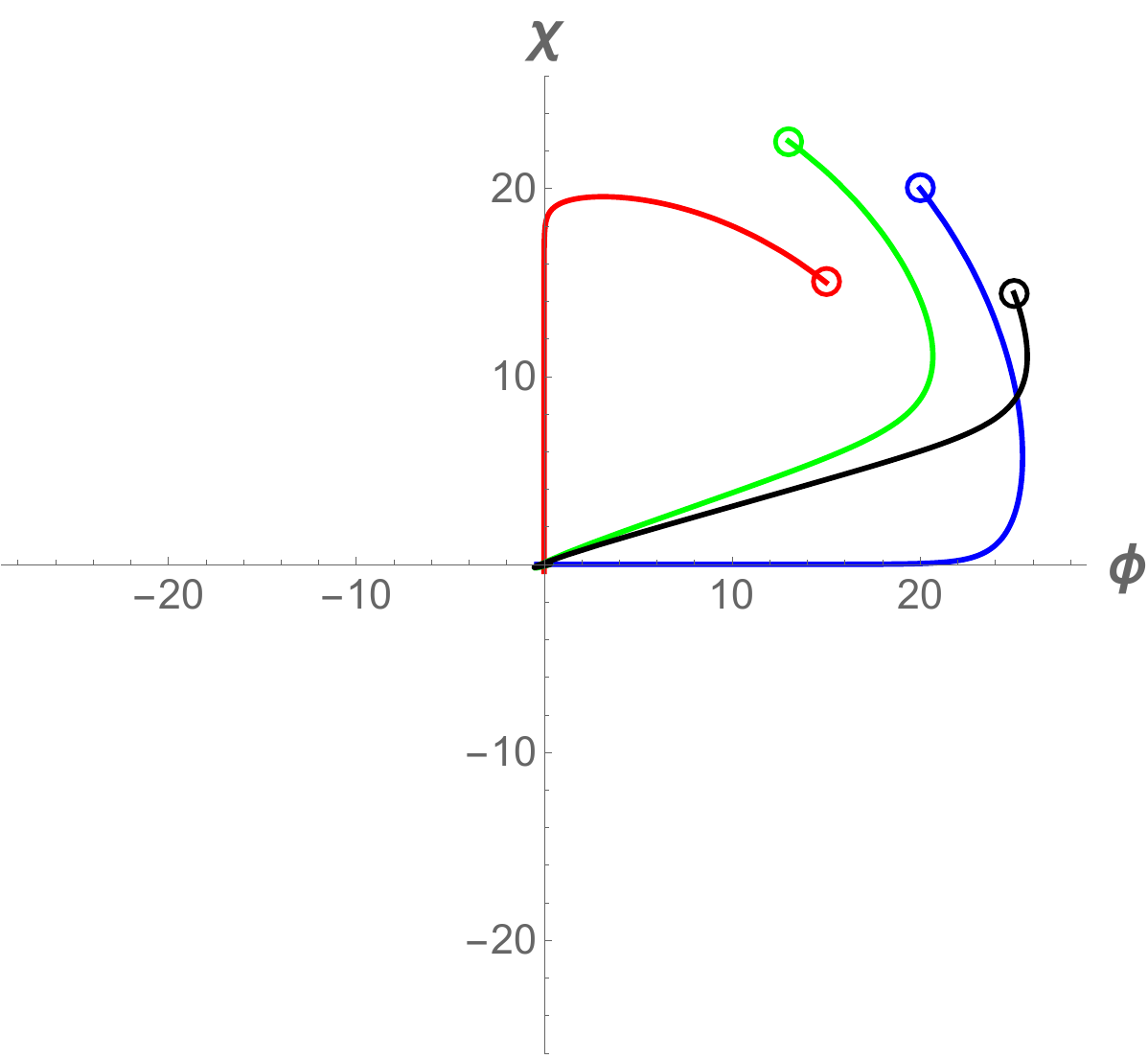}
\caption{ \small \baselineskip 11pt Field trajectories for different couplings and initial conditions. Open circles indicate fields' initial values (in units of $M_{\rm pl}$). We set the fields' initial velocities to zero and vary the initial angle in field space, $\theta_0 \equiv {\rm arctan} (\chi_0 / \phi_0)$. For the figure on the left, we set $\xi_\phi = 10^3$ and $\lambda_\phi = 10^{-2}$; for the figure on the right, we set $\xi_\phi = 10^{-1}$ and $\lambda_\phi = 10^{-10}$. In both figures, the other parameters $\{ \xi_\chi , \lambda_\chi, g, \theta_0 \}$ are: $\{ 1.2 \xi_\phi , 0.75 \lambda_\phi, \lambda_\phi, \pi / 4 \}$ (red); $\{ 0.8 \xi_\phi , \lambda_\phi, \lambda_\phi , \pi / 4 \}$ (blue); $\{ 0.8 \xi_\phi, \lambda_\phi, 0.75 \lambda_\phi , \pi / 3 \}$ (green); $\{ 0.8 \xi_\phi , 1.2 \lambda_\phi, 0.75 \lambda_\phi , \pi / 6 \}$ (black). In each case, the initial transient motion damps out within a few efolds, yielding effectively single-field evolution for (at least) the final 65 efolds of inflation. Moreover, as demonstrated in Refs.~\cite{GKS,KS}, large field velocities at the start of inflation redshift away very quickly and do not significantly alter the single-field attractor behavior during inflation. Such large initial field velocities therefore have no impact on conditions at the start of preheating.  }
\label{SFA}
\end{figure}

Given the distinct ridge-valley structure of the effective potential in the Einstein frame, these models display strong single-field attractor behavior during inflation, across a wide range of couplings and initial conditions \cite{KS}. If the fields happen to begin evolving along the top of a ridge, they will eventually fall into a neighboring valley at a rate that depends on the local curvature of the potential \cite{KMS, SSK}. Once the fields fall into a valley, Hubble drag quickly damps out any transverse motions in field space within a few efolds, after which the system evolves with virtually no turning in field space for the remainder of inflation \cite{KMS,GKS,KS,SSK}. As shown in Fig. \ref{SFA}, the single-field attractor behavior is as generic in the limit $\xi_I < 1$ as it is for $\xi_I \gg 1$. For all of the trajectories shown, the fields settle into a single-field attractor prior to the last 65 efolds of inflation.

\begin{figure}
\centering
\includegraphics[width=0.48 \textwidth]{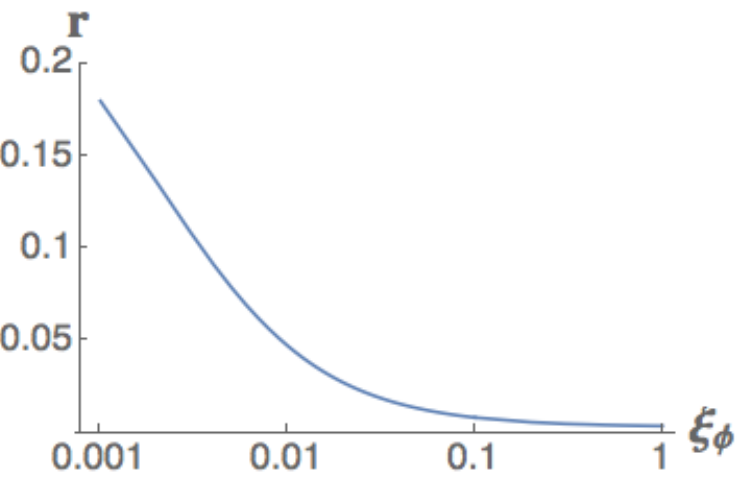} \quad \includegraphics[width=0.48 \textwidth ]{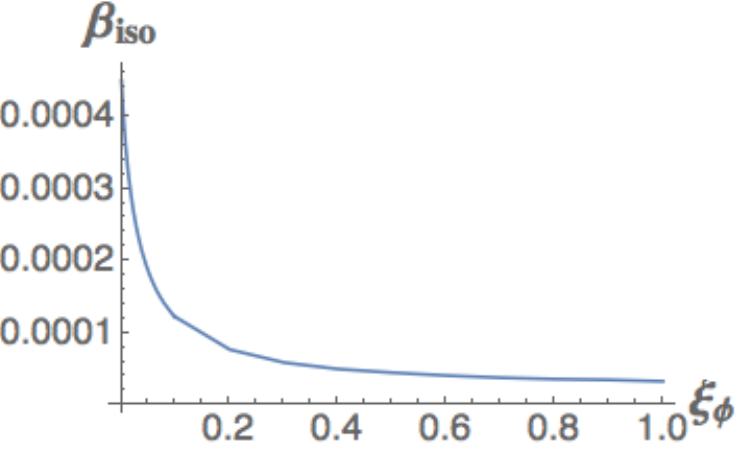}  
\caption{ \small \baselineskip 11pt The tensor-to-scalar ratio ({\it left}) and the fraction of isocurvature modes ({\it right}) as a function of the nonminimal coupling $\xi_\phi$. The isocurvature fraction is calculated for the symmetric (Higgs-like) case $\lambda_\phi=g=\lambda_\chi$ and $\xi_\phi=\xi_\chi$. }
\label{fig:observables}
\end{figure}

Within a single-field attractor, these models predict values for spectral observables such as the primordial spectral index and its running ($n_s$ and $\alpha$), the ratio of power in tensor to scalar modes ($r$), primordial non-Gaussianity ($f_{\rm NL}$), and the fraction of power in isocurvature rather than adiabatic scalar modes ($\beta_{\rm iso}$) all in excellent agreement with the latest observations \cite{KMS,GKS,KS,SSK}. Fig. \ref{fig:observables} shows the tensor-to-scalar ratio $r$ and the isocurvature fraction $\beta_{\rm iso}$ as a function of the nonminimal coupling. The approach to a constant $\xi_I$-independent value for large $\xi_I$ is evident. The fields will only fail to settle into a single-field attractor during inflation if {\it both} the ratios of certain coupling constants {\it and} the fields' initial conditions are fine-tuned. If the fields happen to begin very close to the top of a ridge, for example, and if the local curvature of the potential in the vicinity of that ridge has been fine-tuned to be small (${\cal D}_{IJ} V/ H^2 \ll 1$), then the system can exhibit significant turning in field space late in inflation \cite{KMS,KS,SSK}. In such fine-tuned cases, the system's evolution during the last 65 efolds of inflation can amplify non-Gaussianities and isocurvature perturbations, which could potentially be observable \cite{KMS,SSK,SebastianRP}. 

In Ref.~\cite{SSK} we analyzed the geometric structure of the attractor in the limit $\xi_I \gg 1$; here we generalize that analysis for arbitrary positive $\xi_I$. As in Ref.~\cite{SSK}, we define convenient combinations of couplings,
\beq
\Lambda_\phi \equiv \lambda_\phi \xi_\chi - g \xi_\phi , \>\>\>\> \Lambda_\chi \equiv \lambda_\chi \xi_\phi - g \xi_\chi , \>\>\>\> \varepsilon \equiv \frac{ \xi_\phi - \xi_\chi}{\xi_\phi} ,
\label{Lambdadef}
\eeq
along with the new rescaled quantities
\beq
\tilde{\Lambda}_\phi \equiv \frac{ \Lambda_\phi}{\lambda_\phi \xi_\phi} = \frac{\xi_\chi}{\xi_\phi} - \frac{g}{\lambda_\phi} , \>\>\>\> \tilde{\Lambda}_\chi \equiv \frac{ \Lambda_\chi}{\lambda_\chi \xi_\chi} = \frac{ \xi_\phi}{\xi_\chi} - \frac{g}{\lambda_\chi} .
\label{tildeLambda}
\eeq
For arbitrary $\xi_I > 0$, we find
\beq
{\cal D}_{\chi \chi}  V \vert_{\chi = 0} = \frac{ \lambda_\phi  \phi^2}{ [1 + \delta^2]^3 [ 1 + (1 + 6 \xi_\phi) \delta^2 ] } \left[ - \tilde{\Lambda}_\phi (1 + 6 \xi_\phi) \left( \delta^2 + \delta^4 \right) - \left( \tilde{\Lambda}_\phi + \varepsilon \right) \delta^2 + \frac{ g}{\lambda_\phi } \right],
\label{DchichiV}
\eeq
where $\delta^2 \equiv \xi_\phi \phi^2 / M_{\rm pl}^2$ as in Eq. (\ref{delta}). In the limit $\xi_I \gg 1$, the quantity $\delta^2 \gg 1$ during inflation, and we find ${\cal D}_{ \chi\chi} V \vert_{\chi = 0} \propto - \Lambda_\phi$ \cite{SSK}. In that limit, whenever $\Lambda_\phi < 0$ the direction $\chi = 0$ remains a local minimum of the potential and the background dynamics will obey strong attractor behavior along the direction $\chi = 0$. For $\xi_I \ll 1$, on the other hand, $\delta^2 \gtrsim 2$ during inflation, as may be seen from the scaling relationships in Eq. (\ref{Nstarsmallxi}), and the orientation $\theta = {\rm arctan} (\chi / \phi)$ of the local minimum depends on the ellipticity, $\varepsilon$, and the ratio $g / \lambda_\phi$ in addition to the sign of $\Lambda_\phi$. Even in these cases, the existence of attractor solutions remains generic (as shown in Fig. \ref{SFA}), only the orientation of the attractor in field space changes. For $\xi_I \ll 1$ there are special regions of parameter space for the coupling values where the topography of the potential can change during inflation, meaning that a ridge can turn into a valley as the inflaton rolls. Depending on the curvature, a waterfall-type transition may occur \cite{hybrid}. 

The orientation of the valley of the potential in field space, $\theta = {\rm arctan} (\chi / \phi)$, depends on combinations of couplings $\lambda_I, g$, and $\xi_I$ \cite{SSK}. When studying inflationary dynamics in multifield models, one typically projects physical quantities into adiabatic and isocurvature directions based on the motion of the background fields, $\varphi^I$ \cite{BTW,GWBR,Nibbelink,WandsBartolo,SeeryLidsey,WandsReview,Tegmark}. For our two-field model, we may define the orthogonal unit vectors \cite{KMS,GKS,KS,SSK}
\beq
\hat{\sigma}^I \equiv \frac{ \dot{\varphi}^I }{\dot{\sigma} } , \>\>\> \hat{s}^I \equiv \frac{\omega^I}{\omega}
\label{hatsigmahats}
\eeq
in terms of the magnitude of the background fields' velocity, $\dot{\sigma}$, and their (covariant) turn-rate,
\beq
\dot{\sigma} \equiv \vert \dot{\varphi}^I \vert = \sqrt{ {\cal G}_{IJ} \dot{\varphi}^I \dot{\varphi}^J} , \>\>\> \omega^I \equiv {\cal D}_t \hat{\sigma}^I .
\label{sigma} 
\eeq
We may then project any field-space vector into adiabatic ($\sigma$) and isocurvature ($s$) components,
\beq
A_\sigma \equiv \hat{\sigma}_I A^I , \>\>\> A_s \equiv \hat{s}_I A^I .
\label{adiabaticprojection}
\eeq
Within a single-field attractor, $\omega^I \rightarrow 0$, so that a vector in field space that lies along the adiabatic direction at one time will continue to point along the adiabatic direction at later times. In that case, we may exploit the covariant nature of our framework to perform a rotation in field space, $\varphi^I \rightarrow \varphi^{I \prime }$, such that the valley of the potential lies along the direction $\chi' = 0$. Then the attractor will keep $\chi' \sim \dot{\chi}' \sim 0$, and only $\phi' (t)$ will evolve. With respect to the new field-space coordinates $\{  \phi' ,  \chi' \}$, the adiabatic direction points along $\phi'$ and the isocurvature direction along $ \chi'$. 

We may quantify the strength of the attractor by examining the amount of fine-tuning needed to evade it. We will concentrate on the large-$\xi_I$ regime, as it is enough to show the trend in the attractor's strength as a function of $\xi_I$. Following the analysis of Ref.~\cite{SSK} for the case where the fields $\varphi^I$ start exponentially close to the top of a ridge, we use the linearized equations of motion to study the strength of the attractor. Apart from the fine-tuned curvature of the ridge ($\tilde \Lambda_\phi$), the dynamics of the inflaton field, which is translated into the attractor strength, depend very sensitively on the initial proximity to the top of the ridge. One obvious way to parameterize proximity to the top of the ridge is with the angle in field space, $\theta$. The initial angle is $\theta_0 \approx \chi_0 / \phi_0$ for $\chi_0\ll \phi_0$.
Our criterion will be the following: for the same dimensionless ridge curvature $\tilde \Lambda_\phi$ and the same initial proximity to the ridge $\theta_0$, the strength of the attractor is defined through the number of efolds $N\le 60$ it takes for the inflaton field to develop a large angle in field space, $\theta \simeq 1$.

Following the linearized analysis of Ref.~\cite{SSK}, we take the dominant field $\phi$ to follow the single-field slow-roll solution, which is consistent to linear order in $\chi$
\beq
\dot \phi_{\rm SR} = -{ \sqrt{\lambda_\phi} M_{\rm pl}^3 \over 3 \sqrt 3 \xi_\phi^2 \phi} ,
\eeq
which can be trivially solved to give
\beq
\phi = \sqrt{\phi_0^2 - {4\over 3}{M_{\rm pl}^2 \over \xi_\phi} N} ,
\eeq 
where $\phi(N=0) = \phi_0$ at the start of inflation and we take the Hubble term to be constant during slow-roll,
\beq
H \simeq \sqrt {\lambda_\phi \over 12 \xi_\phi^2} \> M_{\rm pl} \, .
\eeq
The linearized equation of motion for the secondary field $\chi$, when starting near the top of a smooth ridge ($\theta_0\ll 1 ,\tilde \Lambda_\phi \ll1$), is
\beq
\ddot \chi + 3H\dot \chi - {\tilde \Lambda_\phi M_{\rm pl}^2 \over \xi_\phi} \chi \simeq 0 ,
\eeq
and the solution (for $H =$ constant) is
\beq
\chi(N) \simeq   \chi_0 \, \exp \left [ \left (      -{3 \over 2}  +\sqrt{ {9 \over 4}+{12 \tilde \Lambda_\phi   \xi_\phi }  } \right )  N\right ] .
\eeq
The evolution of the field-space angle $\theta$ follows immediately as
\beq
\theta(N)  ={\rm arctan} \left (  \theta_0 { \exp \left [ \left (      -{3\over 2} +\sqrt{ {9\over 4}+12\tilde \Lambda_\phi \xi_\phi } \right )  N\right ] \over  \sqrt{1 - {4\over 3}{ M_{\rm pl} ^2 \over \xi_\phi  \phi_0^2} N}}  \right ).
\label{thetaN}
\eeq
\begin{figure}
\centering
\includegraphics[width=0.48 \textwidth]{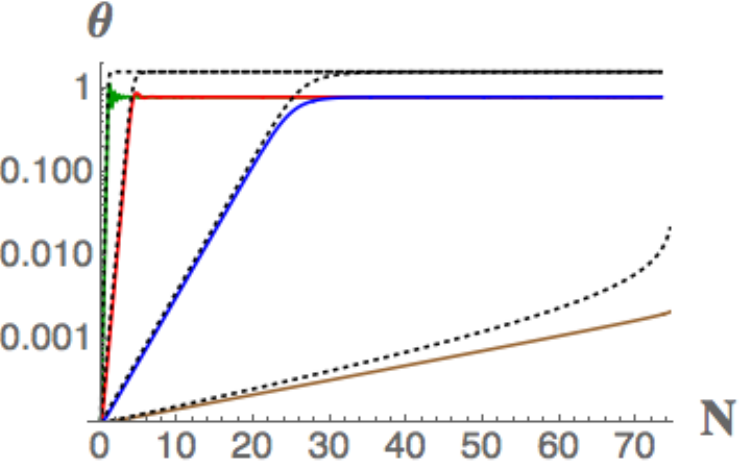} \quad \includegraphics[width=0.48\textwidth]{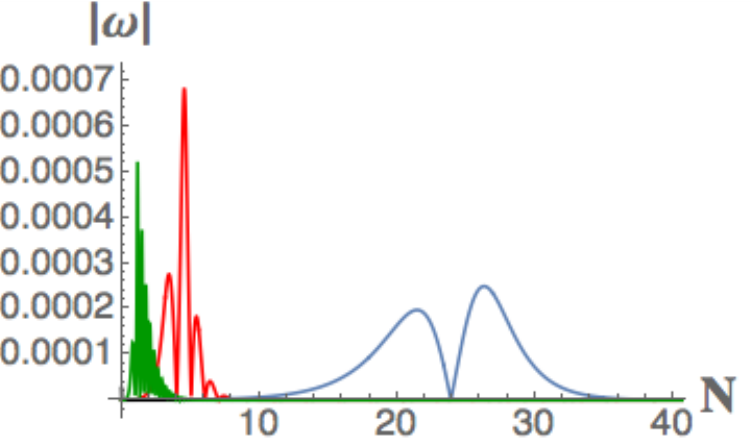}
\caption{ \small \baselineskip 11pt ({\it Left}) Evolution of the angle $\theta$ as a function of number of efolds from the beginning of inflation for $\tilde\Lambda_\phi = 0.001$, $\theta_0 = 10^{-4}$, $\xi_\chi = \xi_\phi$ and $\lambda_\phi=\lambda_\chi$. The values of the nonminimal coupling are $\xi_\phi = 10,10^2,10^3,10^4$ (brown, blue, red, and green respectively). The black dotted lines show the analytic results from Eq. (\ref{thetaN}). ({\it Right}) Evolution of the turn rate $|\omega|\equiv \left |\omega^I \right|$ as a function of the number of efolds from the beginning of inflation for the same parameters and color-coding. The turn-rate for $\xi_\phi=10$ is too small ($|\omega|\lesssim 10^{-7}$) to be visible on this plot.}
\label{fig:attractor_strength}
\end{figure}
As we can easily see from Fig. \ref{fig:attractor_strength}, for the same amount of fine-tuning of the couplings $\tilde \Lambda_\phi$ and initial position $\theta_0$, the attractor gets stronger as $\xi_\phi$ increases. We only consider this fine-tuned regime, since for $\tilde \Lambda = {\cal O}(1)$ or $\theta_0 = {\cal O}(1)$, the approach to the attractor is too fast for the extraction of any reasonable conclusion.
In Fig. \ref{fig:attractor_strength} we also plot the turn-rate $|\omega| \equiv \left | \omega^I \right |$ as a function of time. For $\xi_\phi=10$ and fine-tuned initial conditions, the attractor is too weak and the field remains on the ridge for the duration of the inflationary epoch, leading to a suppressed turn-rate $|\omega|\lesssim 10^{-7}$. For larger values of $\xi_\phi$ we see that the turn-rate spikes at the time when $\theta \simeq 1$, as expected. The turn-rate spikes earlier for larger couplings, indicating again a stronger attractor behavior. In the cases of $\xi_\phi=10^3,10^4$, the attractor is strong enough (meaning that the ridge is steep enough) that the field reaches the valley of the potential while having a significant velocity, which leads it to oscillate around the minimum before settling down to single-field motion. These oscillations perpendicular to the dominant motion of the inflaton can be seen as ``primordial clocks" with possibly interesting observational consequences \cite{ChenClocks}.

Eq.~(\ref{DchichiV}) shows that for asymptotically large field values ($\delta \gg 1$) the ridge-valley nature of the potential is only defined by the sign of $\tilde \Lambda_\phi$, whereas after inflation has ended and the fields have settled into an oscillation pattern close to their minimum, in the limit of $\delta \lesssim 1$, the nature of the extremum is defined by the sign of $g/\lambda_\phi$. There is of course a lot of parameter space between these two extremes, where for example the ellipticity $\varepsilon$ can significantly affect the potential curvature. We will disentangle these effects one-by-one.

\begin{figure}
\centering
\includegraphics[width=0.48\textwidth]{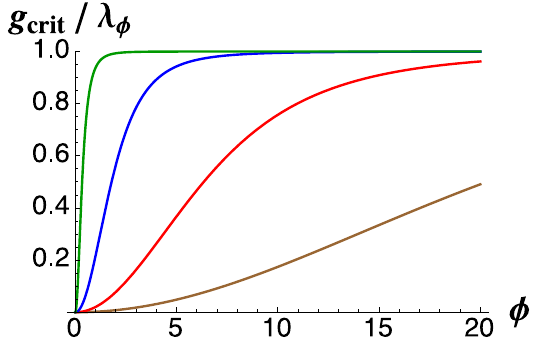} \quad \includegraphics[width=0.48\textwidth]{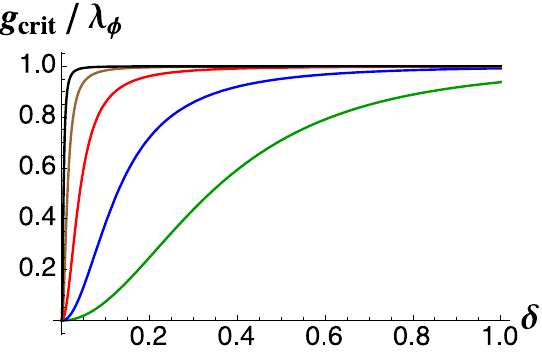} 
\caption{\small \baselineskip 11pt ({\it Left}) The value of $g/\lambda_\phi$ for which ${\cal D}_{\chi\chi} V \vert_{\chi=0}=0$ versus $\phi$ (in units of $M_{\rm pl}$) for $\xi_\phi = \xi_\chi$ and $\xi_\phi =10^{-3},10^{-2},10^{-1},1$ (from bottom to top). ({\it Right}) The same quantity versus $\delta = \sqrt{\xi_\phi} \> \phi / M_{\rm pl}$ for $\xi_\phi =1,10,100,10^3,10^4$ (from bottom to top).
For values of $g / \lambda_\phi$ above each curve, the potential exhibits a valley along $\chi=0$.}
\label{fig:g_xi_phi_valley}
\end{figure}

We start with the case of zero ellipticity, $\varepsilon=0$, or $\xi_\phi = \xi_\chi$, which corresponds to $\tilde \Lambda_\phi = 1 - (g/\lambda_\phi)$. Fig.\ \ref{fig:g_xi_phi_valley} shows how the nature of the extremum at $\chi=0$ varies with all relevant parameters, $ g/\lambda_\phi$, $\xi_\phi$ and $\phi$. A field rolling along an attractor remains along this attractor throughout inflation and preheating. Furthermore, for $\xi_\phi \gtrsim 1$, the condition $g/\lambda_\phi >1$ for the existence of an attractor remains quite accurate.  For smaller $\xi_\phi$, we see that even smaller values of $g/\lambda_\phi $ can provide an attractor along $\chi=0$. Even more interestingly, there are cases in which the extremum can change its nature during inflation. For example, for $\xi_\phi = 10^{-3}$ and $g/\lambda_\phi = 0.2$, we see that the direction $\chi=0$ switches from a ridge to a valley around $\phi \approx 12$ (in units of $M_{\rm pl}$). %

\begin{figure}
\centering
\includegraphics[width=0.48\textwidth]{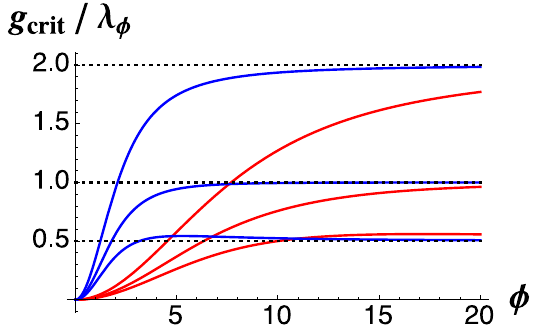} 
\caption{\small \baselineskip 11pt The value of $g/\lambda_\phi$ for which ${\cal D}_{\chi\chi}V \vert_{\chi=0}=0$ versus $\phi$ (in units of $M_{\rm pl}$) for $\xi_\phi=10^{-1}$ (blue) and $\xi_\phi = 10^{-2}$ (red). For each value of $\xi_\phi$, the ellipticity varies as $\varepsilon = -0.5,1,2$ (from top to bottom). For values of $g / \lambda_\phi$ above each curve, the potential exhibits a valley along $\chi=0$.}
\label{fig:g_xi_phi_epsilon_valley}
\end{figure}

Next we consider the effect of an arbitrary ellipticity $\varepsilon \ne 0$. For simplicity, we choose two values of the ellipticity with opposite sign, $\varepsilon = 0.5$ and $\varepsilon = -1$, and compare them to the previous case $\varepsilon =0$. The results are shown in Fig.\ \ref{fig:g_xi_phi_epsilon_valley}. As expected, the values of $g/\lambda_\phi$ are shifted according to the ellipticity, since we can re-write the parameter $\tilde \Lambda_\phi$ as
\beq
\tilde \Lambda_\phi = 1-\varepsilon - {g\over \lambda_\phi} \, . 
\eeq
This means that in the limit where $\tilde \Lambda_\phi$ defines the nature of the extremum (for large $\delta$), the extremum is a minimum for ${g/ \lambda_\phi}>1-\varepsilon $. An interesting phenomenon occurs for positive ellipticity and $g/\lambda_\phi \gtrsim 1-\varepsilon$. In this case, the critical value of $g/\lambda_\phi$ is a non-monotonic function of $\phi$. This means that for a value of $g / \lambda_\phi$ slightly above the critical value, the valley, in which the field is rolling, can turn into a ridge and then into a valley again. This can trigger some genuinely multifield behavior, such as a waterfall transition, similar to hybrid inflation. Density perturbations during a waterfall transition require specialized treatment, due to the lack of a classical field trajectory around which to perturb, and can have interesting observational consequences such as seeding primordial black holes \cite{hybrid}. However, in the context of the family of models that we consider here, such waterfall transitions are rather fine-tuned cases, and we will not pursue them further.

In sum, these models include five coupling constants: $\lambda_\phi, \lambda_\chi, g, \xi_\phi, \xi_\chi$. The Hubble scale during inflation is fixed by the combination $\lambda_\phi/\xi_\phi^2 \simeq 12 H^2/M_{\rm Pl}^2$ (assuming the field is rolling along $\chi=0$). We may reorganize the couplings in terms of the three nontrivial combinations $\Lambda_\phi, \Lambda_\chi, \varepsilon$, introduced in Eq.~(\ref{Lambdadef}). Except for exponentially fine-tuned cases --- fine-tuned in both parameter space and the fields' initial conditions --- the predictions for CMB observables from these models follow the Starobinsky attractor for $\xi_\phi \gtrsim 10$ and essentially any values of the remaining parameter combinations, $\Lambda_\phi, \Lambda_\chi, \varepsilon$, as discussed in detail in Refs.~\cite{KS, SSK, KMS}.

\begin{figure}
\centering
\includegraphics[width=0.6 \textwidth]{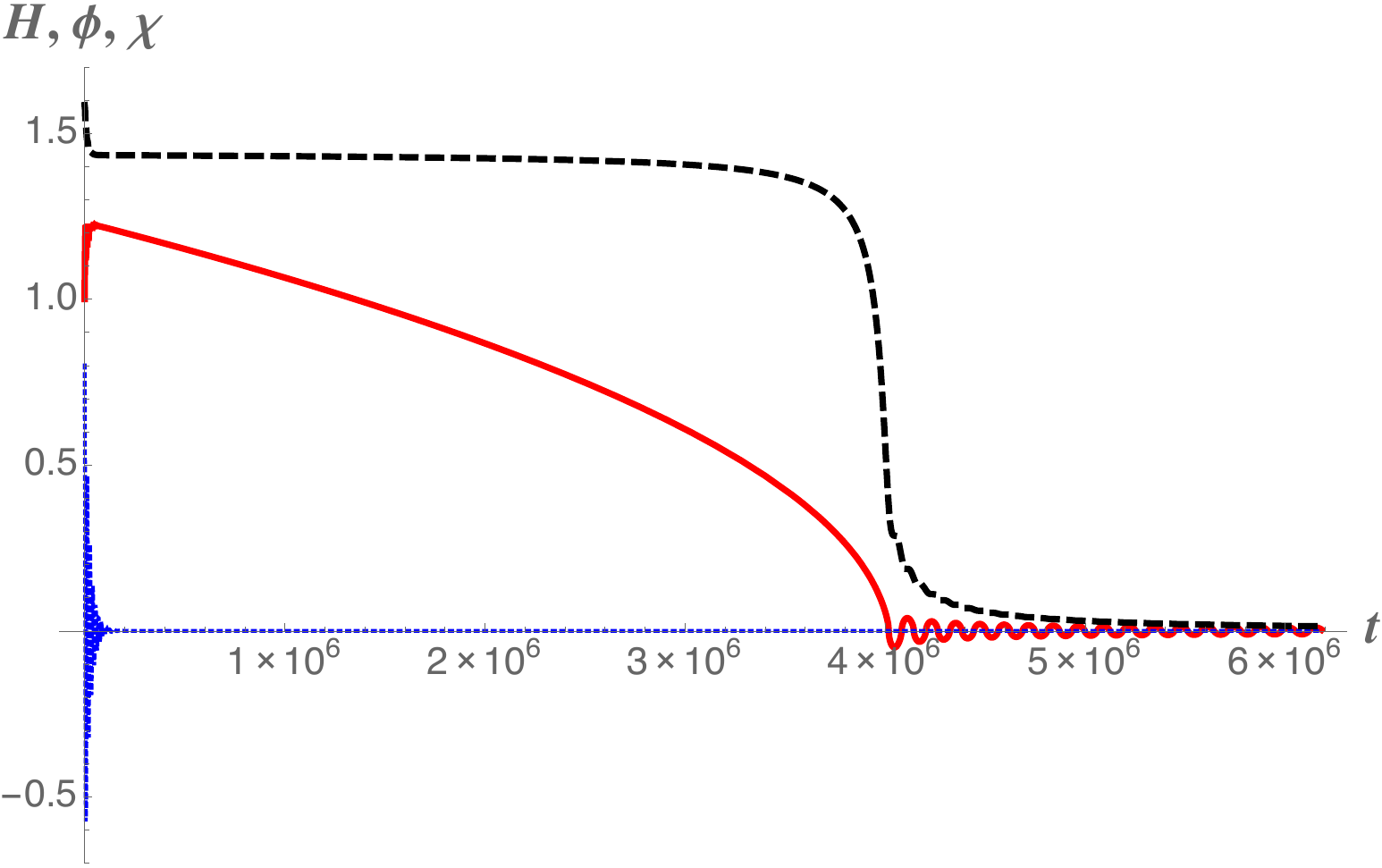} 
\caption{ \small \baselineskip 11pt The evolution of $H(t)$ (black dashed line), $\phi (t)$ (red solid line), and $\chi (t)$ (blue dotted line) during and after inflation, in units of $M_{\rm pl}$. The evolution shown here is for $\xi_\chi = 0.8 \> \xi_\phi, \lambda_\chi = 1.25\> \lambda_\phi$, and $g = \lambda_\phi$, with $\xi_\phi = 10^2$, $\lambda_\phi = 10^{-4}$, and initial conditions $\phi (t_0) = 1$, $\chi (t_0) = 0.8$, $\dot{\phi} (t_0) = \dot{\chi} (t_0) = 0$. (We plot $5 \times 10^4 \> H$ so its magnitude is comparable to $\phi$.) With these parameters and initial conditions, inflation lasts for $N_{\rm tot} = 111.6$ efolds until $t_{\rm end} = 3.99 \times 10^6$. The system rapidly falls into a valley along $\chi = 0$ within the first 3 efolds of inflation, after which $\chi (t)$ remains fixed at $\chi \sim 0$. After $t_{\rm end}$, $\phi (t)$ oscillates around the global minimum of the potential. }
\label{phiosc}
\end{figure}

Inflation ends when the scale factor stops accelerating, $\ddot{a} (t_{\rm end}) = 0$, which is equivalent to $\epsilon (t_{\rm end}) = 1$. (As a reminder, $\epsilon \equiv - \dot{H} / H^2$ should not be confused with the ellipticity parameter, $\varepsilon \equiv (\xi_\phi - \xi_\chi) / \xi_\phi$.) After $t_{\rm end}$, the background fields $\varphi^I (t)$ oscillate around the global minimum of the potential, governed by Eq. (\ref{eomvarphi}). If (as is generic) the system settles into the single-field attractor before the end of inflation, then the motion of $\varphi^I (t)$ in the direction of the potential's valley remains suppressed even after inflation. For example, if the system evolves along a valley in the $\chi = 0$ direction during inflation, then $\chi \sim \dot{\chi} \sim 0$ at $t_{\rm end}$ and Eq. (\ref{eomvarphi}) will maintain $\chi \sim \dot{\chi} \sim 0$ for times $t > t_{\rm end}$, as shown in Fig. \ref{phiosc}. Such attractor behavior after $t_{\rm end}$ persists for at least as long as backreaction from perturbations may be neglected, consistent with the linearized treatment of Eq. (\ref{phivarphi}). Thus the strong attractor behavior that was identified in Refs.~\cite{KMS,GKS,KS,SSK} is characteristic of the preheating phase as well.  

The persistence of the attractor behavior after the end of inflation has important implications for preheating. In particular, although the unit vectors $\hat{\sigma}^I$ and $\hat{s}^I$ may become ill-defined when the motion of $\varphi^I (t)$ is no longer monotonic, the orientation of the attractor in field space, $\theta = {\rm arctan} (\chi / \phi)$, remains unchanged after inflation. Upon performing a rotation $\varphi^I \rightarrow \varphi^{I \prime}$ such that $\chi' = 0$ lies along the direction of the attractor, then only one field, $\phi' (t)$, oscillates after $t_{\rm end}$. With only one background field oscillating, there is no ``de-phasing" of the background fields' oscillations, as is typical for multifield models with minimal couplings \cite{AHKK,Barnaby,Battefeld}. As shown in Refs.~\cite{MultiPreheat2,MultiPreheat3}, these attractor models therefore predict robust, resonant amplification of fluctuations across wide regions of parameter space. 

Within a single-field attractor, both the field-space metric, ${\cal G}_{IJ}$, and the mass-squared tensor, ${\cal M}_{IJ}$ of Eq. (\ref{MIJ}), become effectively diagonal. Upon rotating $\varphi^I \rightarrow \varphi^{I \prime}$ as needed so that the attractor lies along the direction $\chi' = 0$, then ${\cal G}_{\phi' \chi'} \sim {\cal G}^{\phi' \chi'} \sim 0$ and ${\cal M}^{\phi'}_{\> \> \chi'} \sim {\cal M}^{\chi'}_{\>\> \phi'} \sim 0$. As we will see in Section \ref{EvolutionFluctuations}, this feature greatly simplifies the analysis of the fluctuations. Given that we may always perform such a field-space rotation, for most of the following analysis we restrict attention to cases in which the attractor lies along the direction $\chi = 0$, with no loss of generality. In Section \ref{Rotation} we demonstrate that our results remain robust even for cases in which the attractor lies along some other direction $\theta$ in field space.

\subsection{End of Inflation and Effective Equation of State}
\label{OscillationsEES}

\begin{figure}
\centering
\includegraphics[width=0.6\textwidth]{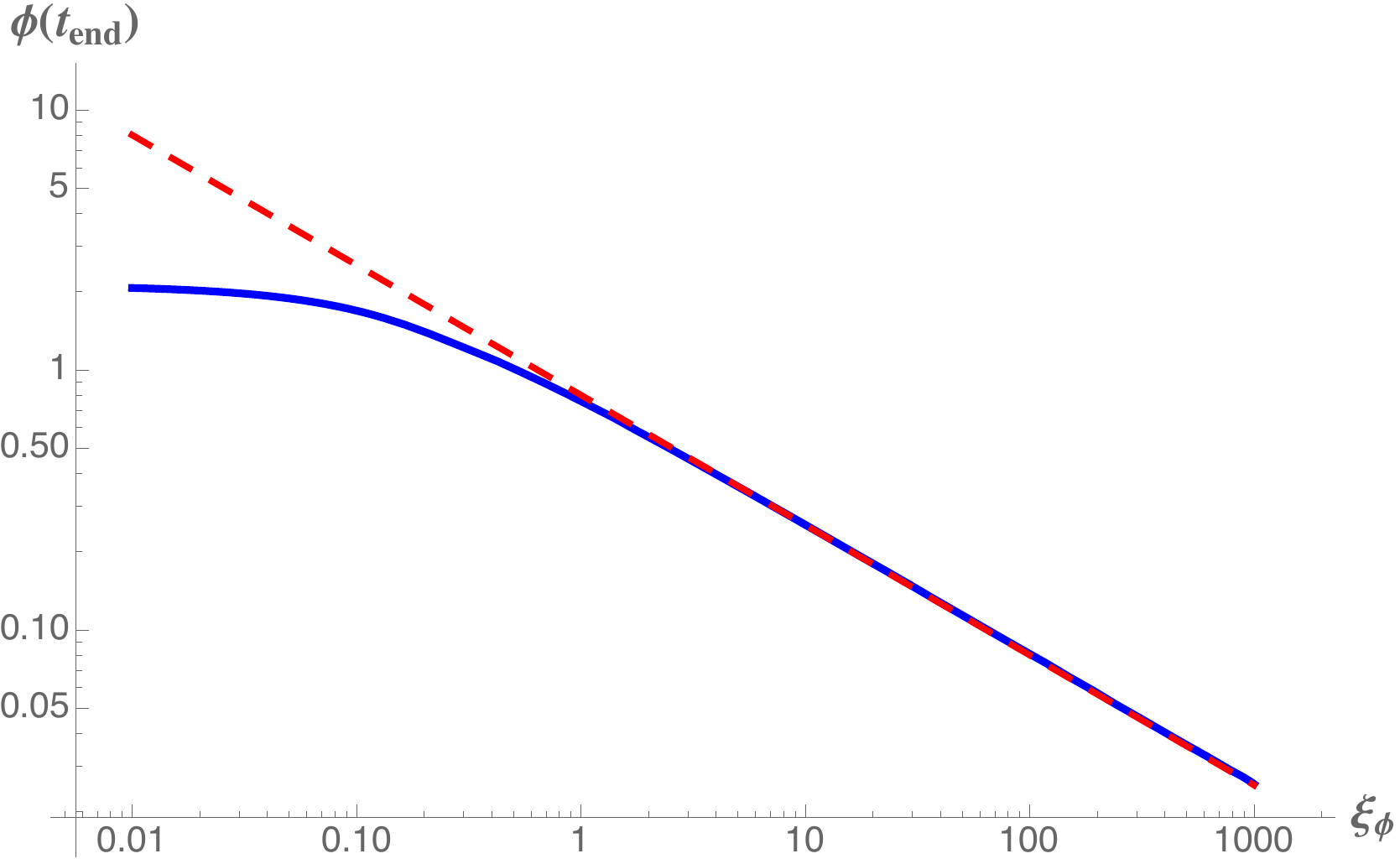} 
\caption{ \small \baselineskip 11pt Within the single-field attractor, the value of $\phi (t_{\rm end})$ depends only on $\xi_\phi$. The blue curve shows the numerical evaluation of $\phi (t_{\rm end})$ (in units of $M_{\rm pl}$), while the red dashed curve shows $0.8 / \sqrt{\xi_\phi}$. }
\label{phiendfig}
\end{figure}

Within the single-field attractor, we may readily study how $\phi (t_{\rm end})$ depends on the coupling constants. First we note that in the single-field attractor (assumed to lie along a $\chi = 0$ valley), the evolution of $\phi (t)$ becomes independent of $\lambda_\chi$, $g$, and $\xi_\chi$. Furthermore, we may rescale $t \rightarrow \tau \equiv \sqrt{\lambda_\phi} \> t$ without affecting the dynamics: $N = \int H dt = \int {\cal H} d\tau$ remains unchanged, as does $\epsilon = - {\cal H}^\prime / {\cal H}^2 = - \dot{H} / H^2$ (where ${\cal H} \equiv a' / a $ and primes denote $d / d\tau$). Therefore $\phi (\tau_{\rm end}) = \phi (t_{\rm end})$. Thus in the single-field attractor, the value of $\phi$ at the end of inflation depends only on $\xi_\phi$. In the limit $\xi_\phi \gg 1$, we expect inflation to end when $\xi_\phi \phi^2 (t_{\rm end}) \simeq M_{\rm pl}^2$, which is indeed the behavior we observe. As shown in Fig. \ref{phiendfig}, $\phi (t_{\rm end})$ is very well fit by $\phi (t_{\rm end}) = 0.8 \> M_{\rm pl} / \sqrt{\xi_\phi}$ for $\xi_\phi \geq 1$, whereas $\phi (t_{\rm end} ) \rightarrow 2.1 \> M_{\rm pl}$ in the limit $\xi_\phi \ll 1$, approaching the result of a minimally coupled $\phi^4$ model. The value $\phi (t_{\rm end})$ sets the initial amplitude of oscillations at the start of preheating. 

We may estimate the effective equation of state during the preheating phase by using the virial theorem \cite{TurnerVirial}. The total kinetic energy for the system (to background order) is \cite{KMS}
\beq
\frac{1}{2} \dot{\sigma}^2 \equiv \frac{1}{2} {\cal G}_{IJ} \dot{\varphi}^I \dot{\varphi}^J ,
\label{sigma}
\eeq
and the energy density and pressure are given by
\beq
\begin{split}
\rho &= \frac{1}{2} \dot{\sigma}^2 + V (\varphi^I ) , \\
p &= \frac{1}{2} \dot{\sigma}^2 - V (\varphi^I ) .
\end{split}
\label{rhopbackground}
\eeq
If we assume an equation of state of the form $p = w \rho$, then we find
\beq
w = \frac{ \dot{\sigma}^2 - 2 V }{\dot{\sigma}^2 + 2 V} 
\label{w}
\eeq
to background order. Using Eqs. (\ref{Friedmann}), (\ref{sigma}), and (\ref{rhopbackground}), we may rewrite Eq. (\ref{epsilon}) as $\epsilon = 3\dot{\sigma}^2 / (\dot{\sigma}^2 + 2 V)$. At $t_{\rm end}$, before the oscillations have begun, we have $\epsilon = 1$ and therefore $w = -1/3$, independent of couplings.

To estimate $w$ once the background fields begin to oscillate, we define a covariant expression for the virial, $q$, 
\beq
q \equiv {\cal G}_{IJ} \dot{\varphi}^I \varphi^J .
\label{qdef}
\eeq
Upon using $\partial {\cal G}_{IJ} / \partial t = ( \partial_K {\cal G}_{IJ} ) \dot{\varphi}^K$ and the usual relations among the Christoffel symbols $\Gamma^I_{\> JK}$, we find
\beq
\dot{q} = \dot{\sigma}^2 - V_{, J} \varphi^J + \frac{1}{2} \left( \partial_K {\cal G}_{IJ} \right) \dot{\varphi}^I \dot{\varphi}^J \varphi^K .
\label{virial1}
\eeq
Eq. (\ref{virial1}) is analogous to applications of the virial theorem in general relativity, in which corrections to the Newtonian result enter as gradients of the metric components \cite{Gourgoulhon}. For trajectories within the single-field attractor (with $\chi \sim \dot{\chi} \sim 0$), we have $\dot{\sigma}^2 \simeq {\cal G}_{\phi \phi} \dot{\phi}^2$ and Eq. (\ref{virial1}) becomes
\beq
\dot{q} \simeq \dot{\sigma}^2 \left[ 1 + \frac{1}{2} \phi \partial_\phi \ln {\cal G}_{\phi \phi} \right] - V_{, J} \varphi^J .
\label{virial2}
\eeq
From Eqs. (\ref{VE}) and (\ref{VJphichi}), we further find
\beq
V_{, J} \varphi^J = 2 M_{\rm pl}^2 \frac{ V}{f} ,
\label{gradV}
\eeq
where $f$ is the nonminimal-coupling function of Eq. (\ref{f2field}). Upon time-averaging over several oscillations we have $\langle \dot{q} \rangle = 0$, and hence 
\beq
\langle \dot \sigma^2 \rangle + {1\over 2} \langle \dot \sigma^2 \cdot  \phi \, \partial_\phi  \ln {\cal G}_{\phi\phi} \rangle = 2M_{\rm pl}^2 \langle {V/ f} \rangle ,
\eeq
where the second term on the left-hand side is the contribution of the stretched field-space manifold. 
The equation of state can be calculated by noting that energy conservation requires (if one neglects Hubble friction)
\beq
\dot \sigma^2 + 2V = 2V_{\rm max} ,
\eeq
which allows Eq.~\eqref{w} to be written solely in terms of $\phi$ and not $\dot \phi$.

\begin{figure}
\centering
\includegraphics[width=0.6\textwidth]{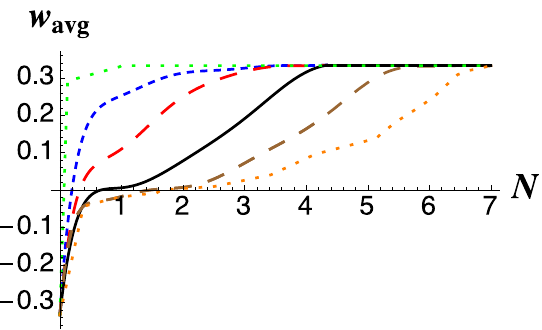}
\caption{\small \baselineskip 11pt The equation of state, $w$ from Eq. (\ref{w}), averaged over several oscillations of $\phi (t)$, as a function of efolds, $N$, after the end of inflation. From bottom to top: $\xi_\phi = 10^4$ (orange dotted line), $\xi_\phi = 10^3$ (brown dashed line), $\xi_\phi = 10^2$ (black line), $\xi_\phi = 10$ (red dashed line), $\xi_\phi = 1$ (blue short-dashed line), and $\xi_\phi = 0.1$ (green dotted line). All simulations used $\xi_\chi = 0.8 \> \xi_\phi$, $\lambda_\chi = 1.25 \> \lambda_\phi$, and $g = \lambda_\phi$. Initial conditions at the start of inflation were set as $\theta_0 = {\rm arctan} (\chi_0 / \phi_0) = \pi / 6$; in each case, the fields settled into the single-field attractor along $\chi \sim 0$ before the end of inflation. }
\label{wavgfig}
\end{figure}

After $t_{\rm end}$, $\phi (t)$ begins to oscillate with an initial amplitude $\phi (t_{\rm end}) \sim M_{\rm pl} / \sqrt{\xi_\phi}$ for $\xi_\phi \gtrsim 1$; at later times, its amplitude falls due to both the expansion of the universe and the transfer of energy to decay products. Fig. \ref{wavgfig} shows the equation of state $w_{\rm avg}$ calculated by solving the background evolution and averaging Eq.~\eqref{w} over several oscillations of $\phi(t)$, starting at the end of inflation, when $w=-1/3$. We see that for large nonminimal couplings, the equation of state spends more time around $w_{\rm avg} \approx 0$, as the universe continues to expand, while eventually reaching $w_{\rm avg} = 1/3$ at late times. Early in the oscillation phase, in other words, the conformal stretching of the Einstein-frame potential makes the background field behave more like a minimally coupled field in a quadratic potential, $V (\phi) = {1 \over 2} m^2 \phi^2$, than a quartic potential, $V (\phi) = {\lambda \over 4} \phi^4$. At late times, however, the system behaves like radiation, as in the minimally coupled case. Calculated to background order, $w_{\rm avg}$ reaches $1/3$ within several efolds after the end of inflation across the range $10^{-1} \leq \xi_\phi \leq 10^4$.

\subsection{Background-Field Oscillations}
\label{Oscillations}

To facilitate comparison with the well-studied case of a minimally coupled field with quartic self-coupling, in this subsection we neglect Hubble expansion during the oscillating phase. This approximation becomes more reliable as the frequency of oscillation $\omega$ grows significantly larger than $H$; in our case, we find a modest separation of time-scales, with $\omega / H > 1$ across a wide range of $\xi_\phi$. (One may incorporate effects from the expansion of the universe perturbatively \cite{KaiserPreh1}, though the $H \sim 0$ limit will suffice for our purposes here.) 

\begin{figure}
\centering
\includegraphics[width=0.48\textwidth]{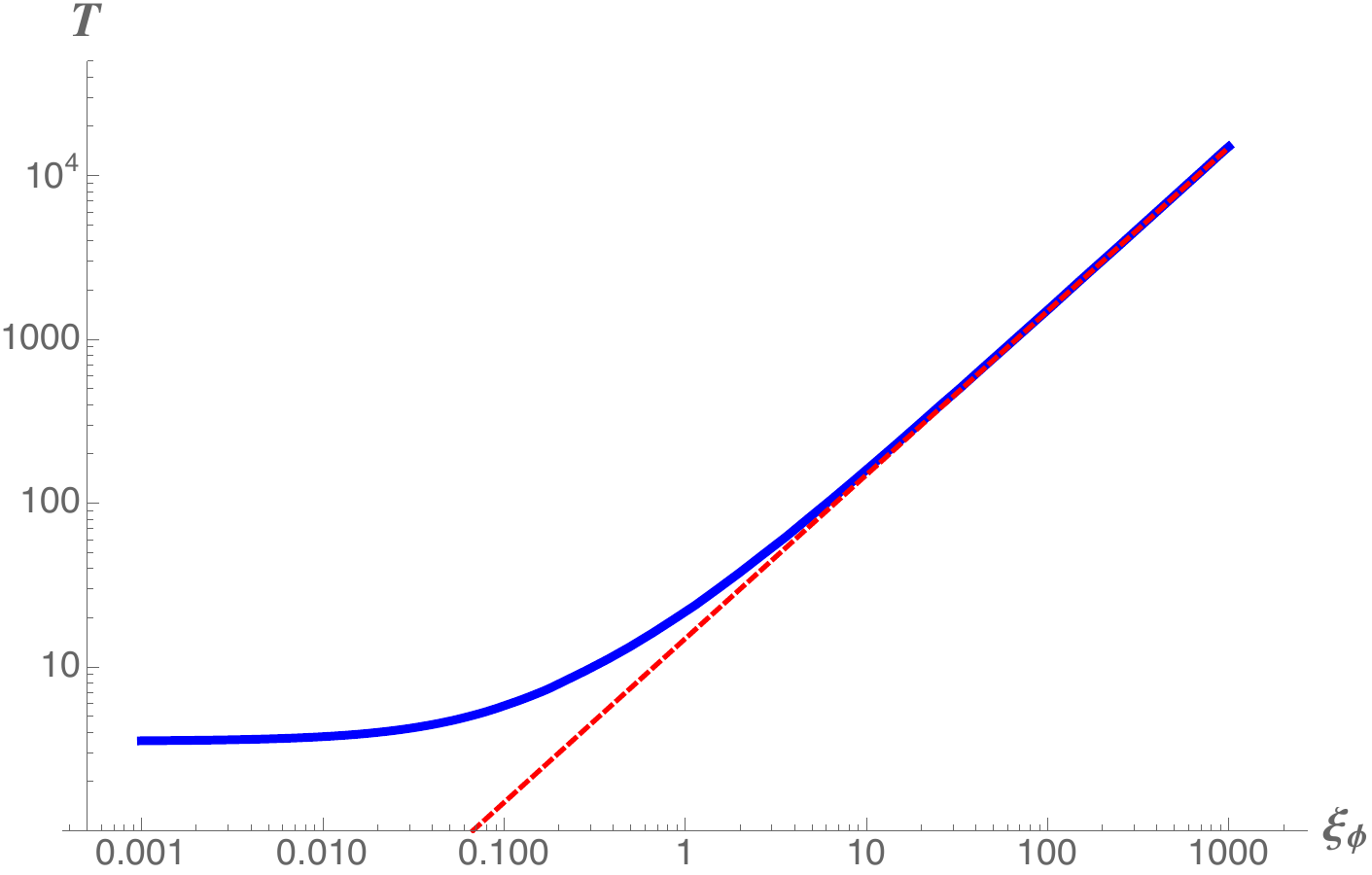} \quad \includegraphics[width=0.48\textwidth]{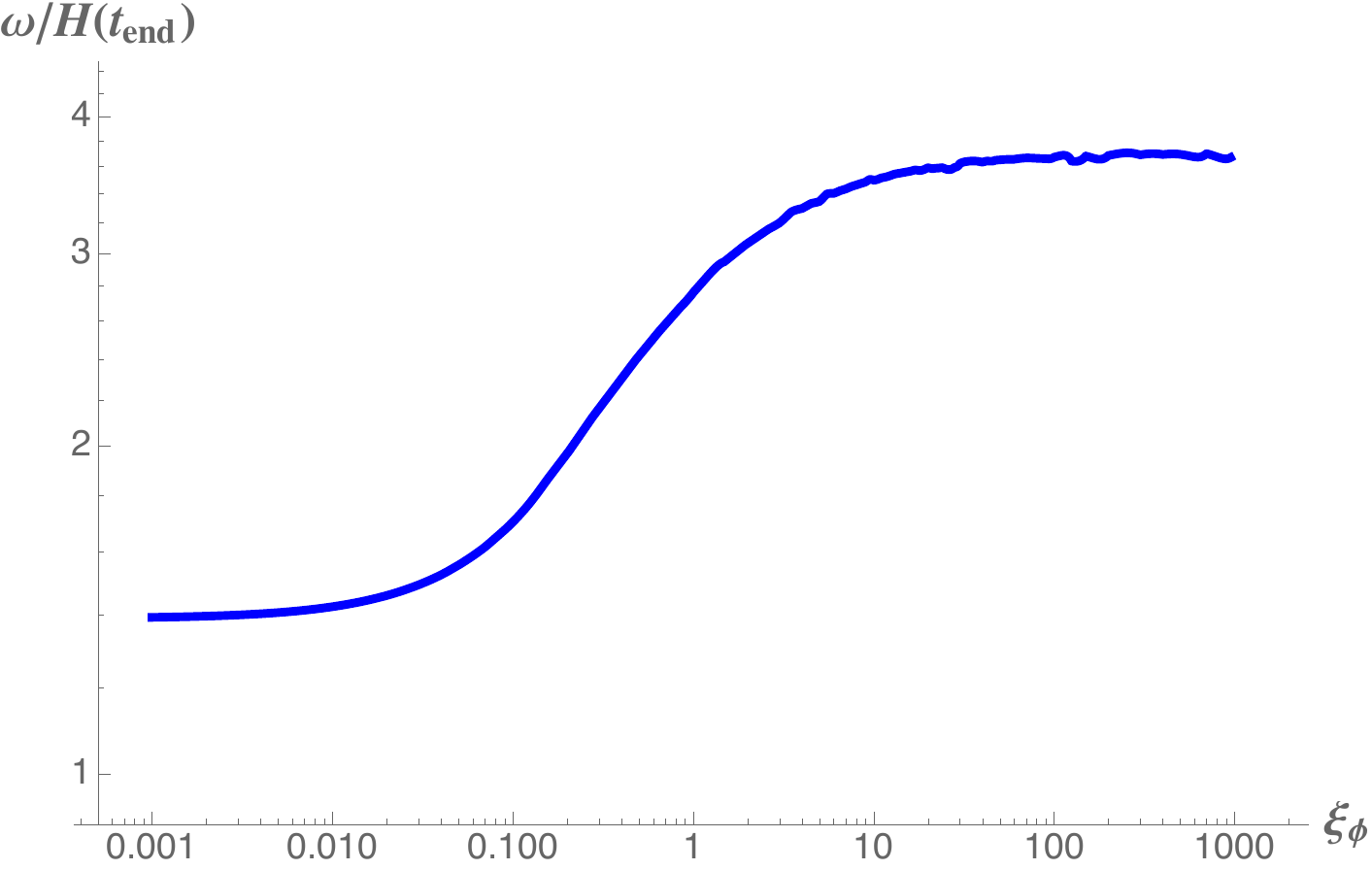}
\caption{ \small \baselineskip 11pt  ({\it Left}) The period of $\phi ( \tau)$'s oscillations, $T$ (in units of $M_{\rm pl}$), as a function of $\xi_\phi$, within the single-field attractor. For large $\xi_\phi$, $T$ grows linearly with $\xi_\phi$, asymptoting to $T \rightarrow 14.8 \xi_\phi / M_{\rm pl}$ (red dashed line). ({\it Right}) The ratio of the frequency of $\phi$'s oscillations, $\omega = 2 \pi / T$, to the Hubble scale at the end of inflation, $H (t_{\rm end})$. For large $\xi_\phi$, both $\omega$ and $H (t_{\rm end})$ scale as $1/\xi_\phi$, yielding $\omega / H (t_{\rm end}) \simeq 4$. }
\label{Tfig}
\end{figure}

Within the single-field attractor, in the limit $H \rightarrow 0$ and neglecting backreaction from produced particles, Eq. (\ref{eomvarphi}) becomes
\beq
\ddot{\phi} + \Gamma^\phi_{\> \phi \phi} \dot{\phi}^2 + {\cal G}^{\phi \phi} V_{, \phi} \simeq 0 .
\label{phieomnoH}
\eeq
We rescale $\tau \equiv \sqrt{\lambda_\phi} \> t$, so that the dynamics depend only on $\xi_\phi$. After $\tau_{\rm end}$, $\phi (\tau)$ oscillates periodically with period given by
\beq
T  = 2 \int_{- \phi_0}^{\phi_0} d \phi \sqrt{ \frac{ {\cal G}_{\phi \phi} }{2 V (\phi_0) - 2 V (\phi) }  }.
\label{T}
\eeq
(In this subsection we label $\phi_0 = \phi (\tau_{\rm end})$ as the amplitude of the field at the start of preheating, rather than the start of inflation.) As shown in Fig. \ref{Tfig}, the period scales approximately linearly with $\xi_\phi$ for $\xi_\phi > 1$, and hence the frequency of oscillations $\omega = 2 \pi / T$ scales like $1/\xi_\phi$. The Hubble scale at the end of inflation $H (t_{\rm end})$ also scales like $1/\xi_\phi$ in the limit of large $\xi_\phi$. We find $\omega / H (t_{\rm end}) > 1$ across the entire range $10^{-3} \leq \xi_\phi \leq 10^3$, with $\omega / H (t_{\rm end}) \sim 3$ at $\xi_\phi = 1$ and $\omega / H (t_{\rm end} ) \rightarrow 4$ for $\xi_\phi \gg 1$.

In the limit $\xi_\phi \gg 1$, the integral for $T$ in Eq. (\ref{T}) may be calculated analytically. For initial data of the form $\phi_0 = \phi (\tau_{\rm end}) = \alpha M_{\rm pl} / \sqrt{\xi_\phi}$ for some constant $\alpha$, and working in the regime $\alpha > 1/\sqrt{6 \xi_\phi}$, we find
\beq
T \rightarrow \frac{ 4 \sqrt{3} \> \xi_\phi}{ M_{\rm pl}} \left[ \pi - {\rm arctan} \left( \frac{ \sqrt{1 + 2 \alpha^2} }{\alpha^2} \right) \right] \frac{1 + \alpha^2}{\sqrt{1 + 2 \alpha^2} } .
\label{Tlargexi}
\eeq
Details of the derivation may be found in Appendix B. Using the best-fit value $\alpha = 0.8$ (see Fig. \ref{phiendfig}) yields $T \rightarrow 14.8 \, \xi_\phi / M_{\rm pl}$ in the limit $\xi_\phi \gg 1$. Meanwhile, in the opposite limit, $\xi_\phi \rightarrow 0$, Eq. (\ref{phieomnoH}) may be solved analytically as a Jacobian elliptic cosine, given the Jordan-frame potential of Eq. (\ref{VJphichi}): $\phi (t) = \phi_0 \> {\rm cn} (\phi_0 \tau , 1 / \sqrt{2})$ \cite{Boyanovsky,KaiserPreh2,GKLS}. The function ${\rm cn} \> (x, \kappa)$ is periodic with period $4 K (\kappa)$, where $K (\kappa)$ is the complete elliptic integral of the first kind \cite{AbramSteg}. Given $\kappa = 1/\sqrt{2}$ and $\phi_0 = 2.1 \> M_{\rm pl}$ for $\xi_\phi = 0$, we find $T \rightarrow 4 K (1/\sqrt{2}) / \phi_0 = 3.9 / M_{\rm pl}$, a good match to the $\xi_\phi \ll 1$ behavior of Fig. \ref{Tfig}. 

More generally, the terms in Eq. (\ref{phieomnoH}) that arise from the nontrivial field-space metric produce a richer structure for $\phi$'s oscillations, with greater numbers of non-negligible harmonics, compared to the $\xi_\phi = 0$ case. In Ref.~\cite{MultiPreheat2} we study this nontrivial harmonic structure and analyze its impact on the structure of the resonances for the coupled fluctuations.

\section{Evolution of the Fluctuations}
\label{EvolutionFluctuations}

In order to study the evolution of the fluctuations $Q^I$ during preheating, we expand the action to second order in both field and metric perturbations, calculate the energy density, and perform a (covariant) mode expansion. These steps enable us to relate the number density of particles for each species to an adiabatic parameter, generalizing the usual single-field expression. The adiabatic parameters may be used to identify regions of parameter space in which the system departs strongly from adiabatic evolution, indicating explosive particle production. We identify important differences in the behavior of the system for three distinct regimes: $\xi_I < {\cal O} (1)$, $\xi_I \sim {\cal O} (1 - 10)$, and $\xi_I \geq {\cal O} (100)$, which we explore further in Refs.~\cite{MultiPreheat2,MultiPreheat3}. These three regimes correspond to what one might expect, {\it a priori}, on perturbative grounds: $\xi_I \rightarrow 0$ (semiclassical analysis), $\xi_I \sim 1$ (nontrivial quantum corrections), and $\xi_I \rightarrow \infty$ (nonperturbative regime).

\subsection{Mode Expansion and Adiabatic Parameters}
\label{ActionQ}

Following the method of Ref.~\cite{GongTanaka} applied to the action in Eq. (\ref{SE}), we may expand the action to second order in the doubly-covariant fluctuation $Q^I$. We find (see also Refs.~\cite{Langlois,KMS,SebastianRP})
\beq
S_2^{(Q)} = \int d^3 x \> dt \> a^3 (t) \left[ - \frac{1}{2} \bar{g}^{\mu\nu} {\cal G}_{IJ} {\cal D}_\mu Q^I {\cal D}_\nu Q^J - \frac{1}{2} {\cal M}_{IJ} Q^I Q^J \right] ,
\label{S2}
\eeq
where $\bar{g}_{\mu\nu}$ is the background spacetime metric, ${\cal M}_{IJ}$ is given in Eq. (\ref{MIJ}), and ${\cal G}_{IJ}$ and ${\cal M}_{IJ}$ are evaluated to background order in the fields, $\varphi^I$. Next we rescale the fluctuations, $Q^I (x^\mu) \rightarrow X^I (x^\mu) / a(t)$ and introduce conformal time, $d\eta = dt / a(t)$, so that the background spacetime line-element may be written $ds^2 = a^2 (\eta) \> \eta_{\mu\nu} dx^\mu dx^\nu$, in terms of the Minkowski spacetime metric $\eta_{\mu\nu}$. Upon integrating by parts, we may rewrite Eq. (\ref{S2}) in the form
\beq
S_2^{(X)} = \int d^3 x \> d\eta \left[ - \frac{1}{2} \eta^{\mu\nu} {\cal G}_{IJ} {\cal D}_\mu X^I {\cal D}_\nu X^J - \frac{1}{2} {\mathbb M}_{IJ}  X^I X^J \right]
\label{S2eta}
\eeq
where 
\beq
{\mathbb M}_{IJ} \equiv a^2 \left( {\cal M}_{IJ} - \frac{1}{6} {\cal G}_{IJ} R \right)
\label{arcMIJ}
\eeq
and $R$ is the spacetime Ricci scalar. We have used the relation $R = 6 a'' / a^{3}$, and in this section we will use primes to denote $d / d\eta$. Note that for an equation of state $w_{\rm avg} \simeq 0$ then $a (t) \sim t^{2/3}$ and $a (\eta) \sim \eta^2$, while for $w_{\rm avg} = 1/3$ then $a (t) \sim t^{1/2}$ and $a (\eta) \sim \eta$.

From Eq. (\ref{S2eta}) we may construct an energy-momentum tensor for the fluctuations,
\beq
T_{\mu\nu}^{(X)} = {\cal G}_{IJ} {\cal D}_\mu X^I {\cal D}_\nu X^J - \frac{1}{2} \eta_{\mu\nu} \left[ \eta^{\alpha \beta} {\cal G}_{IJ} {\cal D}_\alpha X^I {\cal D}_\beta X^J + {\mathbb M}_{IJ} X^I X^J \right] .
\label{TmnQ}
\eeq
The energy density is given by the $00$ component of $T_{\mu\nu}^{(X)}$. The background spacetime metric is spatially flat, so we may easily perform a Fourier transform of a given quantity, $F(x^\mu) = (2 \pi)^{-3/2} \int d^3 k \> F_k ( \eta ) e^{i {\bf k} \cdot {\bf x} }$. The energy density of the fluctuations per Fourier mode then takes the form
\beq
\rho_k^{(X)} = \frac{1}{2} {\cal G}_{IJ} {\cal D}_\eta X_k^I {\cal D}_\eta X_k^J + \frac{1}{2} \left[ \omega_k^2 (\eta) \right]_{IJ} X_k^I X_k^J + {\cal O} (X^3),
\label{rhoQ1}
\eeq
where we have defined
\beq
\left[ \omega_k^2 (\eta ) \right]_{IJ} \equiv k^2 {\cal G}_{IJ} + {\mathbb M}_{IJ} .
\label{omegaIJ}
\eeq
Upon using the equation of motion for $Q^I$, Eq. (\ref{eomQ}), and the relation $Q^I = X^I / a$, we may rewrite Eq. (\ref{rhoQ1}) in the form
\beq
\rho_k^{(X)} = \frac{1}{2} {\cal G}_{IJ} \left[ \left( {\cal D}_\eta X^I \right) \left( {\cal D}_\eta X^J \right) - \left( {\cal D}_\eta^2 X^I \right) X^J  \right] .
\label{rhoQ2}
\eeq

Next we quantize the fluctuations, $X^I \rightarrow \hat{X}^I$, and expand them in a series of creation and annihilation operators in a way that respects the nontrivial field-space manifold \cite{AHKK,WeinbergCosmology},
\beq
\hat{X}^I (x^\mu) = \int \frac{ d^3 k}{( 2 \pi)^{3/2} } \sum_b \left[ u_b^I (k, \eta )\> \hat{a}_{ {\bf k} b}\> e^{i {\bf k} \cdot {\bf x} } + u_b^{I*} (k, \eta ) \> \hat{a}_{ {\bf k} b}^\dagger \> e^{- i {\bf k} \cdot {\bf x} } \right] ,
\label{hatQ}
\eeq
where the index $b = 1, 2, ... , N$. The operators obey
\beq
\hat{a}_{ {\bf k} b} \vert 0 \rangle =  0, \>\> \langle 0 \vert \hat{a}_{ {\bf k} b}^\dagger = 0
\label{avac}
\eeq
for all ${\bf k}$ and $b$, and
\beq
\begin{split}
\Big[ \hat{a}_{ {\bf k} b} , \hat{a}_{ {\bf q} c} \Big] &= \Big[ \hat{a}^\dagger_{ {\bf k} b} , \hat{a}^\dagger_{ {\bf q} c} \Big] = 0 , \\
\Big[ \hat{a}_{ {\bf k} b} , \hat{a}_{ {\bf q} c}^\dagger \Big] &= \delta^{(3)} ( {\bf k } - {\bf q} ) \delta_{bc} .
\end{split}
\label{acommutation}
\eeq
Each of the mode functions satisfies the equation of motion,
\beq
{\cal D}_\eta^2 u_b^I  + \left[ \omega_k^2 (\eta ) \right]^I_{\> J} u_b^J = 0 .
\label{eomubI}
\eeq
As discussed in Ref.~\cite{AHKK}, we have $N$ linear, second-order differential equations (one for each $\hat{X}^I$), which yield $2N$ linearly independent solutions. By parameterizing the fluctuations as in Eq. (\ref{hatQ}), we have introduced $N^2$ complex mode functions $u_b^{\> I} (k, \eta)$, and hence $2N^2$ real-valued scalar functions, $u_b^{\> I} = {\rm Re} \left[ u_b^{\> I} \right] + {\rm Im} \left[ u_b^{\> I} \right]$. But $N$-tuples of the complex mode functions are coupled to each other by Eq. (\ref{eomubI}), which yields $2N (N - 1)$ constraints, leaving exactly $2N^2 - 2 N (N - 1) = 2N$ independent solutions.

We parameterize the mode functions as \cite{AHKK,WeinbergCosmology}
\beq
u_b^I (k, \eta) = h_{(b, I)} (k, \eta) \> e_b^{\> I} (\eta ) ,
\label{ufe}
\eeq
where the $h_{(b, I)}$ are complex scalar functions and the $e_b^{\> I} (\eta )$ are vielbeins of the field-space metric,
\beq
\delta^{bc} e_b^{\> I} ( \eta ) e_c^{\> J} (\eta ) = {\cal G}^{IJ} (\eta ) .
\label{vielbein}
\eeq
Note that the components of the vielbeins are purely real, and, unlike the unit vectors $\hat{\sigma}^I, \hat{s}^I$ defined in Eq. (\ref{hatsigmahats}), the $e_b^{\> I}$ are well-behaved during preheating. (Explicit expressions for the $e_b^{\> I}$ for our two-field model may be found in Appendix A.) The subscripts $(b, I)$ on $h$ are labels only, not vector indices. We then find
\beq
\langle 0 \vert \hat{X}^I (x) \hat{X}^J (x) \vert 0 \rangle = \int \frac{ d^3 k}{ ( 2 \pi)^3 } \delta^{bc} u_b^{\> I} u_c^{\> J*} ,
\label{QQvevIJ}
\eeq
upon using Eqs. (\ref{avac}), (\ref{acommutation}), and (\ref{vielbein}). As emphasized in Refs.~\cite{AHKK,Mustafa}, the cross products, with $I \neq J$, need not vanish.

The vielbeins ``absorb" most of the added structure from the nontrivial field-space manifold, enabling us to manipulate (mostly) ordinary scalar functions. As usual, we raise and lower field-space indices $I, J$ with ${\cal G}_{IJ}$, and we raise and lower internal indices $b, c$ with $\delta_{bc}$. We may also use the vielbeins to ``trade" between field-space indices and internal indices. For an arbitrary vector $A^I$ we may write
\beq
A^b = e^b_{\> I} A^I , \>\> A^I = e_b^{\> I} A^b ,
\label{tradeindex}
\eeq
while Eq. (\ref{vielbein}) implies 
\beq
\begin{split}
e^b_{\> I} e_c^{\> I} &= \delta^b_{\> c} , \\
e_b^{\> I} e^b_{\> J} &= \delta^I_{\> J}. 
\end{split}
\label{eedelta}
\eeq
The covariant derivative of the vielbein with respect to ${\cal G}_{IJ}$ is given in terms of the spin connection, $\omega^{bc}_{\>\>\> I}$,
\beq
{\cal D}_I e^b_{\> J} = - \omega^{bc}_{\> \>\> I} e_{cJ} ,
\label{De}
\eeq
where $\omega^{bc}_{\>\>\> I}$ is antisymmetric in its internal indices, $\omega^{bc}_{\>\>\> I} = - \omega^{cb}_{\>\>\> I}$ \cite{Carroll}. Because of the antisymmetry of the spin connection, the (covariant) directional derivative with respect to conformal time vanishes,
\beq
{\cal D}_\eta e^b_{\> J} = 0
\label{Dte}
\eeq
for all $b$ and $J$ \cite{WeinbergCosmology}.

For our two-field model, with $\{ I, J \} = \{ 1, 2 \}$, we may write out the mode expansions more explicitly. We assign the field-space indices $1 = \phi$ and $2 = \chi$ and write $\hat{a}_{{ \bf k} b} = \hat{b}_{\bf k}$ for $b = 1$, $\hat{a}_{ {\bf k} b} = \hat{c}_{ {\bf k} }$ for $b = 2$. We also label $h_{(1, \phi)} = v_k ( \eta )$, $h_{(2, \phi)} = w_k (\eta )$, $h_{(1, \chi)} = y_k (\eta )$, and $h_{(2,\chi)} = z_k (\eta )$, so that Eq. (\ref{hatQ}) becomes
\beq
\begin{split}
\hat{X}^\phi (x^\mu) &= \int \frac{d^3 k }{ ( 2 \pi)^{3/2} } \left[ \left( v_k e_1^{\> \phi} \hat{b}_{\bf k} + w_k e_2^{ \> \phi} \hat{c}_{\bf k} \right) e^{i {\bf k} \cdot {\bf x} } + \left( v_k^* e_1^{\> \phi} \hat{b}_{ {\bf k}}^\dagger + w_k^* e_2^{\> \phi} \hat{c}_{ \bf k}^\dagger \right) e^{ - i {\bf k} \cdot {\bf x} } \right] , \\
\hat{X}^\chi (x^\mu) &= \int \frac{ d^3 k }{ ( 2 \pi)^{3/2} } \left[ \left(  y_k e_1^{\> \chi} \hat{b}_{\bf k} + z_k e_2^{ \> \chi} \hat{c}_{ \bf k} \right) e^{ i {\bf k} \cdot {\bf x} } + \left( y_k^* e_1^{\> \chi} \hat{b}_{ \bf k}^\dagger + z_k^* e_2^{\> \chi} \hat{c}_{\bf k}^\dagger \right) e^{- i {\bf k} \cdot {\bf x} } \right] .
\end{split}
\label{QphiQchimodes}
\eeq
Eq. (\ref{eomubI}) couples $v_k$ with $y_k$ and $w_k$ with $z_k$:
\beq
\begin{split}
\left( v''_k + \Omega_{(\phi)}^2 v_k \right) e_1^\phi &= - a^2 {\cal M}^\phi_{\>\> \chi} y_k e_1^\chi , \\
\left( w_k'' + \Omega_{(\phi)}^2 w_k \right) e_2^\phi &= - a^2 {\cal M}^\phi_{\>\> \chi} z_k e_2^\chi , \\
\left( y_k'' + \Omega_{(\chi)}^2 y_k \right) e_1^\chi &= - a^2 {\cal M}^\chi_{\>\> \phi} v_k e_1^\phi , \\
\left( z_k'' + \Omega_{(\chi)}^2 z_k \right) e_2^\chi &= - a^2 {\cal M}^\chi_{\>\> \phi} w_k e_2^\phi ,
\end{split}
\label{vwyzeom}
\eeq
where for convenience we have labeled the diagonal components of $[ \omega_k^2 (\eta) ]^I_{\>\> J}$ as
\beq
\begin{split}
\Omega_{(\phi)}^2 (k, \eta) &\equiv k^2 + a^2 m_{\rm eff, \phi}^2 (\eta) , \\
\Omega_{(\chi)}^2 (k, \eta) &\equiv k^2 + a^2 m_{\rm eff \chi}^2 (\eta) ,
\end{split}
\label{Omegadef}
\eeq
in terms of the effective masses
\beq
\begin{split}
m_{\rm eff, \phi}^2 &\equiv {\cal M}^\phi_{\>\> \phi} - \frac{1}{6} R , \\
 m_{\rm eff, \chi}^2 &\equiv {\cal M}^\chi_{\>\> \chi} - \frac{1}{6} R .
 \end{split}
\label{meffI}
\eeq

We are interested in the energy density per mode $k$ of the quantized fluctuations, which we parameterize as
\beq
\langle \hat{\rho}^{(X)} (x^\mu) \rangle =  \int \frac{ d^3 k}{ (2 \pi)^3} \rho_k^{(X) {\rm vev} }  (\eta) .
\label{rhokdef}
\eeq
Upon using Eqs. (\ref{rhoQ2}), (\ref{QQvevIJ}), and (\ref{Dte}) we find
\beq
\begin{split}
\rho_k^{(X) {\rm vev} } &= \frac{1}{2} {\cal G}_{IJ} \sum_b \sum_c  \Big\{ \delta^{bc} \left[ {h}^\prime_{(b,I)} {h}^{* \prime}_{(c,J)}  - {h}^{\prime\prime}_{(b, I)} h_{(c, J)}^*  \right] e_b^{\> I} e_c^{\> J} \Big\} \\
&=   \rho_k^{(\phi)} + \rho_k^{(\chi)} + \rho_k^{( {\rm int} ) }  ,
\end{split}
\label{rhoQvev1}
\eeq
with
\beq
\begin{split}
\rho_k^{(\phi)} &= \frac{1}{2}  {\cal G}_{\phi\phi}  \Big\{  \left( \vert v^\prime_k \vert^2 - v^{\prime \prime}_k v_k^* \right) e_1^\phi e_1^\phi + \left( \vert w_k' \vert^2 - w_k'' w_k^* \right) e_2^\phi e_2^\chi \Big\} , \\
\rho_k^{(\chi)} &= \frac{1}{2} {\cal G}_{\chi\chi} \Big\{ \left( \vert y_k' \vert^2 - y_k'' y_k^* \right) e_1^\chi e_1^\chi + \left( \vert z^\prime_k \vert^2 - z^{\prime \prime}_k z_k^* \right) e_2^\chi e_2^\chi \Big\} \\
\rho_k^{ ( {\rm int} )} &=   {\cal G}_{\phi \chi}  \Big\{ \left( v^\prime_k y^{*\prime}_k - v_k'' y_k^* \right) e_1^\phi e_1^\chi + \left( y_k' v_k^{*\prime} - y_k'' v_k^* \right) e_1^\chi e_1^\phi \\
&\quad\quad \quad\quad\quad + \left( w_k' z_k^{*\prime} - w_k'' z_k^* \right) e_2^\phi e_2^\chi + \left( z_k' w_k^{*\prime} - z_k'' w_k^* \right) e_2^\chi e_2^\phi \Big\} .
\end{split}
\label{rhokphirhokchi}
\eeq
One may use the equations of motion in Eq. (\ref{vwyzeom}) to demonstrate that the expressions in Eq. (\ref{rhokphirhokchi}) are purely real. The number density per mode of quanta of a given field $I$ ($\phi$ or $\chi$) may be related to the energy density by 
\beq
\begin{split}
n_k^{(I)} &= \frac{ \rho_k^{(I)} }{\Omega_{(I)} } - \frac{1}{2} .
\end{split}
\label{nkdef}
\eeq
The number density per mode for each species $I = \phi, \chi$ will be well-defined in the limit $\rho_k^{(\rm int )} \ll \rho_k^{(I)}$.

We noted in Section \ref{AttractorBehavior} that within a single-field attractor (along the direction $\chi = 0$), the cross-terms in both ${\cal G}_{IJ}$ and ${\cal M}^I_{\>\>J}$ vanish. In that case, the vielbeins also become diagonal, 
\beq
e_b^I \rightarrow \left( \begin{array}{cc} e_1^\phi & 0 \\ 0 & e_2^\chi \end{array} \right) ,
\label{ebISFA}
\eeq
with $e_2^\phi \sim e_1^\chi \sim 0$, $e_1^\phi e_1^\phi \simeq {\cal G}^{\phi \phi}$, $e_2^\chi e_2^\chi \simeq {\cal G}^{\chi\chi}$, and ${\cal G}_{\phi \phi} {\cal G}^{\phi \phi} = {\cal G}_{\chi\chi} {\cal G}^{\chi\chi} = 1+ {\cal O} (\chi^2)$. Then the fluctuations $\hat{X}^I$ simplify considerably: $\hat{X}^\phi$ is expanded only in the $\hat{b}_{\bf k} , \hat{b}_{\bf k}^\dagger$ operators, and $\hat{X}^\chi$ only in the $\hat{c}_{\bf k} , \hat{c}_{\bf k}^\dagger$ operators. Given both ${\cal M}^\phi_{\>\> \chi} \sim {\cal M}^\chi_{\>\> \phi} \sim 0$ and $e_2^\phi \sim e_1^\chi \sim 0$, moreover, the scalar mode functions decouple: the functions $v_k (\eta)$ and $z_k (\eta)$ satisfy source-free equations of motion, while $w_k (\eta) \sim y_k (\eta) \sim 0$. Within the attractor, the expressions in Eq. (\ref{rhokphirhokchi}) simplify as well:
\beq
\begin{split}
\rho_k^{(\phi)} &\rightarrow \frac{1}{2}  \left( \vert v_k' \vert^2 - v_k'' v_k^* \right) + {\cal O} (\chi^2) , \\
\rho_k^{(\chi)} &\rightarrow \frac{1}{2}  \left( \vert z_k' \vert^2 - z_k'' z_k^* \right)  + {\cal O} (\chi^2) , \\
\rho_k^{({\rm int} ) } &\rightarrow {\cal O} (\chi^2) \sim 0 .
\end{split}
\label{rhokSFA}
\eeq
Since $\rho_k^{(\rm int)}$ remains subdominant within the single-field attractor, the notion of particle number for each species is well-defined in that limit, and we may relate $\rho_k^{(\phi)}$ and $\rho_k^{(\chi)}$ to the corresponding number densities of produced particles.

To calculate the number density of created particles and relate those expressions to adiabatic parameters, we generalize the familiar result from studies of single-field models with minimal couplings. (See also Refs.~\cite{KLS,BKM,Barnaby,HertzbergKarouby}.) Within the single-field attractor, the coupled equations of motion in Eq. (\ref{vwyzeom}) reduce to
\beq
\begin{split}
v_k^{\prime\prime} + \Omega^2_{(\phi)} (k, \eta) \>v_k &\simeq 0 , \\
z_k^{\prime\prime} + \Omega^2_{(\chi)} (k, \eta) \>z_k &\simeq 0 .
\end{split}
\label{vzeomhomog}
\eeq
We are interested in how efficiently the background fields $\varphi^I$ transfer energy to the fluctuations after the end of inflation, so we quantize the fluctuations with respect to the adiabatic vacuum $\vert 0 (t_{\rm end}) \rangle$, that is, the state that instantaneously minimizes the system's energy density at $t_{\rm end}$ \cite{BirrellDavies,ParkerToms,AHKK}. We then posit solutions to Eq. (\ref{vzeomhomog}) of the form
\beq
\begin{split}
v_k (\eta) &= \frac{1}{ \sqrt{ 2 W_{(\phi )} (k, \eta) } } \exp\left[ -i \int^\eta d\eta' \> W_{( \phi )} (k, \eta' ) \right] , \\
z_k (\eta) &= \frac{1}{ \sqrt{ 2 W_{ (\chi )} (k, \eta) } } \exp \left[ - i \int^\eta d\eta' \> W_{( \chi )} (k, \eta') \right] , 
\end{split}
\label{vwzy}
\eeq
in terms of the (as yet unspecified) real-valued functions $W_{(I)} (k, \eta)$. The choice of adiabatic vacuum corresponds to the boundary conditions $W_{(\phi)} (k, \eta_{\rm end}) = \Omega_{(\phi)} (k, \eta_{\rm end})$ and $W_{(\chi)} (k, \eta_{\rm end}) = \Omega_{(\phi)} (k, \eta_{\rm end})$. Given the ansatz in Eq. (\ref{vwzy}), the expressions in Eq. (\ref{rhokphirhokchi}) for the energy density per mode take the form
\beq
\rho_k^{(\phi)} = \frac{1}{2}  \left[ W_{(\phi)} + \frac{ W_{(\phi)}^{\prime\prime} }{4 W_{(\phi)}^2} - \frac{ W_{( \phi )}^{\prime 2} }{4 W_{(\phi)}^3 } \right] + {\cal O} (\chi^2), 
\label{rhokphi1}
\eeq
and likewise for $\rho_k^{(\chi)}$ in terms of $W_{(\chi)}$ and its derivatives. 

Within the single-field attractor, when $\rho_k^{(\rm int )} \sim 0$ and $\rho_k^{(\phi )}$ and $\rho_k^{(\chi)}$ assume the simple forms in Eq. (\ref{rhokSFA}), the number densities in Eq. (\ref{nkdef}) likewise simplify. We may also use Eq. (\ref{vzeomhomog}) to relate $W_{(\phi)} (k, \eta)$ to $\Omega_{(\phi)} (k, \eta)$, which yields
\beq
W^2_{(\phi)} = \Omega^2_{(\phi)} - \frac{1}{2} \left[ \frac{ W^{\prime\prime}_{(\phi)} }{W_{(\phi)} } - \frac{3}{2} \frac{ W^\prime_{(\phi)} }{W_{(\phi)}^2 } \right] .
\label{WOmega}
\eeq
Away from resonance bands we expect the modes to evolve adiabatically, for which $W_{(\phi)} (k, \eta) \rightarrow \Omega_{(\phi)} (k, \eta) + {\cal O} ({\cal A}_{(\phi)}^2 )$, where 
\beq
{\cal A}_{(\phi)} (k, \eta) \equiv \frac{ \Omega^\prime_{(\phi)} (k, \eta) }{ \Omega_{(\phi)}^2 (k, \eta) } .
\label{Aphidef}
\eeq
As in Ref.~\cite{BirrellDavies}, we may then solve Eq. (\ref{WOmega}) iteratively, in increasing powers of ${\cal A}_{(\phi)}$. Combining Eqs. (\ref{rhokphi1}) - (\ref{Aphidef}), we find
\beq
n_k^{(\phi)} = \frac{1}{16} {\cal A}^2_{(\phi)} + {\cal O} (\chi^2) +  {\cal O} ({\cal A}^3_{(\phi)} ) ,
\label{nkphi}
\eeq
with a comparable expression for $n_k^{(\chi)}$. Much as in familiar cases with minimally coupled fields \cite{AHKK,KLS,BKM,Barnaby}, regions of parameter space in which ${\cal A}_{(I)} (k, \eta) \gg 1$ correspond to strong departures from adiabatic evolution, and hence to bursts of particle production.

\subsection{Resonant Amplification within the Attractor}
\label{Resonances}

The behavior of the adiabatic parameters, ${\cal A}_{(I)} (k, \eta)$, depends upon the effective frequencies, $\Omega_{(I)} (k, \eta)$, which in turn depend upon the effective masses, $m_{ { \rm eff}, I}^2$, defined in Eq. (\ref{meffI}). After the end of inflation, as $\varphi^I (t)$ oscillates, one or more of the $m_{ {\rm eff}, I}^2$ will oscillate as well, which can drive resonant amplification of the coupled fluctuations, $\hat{Q}^I$. We may rewrite Eq. (\ref{Aphidef}) in terms of cosmic time rather than conformal time,
\beq
{\cal A}_{(I)} = \frac{ H^{-3} \partial_t m_{ {\rm eff}, I}^2 + 2 (m_{ {\rm eff}, I} / H )^2}{2 \left[ \ell^2 + (m_{ {\rm eff}, I }/ H)^2 \right]^{3/2} } ,
\label{Aphit}
\eeq
where $\ell \equiv k_{\rm phys} / H = k / (aH)$. In the limit $\ell \ll 1$, we find
\beq
{\cal A}_{(I)} = \frac{\partial_t m_{ {\rm eff}, I}^2}{2 m_{ {\rm eff}, I}^3 } + \frac{ H}{m_{ {\rm eff}, I} } + {\cal O} (\ell^2 ) .
\label{Aphit2}
\eeq
In the limit $k \ll aH$, we expect $\vert {\cal A}_{(I)} \vert \gg 1$ whenever $\partial_t m_{ {\rm eff}, I}^2$ spikes and/or $m_{ { \rm eff}, I}^2$ passes through zero. 

Given the form of Eq. (\ref{meffI}), we may distinguish four separate contributions to $m_{ {\rm eff}, \phi}^2$:
\beq
m_{\rm eff, \phi}^2 = m_{1, \phi}^2 + m_{2, \phi}^2 + m_{3, \phi}^3 + m_{4, \phi}^2 ,
\label{m1234a}
\eeq
where
\beq
\begin{split}
m_{1, \phi}^2 &\equiv {\cal G}^{\phi K} \left( {\cal D}_\phi {\cal D}_K V \right) , \\
m_{2, \phi}^2 &\equiv - {\cal R}^\phi_{\>\> LM \phi} \dot{\varphi}^L \dot{\varphi}^M , \\
m_{3, \phi}^2 &\equiv - \frac{1}{ M_{\rm pl}^2 a^3} \delta^\phi_{\> I } \delta^J_{\> \phi} {\cal D}_t \left( \frac{ a^3}{H} \dot{\varphi}^I \dot{\varphi}_J \right) , \\
m_{4, \phi}^2 &\equiv - \frac{1}{6} R ,
\end{split}
\label{m1234phi}
\eeq
with comparable expressions for the contributions to $m_{\rm eff,\chi}^2$. Note that $m_{1, I}^2$ arises from the gradient of the potential; $m_{2, I}^2$ from the nontrivial field-space manifold; $m_{3, I}^2$ from the coupled metric perturbations; and $m_{4, I}^2$ from the expansion of the background spacetime. The term $m_{2,I}^2$, in particular, has no analogue in models with minimally coupled fields and canonical kinetic terms, and can play important roles in the dynamics during and after inflation \cite{KMS,GKS,KS,SSK,SebastienRPgeom,LeeSasaki,Yokoyama,Langlois,Tegmark,GongTanaka,Seery,GaoShukla}.

We first note that
\beq
m_{4, I}^2 = - \frac{1}{6} R = - \left( \dot{H} + 2 H^2 \right) = (\epsilon - 2) H^2 .
\label{m4I}
\eeq
We observed in Section \ref{OscillationsEES} that $\epsilon = 3\dot{\sigma}^2 / (\dot{\sigma}^2 + 2 V )$, so $0 \leq \epsilon \leq 3$, and hence $m_{4, I}^2 / H^2 = {\cal O} (1)$ regardless of the couplings and of the motion of the background fields $\varphi^I$. Within the single-field attractor (with $\chi \sim \dot{\chi} \sim 0$), many of the other terms in Eq. (\ref{m1234phi}) also become negligible. In particular, 
\beq
\begin{split}
{\cal G}_{\phi\chi} &\sim {\cal G}^{\phi \chi} \sim {\cal O} (\chi) \sim 0 , \\
\Gamma^\phi_{\> \phi \chi} &\sim \Gamma^\chi_{\> \phi \phi} \sim \Gamma^\chi_{\> \chi \chi} \sim {\cal O} (\chi) \sim 0 , \\
V_{, \chi} &\sim V_{, \phi \chi} \sim {\cal O} (\chi) \sim 0 .
\end{split}
\label{GVattractor1}
\eeq
Upon using the expressions for ${\cal G}_{IJ}$, $\Gamma^I_{\> JK}$, and ${\cal R}_{ILMJ}$ in Appendix A, we then find
\beq
\begin{split}
m_{1,\phi}^2 &= {\cal G}^{\phi \phi} \left[ V_{, \phi \phi} - \Gamma^\phi_{\> \phi \phi} V_{, \phi} \right] + {\cal O} (\chi^2) , \\
m_{2, \phi}^2 &= {\cal O} (\chi \dot{\chi} ) \sim 0 , \\
m_{3, \phi}^2 &= - \frac{ {\cal G}_{\phi \phi} }{M_{\rm pl}^2} \left[ (3 + \epsilon ) \dot{\phi}^2 + \frac{2}{H} \dot{\phi} \ddot{\phi} \right] + {\cal O} (\chi \dot{\chi} ) , \\
m_{1, \chi}^2 &= {\cal G}^{\chi\chi} \left[ V_{, \chi\chi} - \Gamma^\phi_{\> \chi\chi} V_{, \phi} \right] + {\cal O} (\chi^2) , \\
m_{2, \chi}^2 &= \frac{1}{2} {\cal R} \> {\cal G}_{\phi \phi} \dot{\phi}^2 + {\cal O} (\chi\dot{\chi} ), \\
m_{3, \chi}^2 &= {\cal O} (\chi \dot{\chi} ) \sim 0 .
\end{split}
\label{m123b}
\eeq
The ${\cal R}$ in $m_{2,\chi}^2$ is the Ricci curvature scalar of the field-space manifold, an explicit expression for which may be found in Eq. (\ref{Ricci2d}) in Appendix A. 

\begin{figure}
\centering
\includegraphics[width=0.48\textwidth]{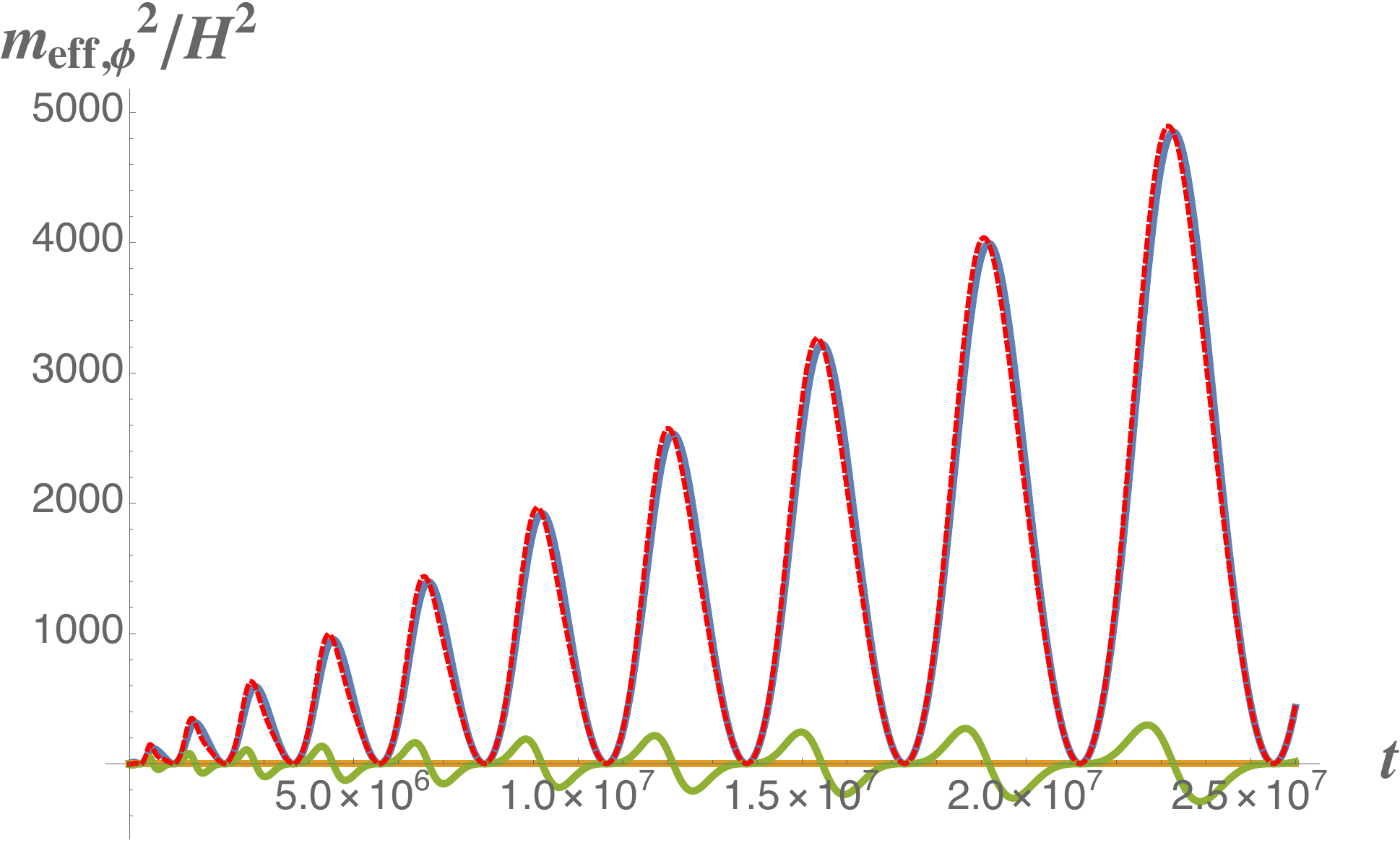} \quad \includegraphics[width=0.48\textwidth]{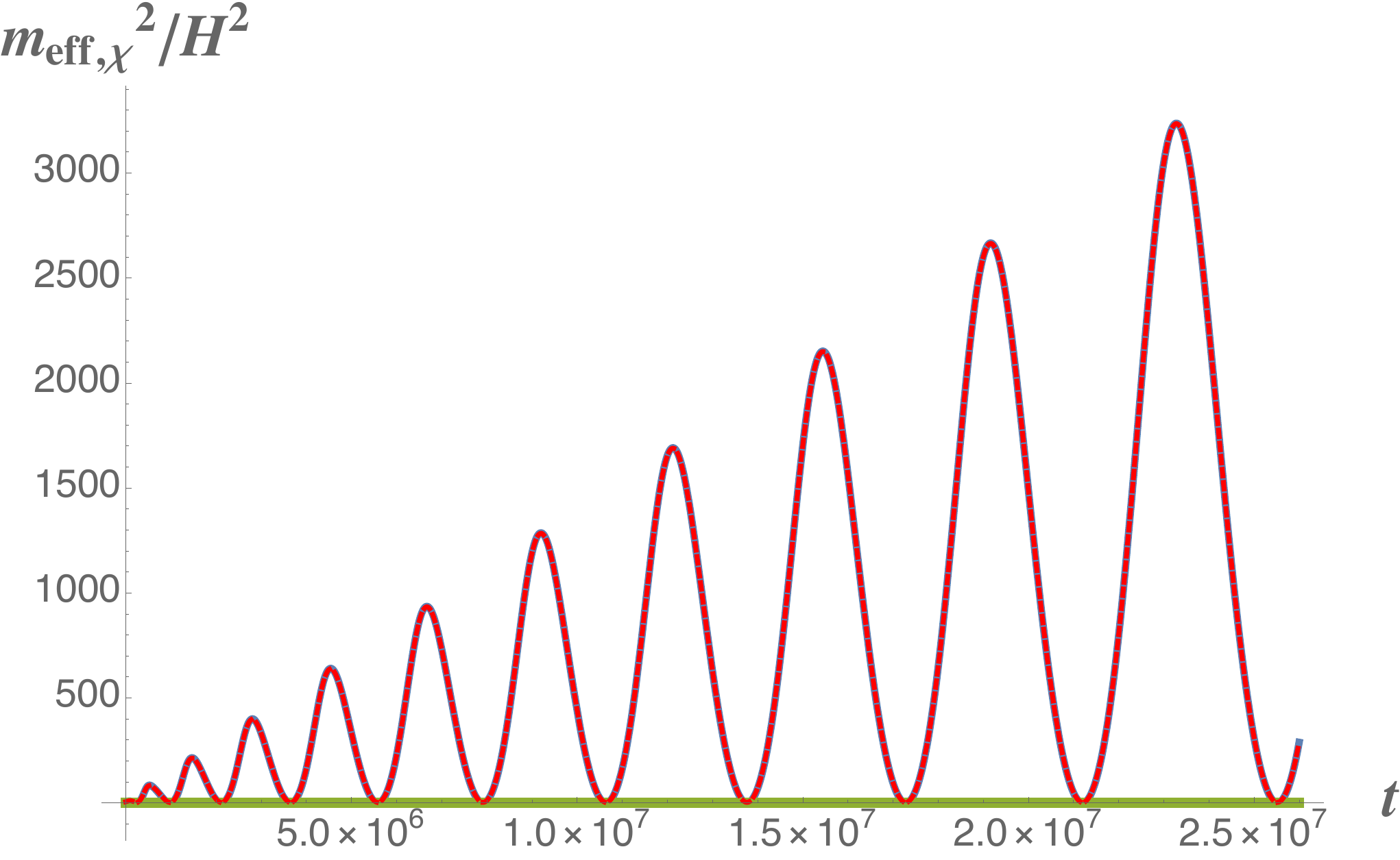} \\ \vskip 10pt
\includegraphics[width=0.48\textwidth]{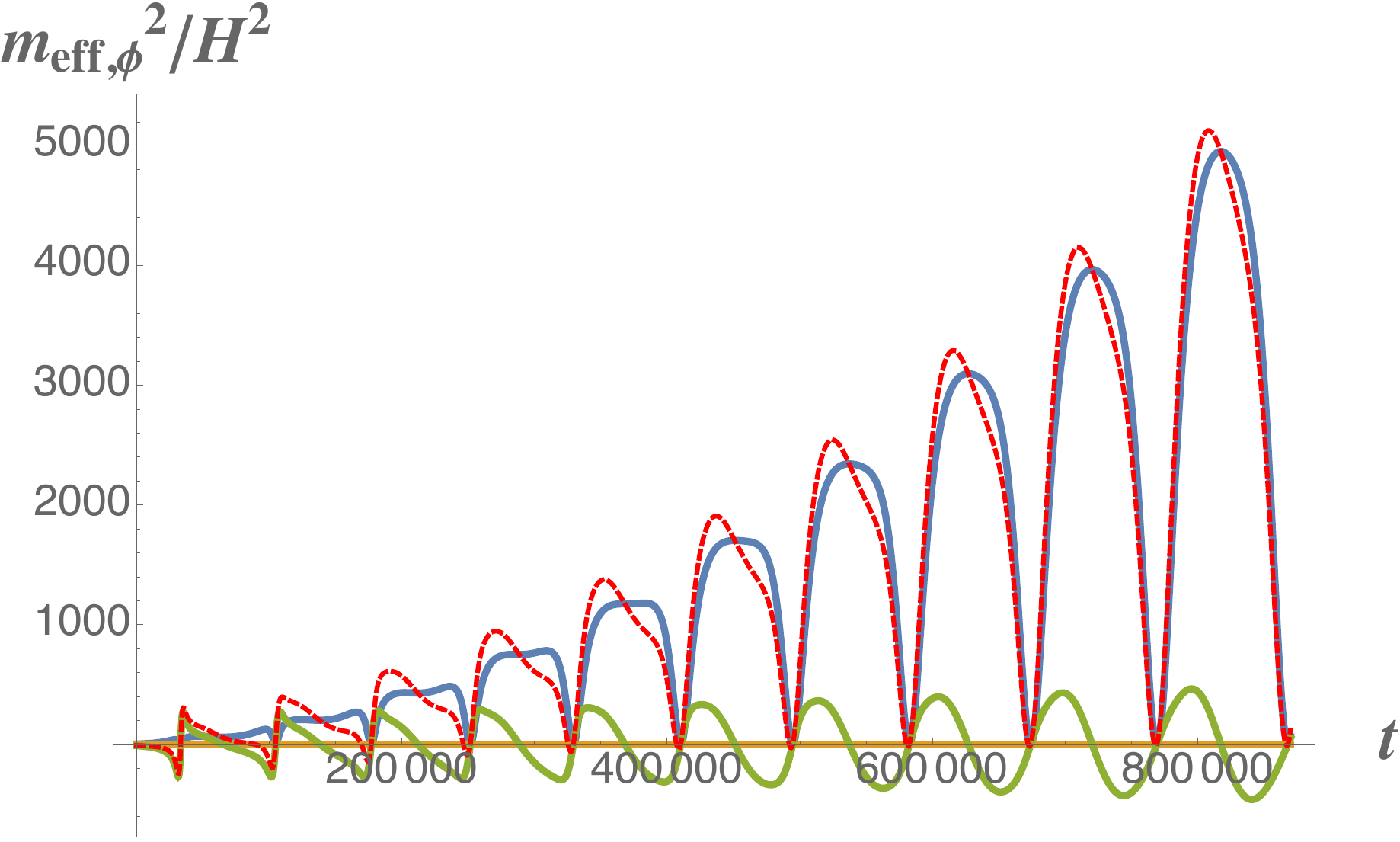}\quad \includegraphics[width=0.48\textwidth]{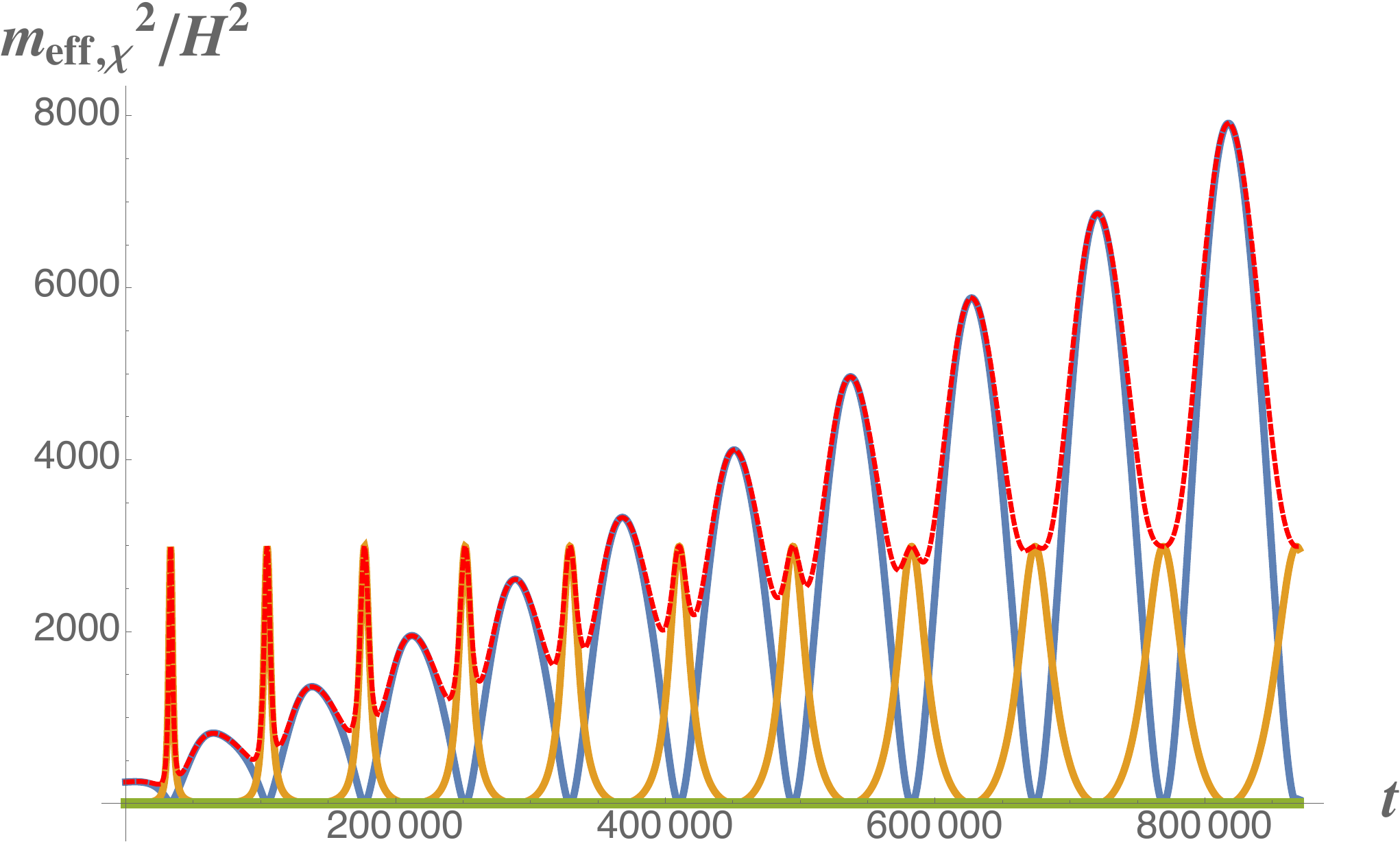} \\ \vskip 10pt
\includegraphics[width=0.48\textwidth]{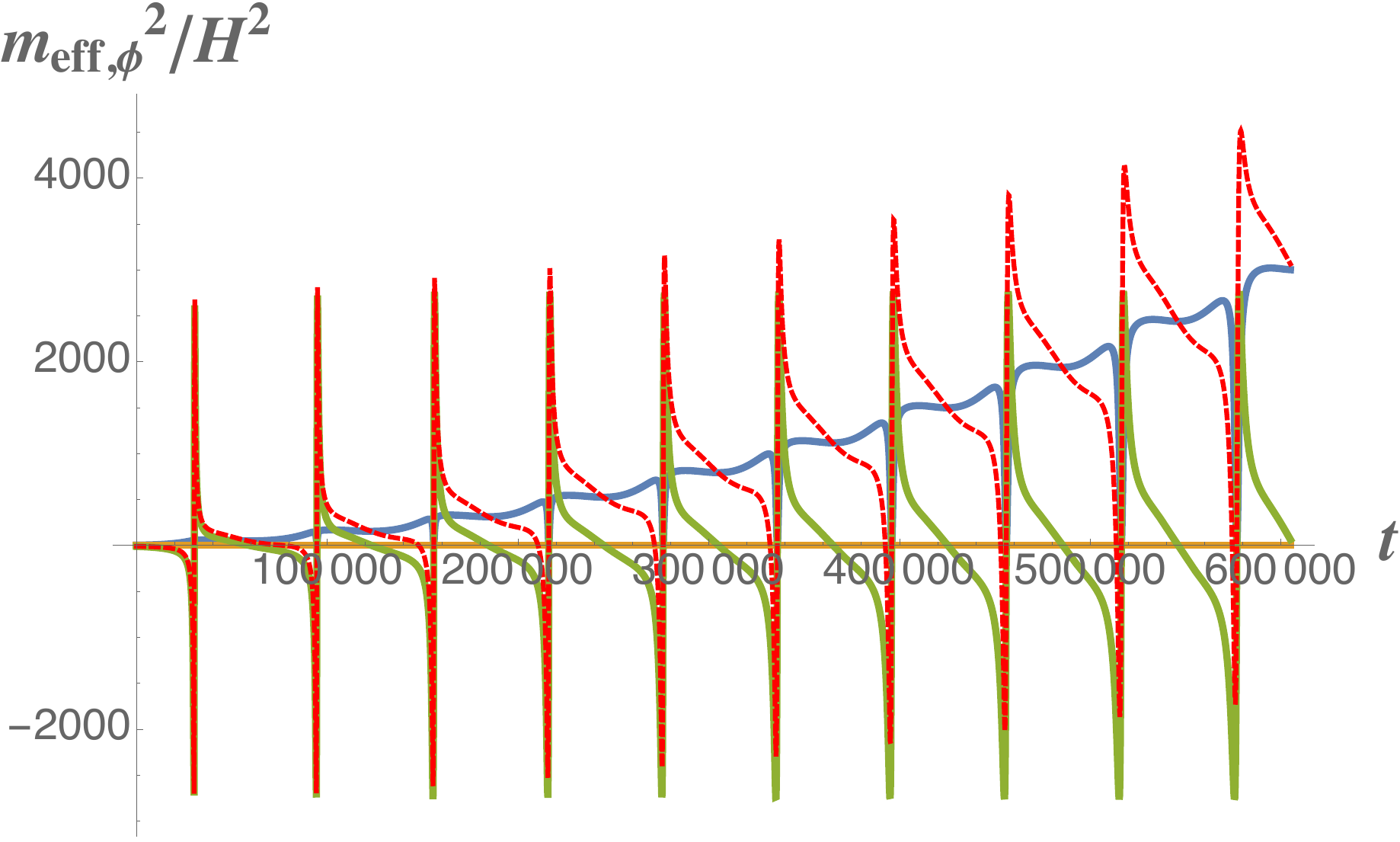} \quad \includegraphics[width=0.48\textwidth]{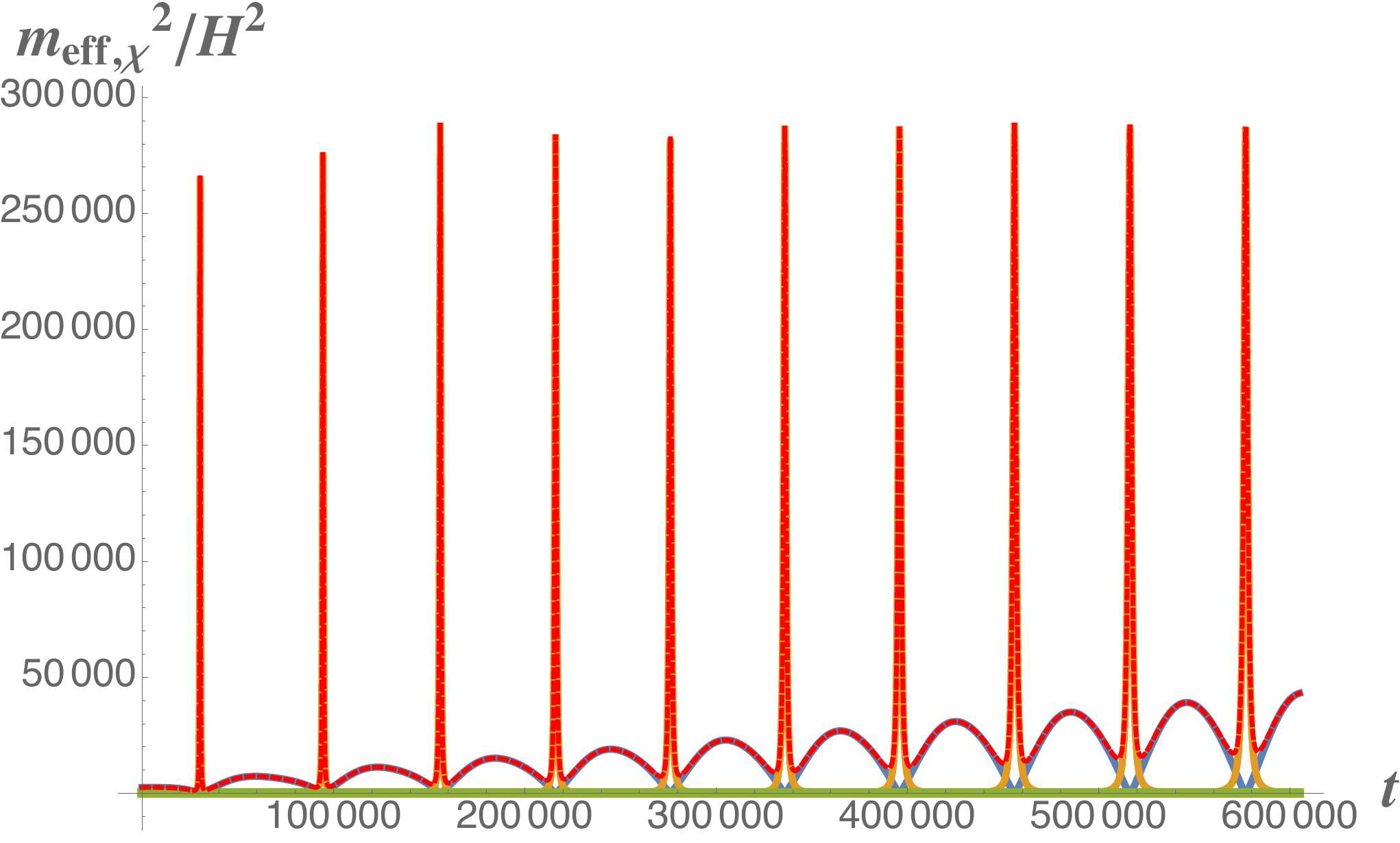}
\caption{ \small \baselineskip 11pt  The contributions to $m_{\rm eff,\phi}^2$ ({\it left}) and $m_{\rm eff,\chi}^2$ ({\it right}) as functions of $t$; inflation ends and preheating begins at $t = 0$. In each plot, we show the individual contributions to $m_{{ \rm eff}, I}^2$: $m_{1,I}^2$ (blue) arising from the potential; $m_{2,I}^2$ (gold) arising from the curved field-space manifold; $m_{3, I}^2$ (green) arising from the coupled metric perturbations. The contribution $m_{4,I}^2$ is not plotted (since it remains so small), though it is included numerically in our solutions for $m_{ {\rm eff}, I}^2 / H^2$ (red dashed). For each plot, we fix $\sqrt{ \lambda_\phi }/ \xi_\phi = 10^{-4}$, $\xi_\chi / \xi_\phi = 0.8$, $\lambda_\chi / \lambda_\phi = 1.25$, and $g / \lambda_\phi = 2$, and vary $\xi_\phi$: $\xi_\phi = 0.1$ ({\it top}); $\xi_\phi = 10$ ({\it middle}); $\xi_\phi = 100$ ({\it bottom}). The quantity $m_{ {\rm eff}, I}^2 / H^2$ grows over time because $H (t)$ falls after the end of inflation. }
\label{mphimchi}
\end{figure}

As shown in Fig.~\ref{mphimchi}, there exist three distinct regimes of interest, depending on whether $\xi_I < {\cal O} (1)$, $\xi_I \sim {\cal O} (1 - 10)$, or $\xi_I \geq {\cal O} (100)$. Both $m_{\rm eff,\phi}^2$ and $m_{\rm eff,\chi}^2$ develop increasingly sharp features with increasing $\xi_I$, an effect studied in Ref.~\cite{Ema} and futher explored in Refs.~\cite{MultiPreheat2,MultiPreheat3}. These sharp features lead to spikes in $\partial_t m_{ {\rm eff}, I}^2$ (and hence in ${\cal A}_{(I)}$) for both adiabatic and isocurvature modes for $\xi_I \geq {\cal O} (100)$, yielding efficient particle production in that limit. Other effects are notable in Fig.~\ref{mphimchi}. For example, for the adiabatic modes, the term arising from the coupled metric perturbations, $m_{3,\phi}^2$, becomes increasingly important as $\xi_I$ becomes large, periodically driving $m_{\rm eff,\phi}^2 < 0$ and hence yielding brief, tachyonic bursts of particle production, an effect we study in more detail in Ref.~\cite{MultiPreheat3}. 

\begin{figure}
\centering
\includegraphics[width=0.6\textwidth]{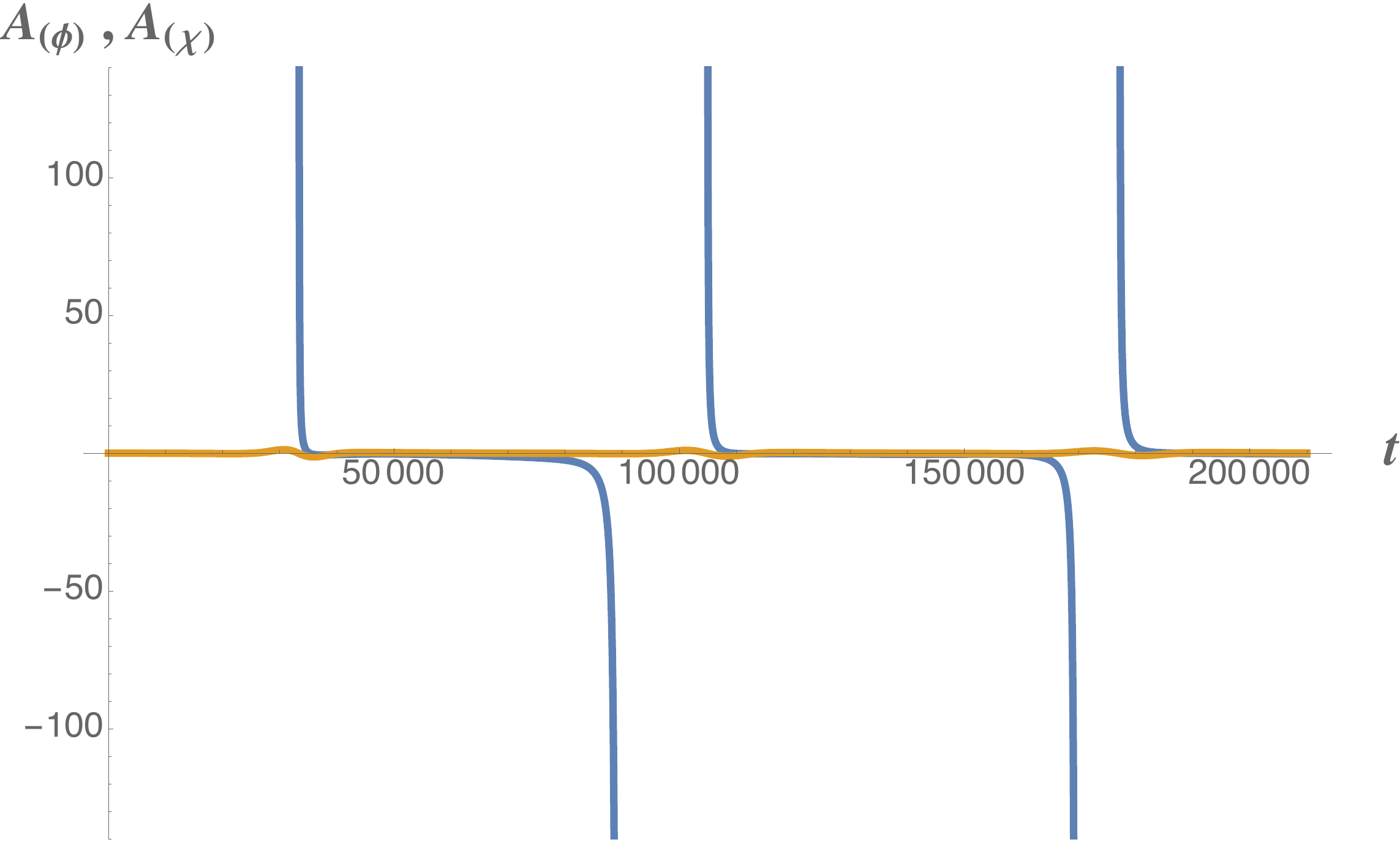}
\caption{\small \baselineskip 11pt The adiabatic parameters ${\cal A}_{(\phi)}$ (blue) and ${\cal A}_{(\chi)}$ (gold) for $k \ll aH$ and $\xi_\phi = 10$, with the ratios of couplings as in Fig.~\ref{mphimchi}. In the intermediate regime, with $\xi_I \sim {\cal O} (10)$, adiabatic perturbations are amplified while isocurvature modes are suppressed.}
\label{Aphichi1}
\end{figure}

On the other hand, for intermediate values of the nonminimal couplings, $\xi_I \sim {\cal O} (10)$, we see that $m_{\rm eff,\chi}^2$ neither becomes sharply peaked nor oscillates to zero. In the intermediate regime, therefore, we expect suppressed amplification of the isocurvature modes. We may understand this suppression analytically. Along the isocurvature direction, $m_{2, \chi}^2 \propto \dot{\phi}^2$ may become comparable in magnitude, but opposite in phase, to $m_{1,\chi}^2 \propto \phi^2$ depending on the magnitude of $\xi_I$. For $\xi_I < 1$, we may expand
\beq
m_{1, \chi}^2 = \frac{ g M_{\rm pl}^2}{\xi_\phi} \delta^2 \left[ 1 - \delta^2 \left( 1 + \frac{ \lambda_\phi}{g} (2 - \varepsilon ) \right) \right] + {\cal O} (\xi_I^2 ) ,
\label{m1chismallxi}
\eeq
where $\delta^2$ is defined in Eq. (\ref{delta}) and the eccentricity $\varepsilon$ is defined in Eq. (\ref{Lambdadef}). In the same limit, we have
\beq
m_{2, \chi}^2 = \left( \frac{ \dot{\phi}^2}{M_{\rm pl}^4} \right) \left[ \xi_\phi + \xi_\chi \right] + {\cal O} (\xi_I^2 ) .
\label{m2chismallxi}
\eeq
For an order-of-magnitude estimate in this limit, we may approximate $\dot{\phi}^2 \sim \omega^2 \phi^2$ and use our results from Section \ref{Oscillations}. For $\xi_I \sim 0.1$, we have $\omega = 2 \pi / T \rightarrow ( 2\pi / 3.9 ) \sqrt{\lambda_\phi} \> M_{\rm pl}$ and hence
\beq
\frac{ m_{2,\chi}^2}{ m_{1, \chi}^2 } \sim \frac{\lambda_\phi}{g} \left( \xi_\phi + \xi_\chi \right) + {\cal O} (\xi_I^2) .
\label{m2m1smallxi}
\eeq
For $\xi_I < 1$, we therefore find a clear separation of scales, $m_{2, \chi}^2 \ll m_{1, \chi}^2$. In that limit, $m_{\rm eff, \chi}^2$ passes near zero as the background field $\phi (t)$ oscillates, as shown in Fig. \ref{mphimchi}b. For $\xi_I \sim 10$, however, we find 
\beq
m_{1, \chi}^2 = - \frac{ \Lambda_\phi}{\xi_\phi^2} M_{\rm pl}^2 \left( \frac{ \delta^2}{1 + \delta^2} \right) + {\cal O} (\xi_I^{-2} ) ,
\label{m1chilargexi}
\eeq
where $\Lambda_\phi$ is defined in Eq. (\ref{Lambdadef}). For $\xi_I \sim 10$, the parameter $\delta^2 \sim {\cal O} (1)$ at the end of inflation. Upon using Eq. (\ref{Ricci2d}), we find
\beq
m_{2, \chi}^2 = \frac{ 6 \xi_\phi \xi_\chi }{M_{\rm pl}^2} \dot{\phi}^2 + {\cal O} (\xi_I) .
\label{m2chilargexi}
\eeq
Again making use of our results in Section \ref{Oscillations} to replace $\dot{\phi}^2 \sim \omega^2 \phi^2$, now with $\xi_I \sim 10$, we have $\omega = 2 \pi / T \rightarrow (2 \pi / 14.8) \sqrt{\lambda_\phi} \> M_{\rm pl} / \xi_\phi$, which yields
\beq
\frac{ m_{2, \chi}^2 }{m_{1, \chi}^2 } \sim \frac{ \lambda_\phi \xi_\chi }{ \vert \Lambda_\phi \vert } \sim {\cal O} (1) .
\label{m2m1largexi}
\eeq
Therefore we do indeed expect $m_{1, \chi}^2$ and $m_{2,\chi}^2$ to remain comparable in magnitude but opposite in phase for $\xi_I \sim {\cal O} (10)$. In that case, $m_{\rm eff, \chi}^2$ never passes through zero, as shown in Fig. \ref{mphimchi}d. Meanwhile, for $\xi_I \geq {\cal O} (100)$, the oscillations of $\phi (t)$ become sufficiently different from the near-harmonic case that $\dot{\phi}^2 \gg \omega^2 \phi^2$ \cite{MultiPreheat2,MultiPreheat3,Ema}, and we find that $m_{2,\chi}^2 \gg m_{1,\chi}^2$, as shown in Fig.~\ref{mphimchi}f. The intermediate-$\xi_I$ regime is thus characterized by efficient growth of adiabatic perturbations, with $\vert {\cal A}_{(\phi)} \vert > 1$, but suppression of isocurvature perturbations, with $\vert {\cal A}_{( \chi)} \vert < 1$, as shown in Fig.~\ref{Aphichi1}.

\subsection{Rotating the Field-Space Coordinates}
\label{Rotation}

\begin{figure}
\centering
\includegraphics[width=3.0in]{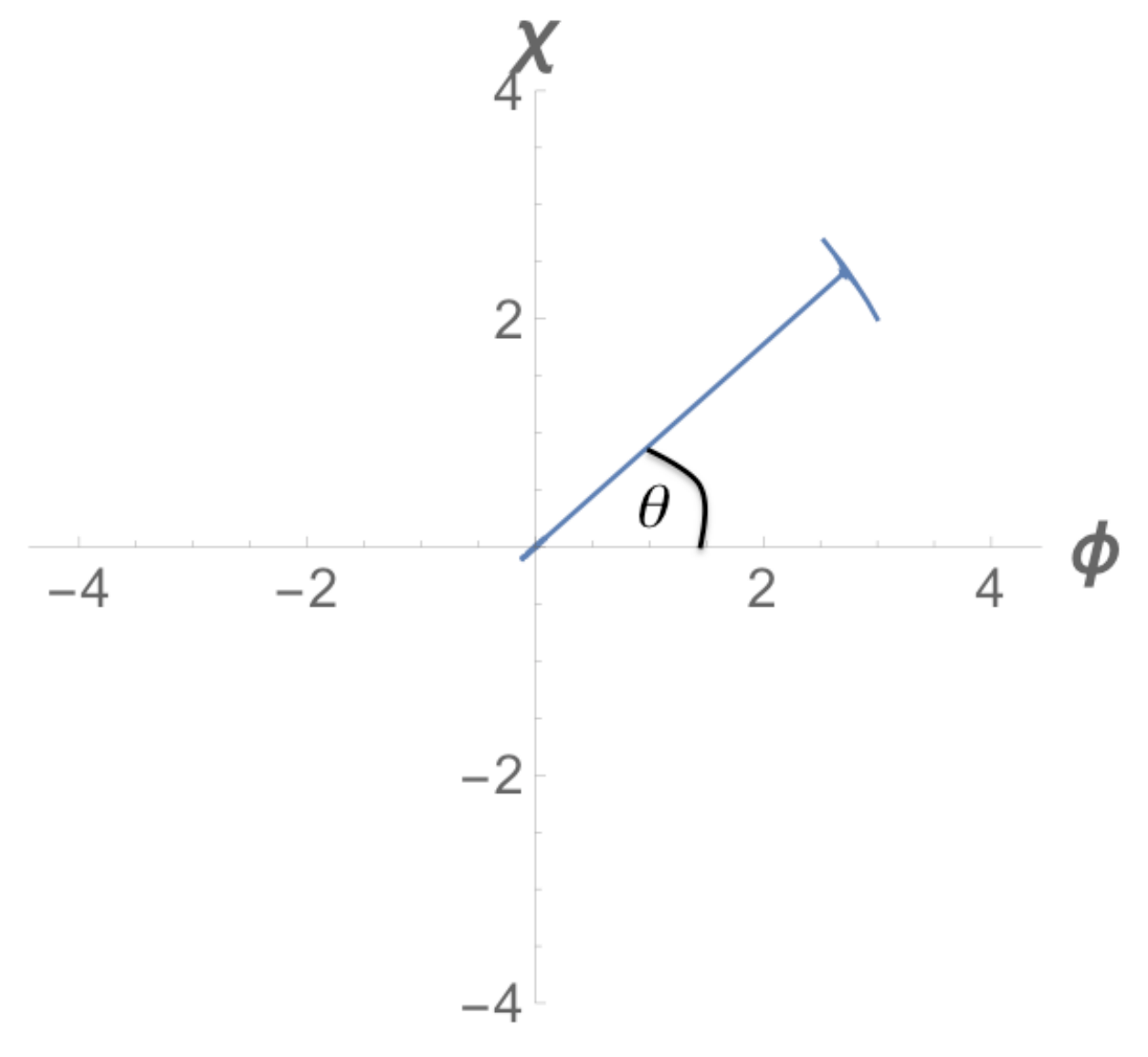}
\caption{\small \baselineskip 10pt For some choices of the coupling constants, the background fields evolve along a single-field trajectory at some angle $\theta$ that does not coincide with either the $\phi$ or $\chi$ axes. Shown here is the case for $\xi_\chi / \xi_\phi = 0.8, \lambda_\chi  / \lambda_\phi = 1.25, g / \lambda_\phi = - 1/2$, with $\xi_\phi = 10, \lambda_\phi = 10^{-6}$. The angle, $\theta = {\rm arctan} (\chi / \phi)$, is independent of time during as well as after inflation.}
\label{chiphinegg}
\end{figure}

\begin{figure}
\centering
\includegraphics[width=0.48\textwidth]{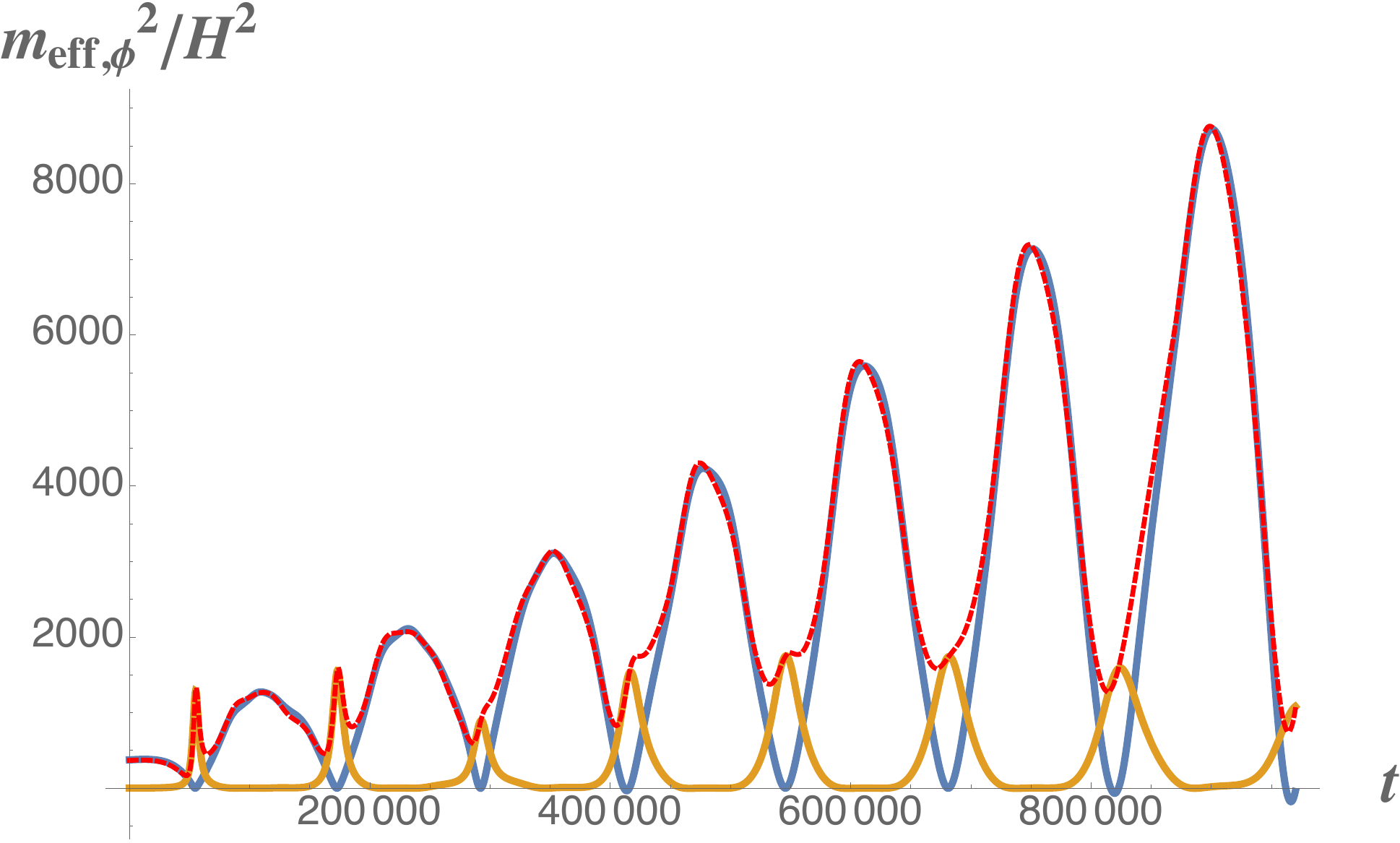} \quad \includegraphics[width=0.48\textwidth]{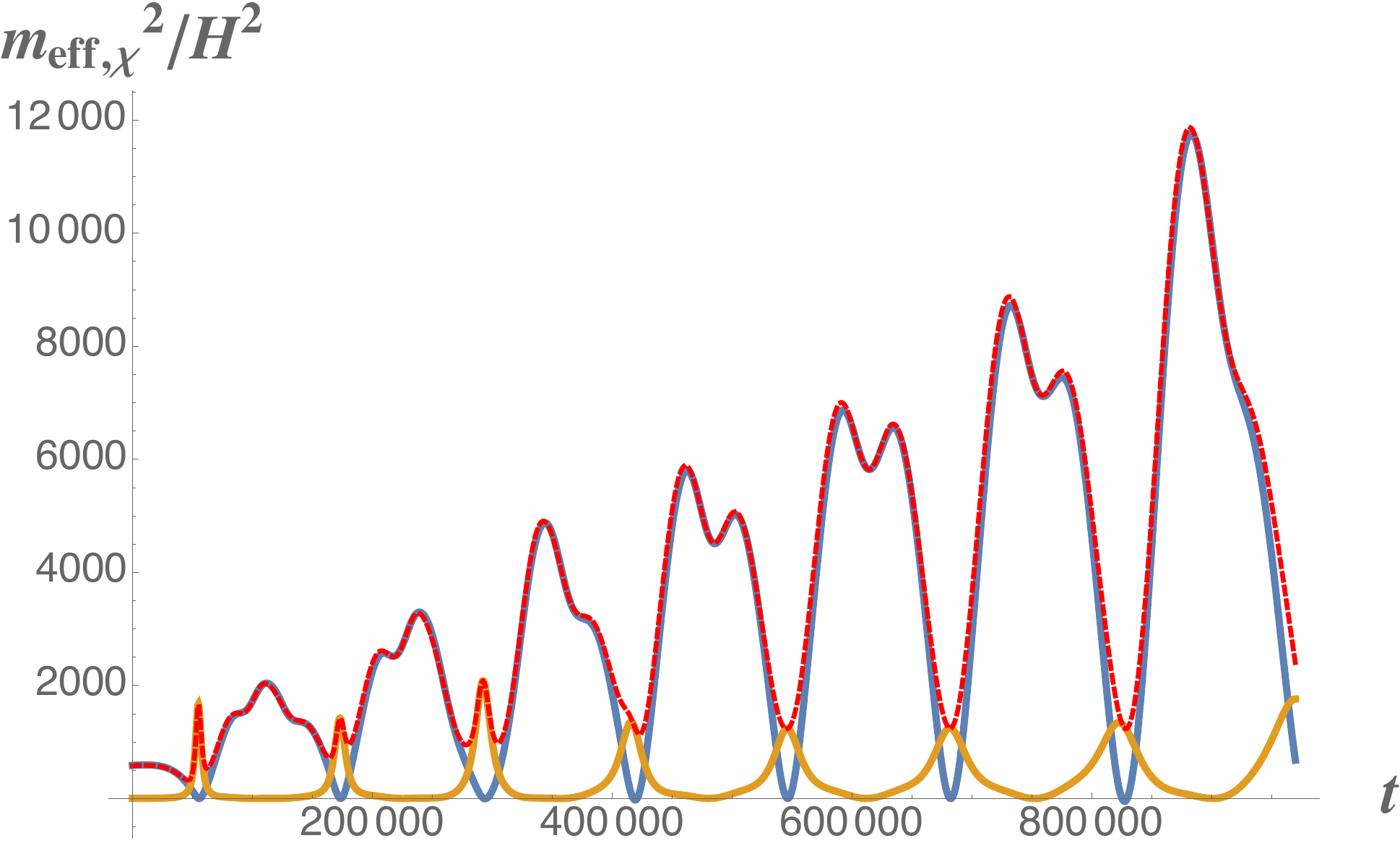}
\caption{\small \baselineskip 11pt ({\it Left}) The terms $m_{1, \phi}^2$ (blue) and $m_{2, \phi}^2$ (gold) compared to $m_{\rm eff, \phi}^2$ (red dashed) for $\xi_\phi = 10$, $g = -1/2$, and the other couplings as in Eq. (\ref{couplings2}). When plotted with respect to the original coordinate bases, $m_{\rm eff, \phi}^2$ no longer oscillates through zero. ({\it Right}) The terms $m_{1, \chi}^2$ (blue) and $m_{2, \chi}^2$ (gold) compared to $m_{\rm eff, \chi}^2$ (red dashed) for $\xi_\phi = 10$, $g = -1/2$, and the other couplings as in Eq. (\ref{couplings2}).}
\label{m1m2phinegg}
\end{figure}

\begin{figure}
\centering
\includegraphics[width=4.0in]{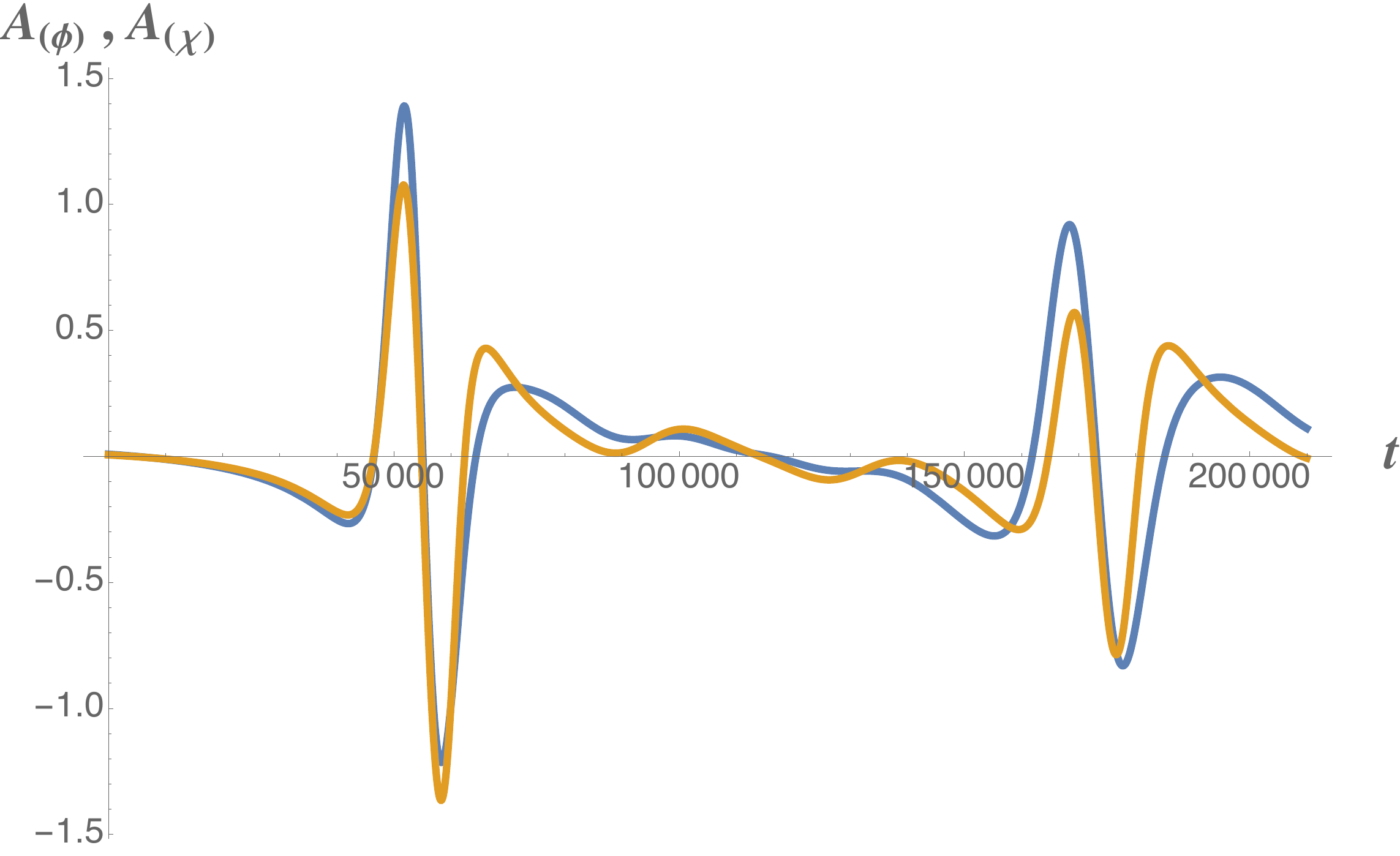}
\caption{\small \baselineskip 10pt The adiabatic parameters ${\cal A}_{(\phi)}$ and ${\cal A}_{(\chi)}$ for the original coordinate bases, with $\xi_\phi = 10$, $g = - 1/2$, and the other couplings as in Eq. (\ref{couplings2}). Because $m_{2, I}^2$ remains comparable in magnitude but opposite in phase to $m_{1, I}^2$, neither ${\cal A}_{(\phi)}$ nor ${\cal A}_{(\chi)}$ grows much larger than 1. }
\label{AphiAchinegg}
\end{figure}

Now consider what happens when we change the couplings so that the single-field attractor lies along some distinct direction in field space. For example, we may select the couplings
\beq
\frac{ \lambda_\chi }{\lambda_\phi}  =  1.25, \>\> \frac{g }{\lambda_\phi} = - 1/2 , \>\> \frac{\xi_\chi }{ \xi_\phi} = 0.8 .
\label{couplings2}
\eeq
For minimally coupled models, $g < 0$ leads to an explosive ``negative coupling instability" for long-wavelength modes \cite{NCI,Sting}. In the presence of nonminimal couplings, however, at least for $\vert g \vert \sim {\cal O} (\lambda_\phi)$, the effect of the negative coupling is to rotate the orientation of the valley of the potential away from the direction $\chi = 0$. See Fig. \ref{chiphinegg}. With the fields' motion ``misaligned" with respect to the original axes of our field-space coordinate system, we find suppression of the resonances along {\it both} of the original axes, since in this case $m_{2, I}^2$ remains comparable in magnitude (but opposite in phase) with $m_{1,I}^2$ for both $m_{\rm eff, \phi}^2$ and $m_{\rm eff, \chi}^2$. See Fig. \ref{m1m2phinegg}. Therefore both ${\cal A}_{(\phi)}$ and ${\cal A}_{(\chi)}$ remain ${\cal O} (1)$, as shown in Fig. \ref{AphiAchinegg}.

\begin{figure}
\centering
\includegraphics[width=0.48\textwidth]{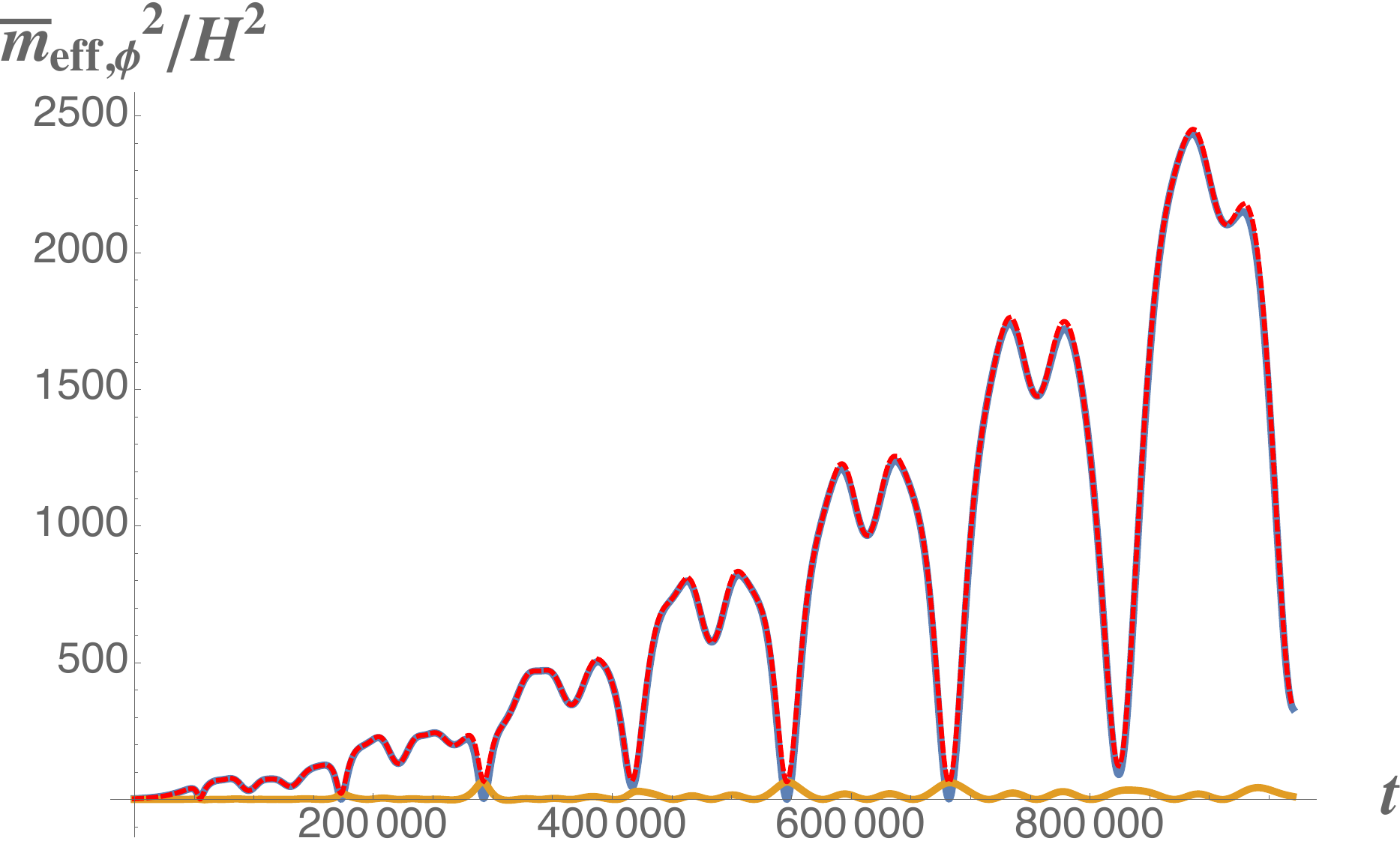} \quad \includegraphics[width=0.48\textwidth]{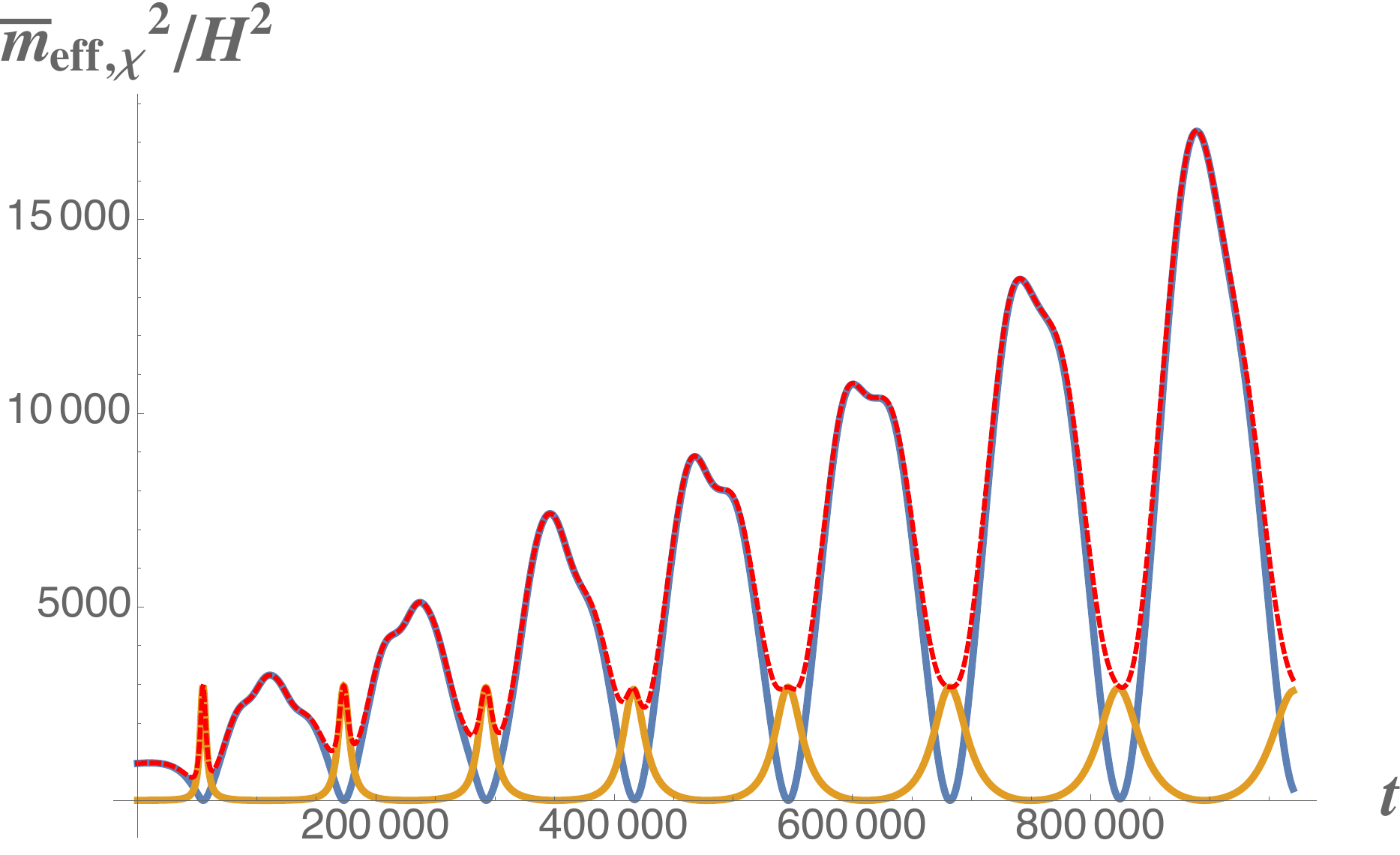}
\caption{ \small \baselineskip 10pt ({\it Left}) The contributions $\bar{m}_{1, \phi}^2$ (blue) and $\bar{m}_{2,\phi}^2$ (gold) to $\bar{m}_{\rm eff,\phi}^2$ (red dashed), upon making the rotation in field space, for $\xi_\phi = 10$, $g = - 1/2$, and the other couplings as in Eq. (\ref{couplings2}). Unlike in Fig. \ref{m1m2phinegg}, here we find the contribution from the field-space manifold, $\bar{m}_{2, \phi}^2$, negligible, and hence $\bar{m}_{\rm eff, \phi}^2 \sim \bar{m}_{1, \phi}^2$ oscillates through zero. ({\it Right}) The contributions $\bar{m}_{1,\chi}^2$ (blue) and $\bar{m}_{2, \chi}^2$ (gold) to $\bar{m}_{\rm eff,\chi}^2$ (red dashed), upon making the rotation in field space, for $\xi_\phi = 10$, $g = -1/2$, and the other couplings as in Eq. (\ref{couplings2}). Just as in the case when the single-field attractor lay along the direction $\chi = 0$, in this case we find $\bar{m}_{1,\chi}^2 \sim \bar{m}_{2,\chi}^2$ but out of phase with each other, so that $\bar{m}_{\rm eff,\chi}^2$ never oscillates through zero.  }
\label{m1m2phibar}
\end{figure}

\begin{figure}
\centering
\includegraphics[width=4.0in]{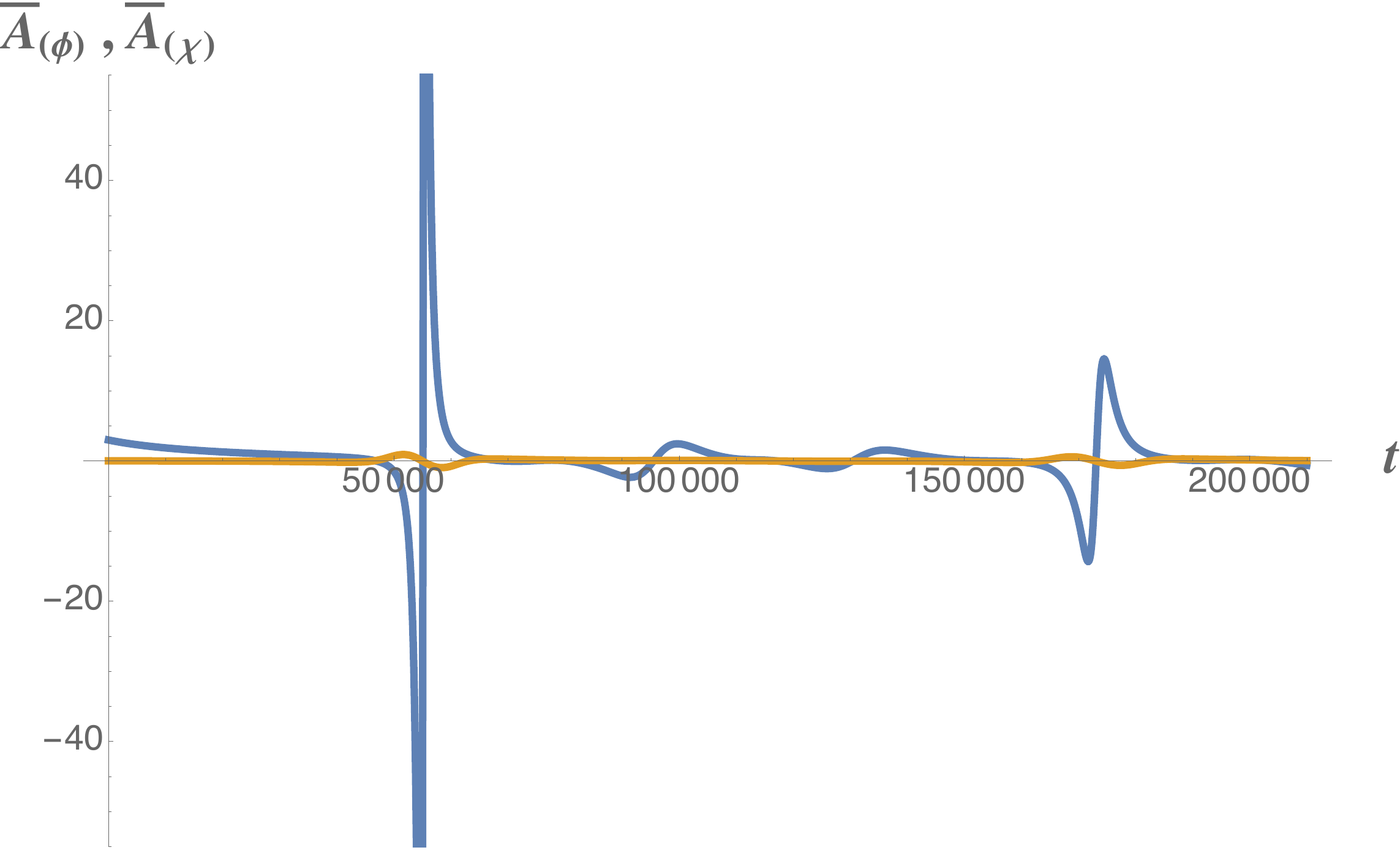}
\caption{\small \baselineskip 10pt The adiabatic parameters $\bar{\cal A}_{(\phi)}$ (blue) and $\bar{\cal A}_{(\chi)}$ (gold) with $\xi_\phi = 10$, $g = -1/2$, and the other couplings as in Eq. (\ref{couplings2}), upon performing the rotation in field space. Here we recover behavior akin to the original example, when the single-field attractor lay along the direction $\chi = 0$: fluctuations along the adiabatic direction become strongly amplified, but those in the isocurvature direction do not. }
\label{AphibarAchibar}
\end{figure}

However, as Fig. \ref{chiphinegg} makes clear, in this case the fields still evolve within a single-field attractor. We may parameterize the motion by a single angle, $\theta \equiv {\rm arctan} (\chi / \phi)$, which, following an initial transient, does not vary over time (even after the end of inflation). That is, when plotted in the original coordinate system, the background fields' motion obeys
\beq
\phi (t) = r(t) \cos \theta , \>\> \chi (t) = r (t) \sin \theta .
\label{phichioriginal}
\eeq
We may then perform a rotation of our coordinates in field space so that the single-field attractor lies along the $\bar{\chi}$ direction, with all motion of the background fields along the $\bar{\phi}$ axis. (In this subsection we denote the rotated coordinate system with an overbar rather than a prime, to avoid confusion with derivatives, $d / d\eta$.) Hence we may write
\beq
\begin{split}
\bar{\phi} &= \phi \cos \theta + \chi \sin\theta , \\
\bar{\chi} &= \chi \cos \theta - \phi \sin \theta .
\end{split}
\label{phibarchibar}
\eeq
Components of the tensor $[ \omega_k^2 ]^I_{\> \>J}$ transform in the usual way under this coordinate transformation:
\beq
\left[ \bar{\omega}_k^2 \right]^I_{\>\> J} = \left( \frac{ \partial \bar{\varphi}^I }{\partial \varphi^K } \right) \left( \frac{ \partial \varphi^L }{ \partial \bar{\varphi}^J } \right) \left[ \omega_k^2 \right]^K_{\>\> L} .
\label{baromegaIJ}
\eeq
In particular, we find
\beq
\begin{split}
\left[ \bar{\omega}_k^2 \right]^\phi_{\>\> \phi} &= \cos^2 \theta \left[ \omega_k^2 \right]^\phi_{\>\> \phi} + \sin\theta \cos \theta \left( \left[ \omega_k^2 \right]^\chi_{\>\> \phi} + \left[ \omega_k^2 \right]^\phi_{\>\> \chi} \right) + \sin^2 \theta \left[ \omega_k^2 \right]^\chi_{\>\> \chi} , \\
\left[ \bar{\omega}_k^2 \right]^\chi_{\>\> \chi} &= \cos^2 \theta \left[ \omega_k^2 \right]^\chi_{\>\> \chi} - \sin \theta \cos \theta \left( \left[ \omega_k^2 \right]^\phi_{\>\> \chi} + \left[ \omega_k^2 \right]^\chi_{\>\> \phi} \right) + \sin^2 \theta \left[ \omega_k^2 \right]^\phi_{\>\>\phi} .
\end{split}
\label{baromegachichi}
\eeq
When plotted with respect to the rotated coordinate system, we recover the type of behavior we had found in Section \ref{Resonances} for a single-field attractor along the direction $\chi = 0$. Fig. \ref{m1m2phibar} shows the dominant contributions to $\bar{m}_{\rm eff,\phi}^2$, revealing that in the rotated coordinate system, the contributions from the field-space manifold become negligible, just as they do for $m_{\rm eff, \phi}^2$ when the single-field attractor lies along the $\chi = 0$ direction (as in Fig.~\ref{mphimchi}). On the other hand, in the rotated coordinate basis, $\bar{m}_{2, \chi}^2$ remains comparable in magnitude to $\bar{m}_{1, \chi}^2$ but with opposite phase, so that $\bar{m}_{\rm eff, \chi}^2$ never oscillates through zero (again like the behavior in Fig.~\ref{mphimchi}). Moreover, if we compute
\beq
\bar{\cal A}_{(I)} = \frac{\partial_t \bar{m}_{ {\rm eff}, I}^2 }{ 2 \left( \bar{m}_{ {\rm eff}, I}^2 \right)^{3/2} } + \frac{H}{ \bar{m}_{{ \rm eff},I} } ,
\label{barAphi}
\eeq
we find behavior akin to the original analysis for the $\chi = 0$ attractor, as shown in Fig. \ref{AphibarAchibar}. Thus we surmise that within {\it any} single-field attractor, in the intermediate regime with $\xi_I \sim {\cal O} (10)$, we find suppression of the resonances for the isocurvature direction and amplification of the fluctuations along the adiabatic direction. This general result holds even though the models we consider do not obey an $O (N)$ symmetry.

\section{Conclusions}
\label{Conclusions}

Realistic models of high-energy physics typically include multiple scalar fields, each with its own nonminimal coupling. In this paper we have demonstrated that preheating after inflation in such models introduces unique features that are distinct from other well-studied models of preheating. 

In particular, nonminimally coupled fields yield a conformally stretched effective potential in the Einstein frame. In previous work we had highlighted a generic feature that arises from such conformal stretching, namely, the existence of strong single-field attractor behavior across a wide range of couplings and initial conditions \cite{KMS, GKS,KS,SSK}. Here we have found two main effects related to the conformal stretching and attractor behavior: the effectively single-field evolution of the background fields $\varphi^I (t)$ persists during the oscillatory phase --- thereby avoiding the ``de-phasing" that is typical of preheating with minimally coupled scalar fields --- and the conformal stretching of the potential alters the time-evolution of $\varphi^I (t)$ as the background field(s) oscillate around the global minimum of the potential. 

The persistence of the single-field attractor during the preheating phase leads to efficient transfer of energy from the background fields to coupled fluctuations. The balance of the transfer to fluctuations in the adiabatic versus isocurvature directions depends on the nonminimal coupling constants. We identify here, and study further in Refs.~\cite{MultiPreheat2,MultiPreheat3}, three distinct regimes, depending on whether $\xi_I < {\cal O} (1)$, $\xi_I \sim {\cal O} (1 - 10)$, or $\xi_I \geq {\cal O} (100)$. The growth of long-wavelength isocurvature modes is suppressed for intermediate couplings, $\xi_I \sim {\cal O} (10)$ --- a new effect arising entirely from the nontrivial field-space manifold, which has no analogue in models with minimally coupled fields. In the large-$\xi_I$ regime, however, appropriate to such models as Higgs inflation \cite{BezrukovShaposhnikov}, the amplification of isocurvature modes becomes very efficient \cite{MultiPreheat2,MultiPreheat3,Ema}. (Naturally, the efficient amplification of isocurvature perturbations after the end of inflation is quite distinct from the amplification of isocurvature perturbations during inflation, which is generically suppressed in these models \cite{SSK}. Modes amplified during inflation would have length-scales today of tens to thousands of Mpc, due to their exponential stretching during inflation; modes amplified after the end of inflation would have exponentially shorter length-scales, and would not affect observables such as $\beta_{\rm iso}$.) 

The efficiency of the reheating stage can have observational consequences, both for the CMB and for the particle content of the universe. The values of the CMB observables $n_s$ and $r$ may be related to the time $N_*$, where $N_*$ is the number of efolds before the end of inflation when perturbations on CMB-relevant length-scales crossed outside the Hubble radius. For models in the family we consider here, these relations are given by $n_s \simeq 1 - 2 / N_* - 3 / N_*^2$ and $r \simeq 12 / N_*^2$ (see, e.g., Ref.~\cite{KS}). Depending on how quickly the universe transitions to a radiation-dominated phase after the end of inflation, the observationally relevant $N_*$ may vary by as much as 10 efolds (see, e.g., Ref.~\cite{AHKK}), shifting the predictions for $r$ by as much as $30\%$ and for $n_s - 1$ by as much as $10\%$. Furthermore, different reheating scenarios can yield different reheat temperatures, which can have other implications, such as washing out lepton or baryon asymmetries that might have been generated at the end of inflation. Such possibilities make it critical to gain an understanding of the reheating process following inflation.

In Refs.~\cite{MultiPreheat2,MultiPreheat3} we exploit the covariant formalism developed here to more thoroughly explore the resonance structure in this family of models as functions of wavenumber, $k$, as well as coupling constants, $\xi_I, \lambda_I$, and $g$. Other effects also deserve further attention. In particular, the conformal stretching of the potential in the Einstein frame could produce metastable oscillons after inflation. The formation of such long-lived, topologically metastable objects could become important after the earliest stages of preheating, impacting the rate at which the system ultimately reaches thermal equilibrium. These and related nonlinear effects could therefore affect the final reheat temperature and the expansion history of the universe after inflation \cite{GleiserOscillons,AminOscillons,ZhouOscillons,Lozanov,Figueroa}. These possibilities remain the subject of further research.

\section*{Appendix A: Field-Space Metric and Related Quantities}
\label{AppendixA}

Given $f (\phi^I)$ in Eq. (\ref{f2field}) for a two-field model, the field-space metric in the Einstein frame, Eq. (\ref{GIJ}), takes the form
\beq
\begin{split}
{\cal G}_{\phi \phi} &= \left( \frac{M_{\rm pl}^2}{2f} \right) \left[ 1 + \frac{3 \xi_\phi^2 \phi^2}{f} \right] , \\
{\cal G}_{\phi \chi} = {\cal G}_{\chi \phi} &= \left( \frac{M_{\rm pl}^2}{2f} \right) \left[ \frac{3 \xi_\phi \xi_\chi \phi \chi}{f} \right] , \\
{\cal G}_{\chi \chi} &= \left( \frac{M_{\rm pl}^2}{2f} \right) \left[ 1 + \frac{3 \xi_\chi^2 \chi^2}{f} \right] .
\end{split}
\label{Gphiphi}
\eeq
The components of the inverse metric are
\beq
\begin{split}
{\cal G}^{\phi \phi} &= \left( \frac{ 2f}{M_{\rm pl}^2} \right) \left[ \frac{2f + 6 \xi_\chi^2 \chi^2}{C} \right] , \\
{\cal G}^{\phi \chi} = {\cal G}^{\chi \phi} &= - \left( \frac{ 2f}{M_{\rm pl}^2} \right) \left[ \frac{6 \xi_\phi \xi_\chi \phi \chi }{C} \right] , \\
{\cal G}^{\chi\chi} &= \left( \frac{ 2f}{M_{\rm pl}^2} \right)  \left[ \frac{2f + 6 \xi_\phi^2 \phi^2}{C} \right] ,
\end{split}
\label{inverseG}
\eeq
where $C (\phi^I)$ is defined as
\beq
\begin{split}
C (\phi, \chi) &\equiv M_{\rm pl}^2 + \xi_\phi (1 + 6 \xi_\phi) \phi^2 + \xi_\chi (1 + 6 \xi_\chi ) \chi^2 \\
&= 2f + 6 \xi_\phi^2 \phi^2 + 6 \xi_\chi^2 \chi^2 .
\end{split}
\label{C}
\eeq

The Christoffel symbols for our field space take the form
\beq
\begin{split}
\Gamma^\phi_{\>\> \phi \phi} &= \frac{\xi_\phi (1 + 6 \xi_\phi ) \phi}{C} - \frac{\xi_\phi \phi}{f} , \\
\Gamma^\phi_{\>\> \chi \phi} = \Gamma^\phi_{\>\> \phi \chi} &= - \frac{\xi_\chi \chi}{2f} , \\
\Gamma^\phi_{\>\> \chi \chi} &= \frac{\xi_\phi (1 + 6 \xi_\chi) \phi}{C} , \\
\Gamma^\chi_{\>\> \phi \phi} &= \frac{\xi_\chi (1 + 6 \xi_\phi) \chi}{C} , \\
\Gamma^\chi_{\>\> \phi \chi} = \Gamma^\chi_{\>\> \chi \phi} &= - \frac{\xi_\phi \phi}{2f} , \\
\Gamma^\chi_{\>\> \chi \chi} &= \frac{\xi_\chi (1 + 6 \xi_\chi ) \chi}{C} - \frac{\xi_\chi \chi}{f}
\end{split}
\label{Gammas}
\eeq

For two-dimensional manifolds we may always write the Riemann tensor in the form
\beq
{\cal R}_{ABCD} = \frac{1}{2} {\cal R} (\phi^I) \left[ {\cal G}_{AC} {\cal G}_{BD} - {\cal G}_{AD} {\cal G}_{BC} \right] ,
\label{Riemann2d}
\eeq
where ${\cal R} (\phi^I)$ is the Ricci scalar. Given the field-space metric of Eq. (\ref{Gphiphi}), we find
\beq
{\cal R} (\phi^I ) = \frac{1}{3 M_{\rm pl}^2 C^2} \left[ (1 + 6 \xi_\phi) (1 + 6 \xi_\chi) (4 f^2 ) - C^2 \right] .
\label{Ricci2d}
\eeq
For the two-field model, we may also solve explicitly for the vielbeins, $e_b^I$, of Eq. (\ref{vielbein}). Defining
\beq
\begin{split}
A &\equiv C - 6 \xi_\phi^2 \phi^2 , \\
B &\equiv C - 6 \xi_\chi^2 \chi^2 , \\
E &\equiv C - 3 \xi_\phi^2 \phi^2 - 3 \xi_\chi^2 \chi^2 , \\
F &\equiv \frac{ \sqrt{ 2 f C} \> \sqrt{E - \sqrt{2 f C}} }{3 \sqrt{2} \> M_{\rm pl} ( \xi_\chi^2 \chi^2 + \xi_\phi^2 \phi^2 ) C } ,
\end{split}
\label{ABEF}
\eeq
then we may satisfy Eq. (\ref{vielbein}) with 
\beq
\begin{split}
e_1^\phi &= F \left(A + \sqrt{2fC} \right) , \\
e_1^\chi &= - 6 F \xi_\phi \xi_\phi \phi \chi , \\
e_2^\phi &= e_1^\chi , \\
e_2^\chi &= F \left( B + \sqrt{ 2 f C} \right) .
\end{split}
\label{ebIphichi}
\eeq
We note that within the single-field attractor along the direction $\chi = 0$, $e_2^\phi \sim e_1^\chi \sim 0$, $e_1^\phi e_1^\phi \rightarrow {\cal G}^{\phi \phi} + {\cal O} (\chi^2)$, and $e_2^\chi e_2^\chi \rightarrow {\cal G}^{\chi\chi} + {\cal O} (\chi^2 )$.

\section*{Appendix B: Period of single-field background oscillations}
\label{AppendixB}

\begin{figure}
\centering
\includegraphics[width=3in]{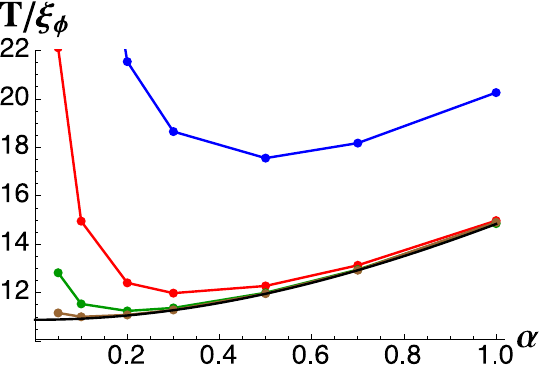}
\caption{ \small \baselineskip 12pt  Period of oscillation, $T$, rescaled by the nonminimal coupling, in units of $(\sqrt{\lambda_\phi} M_{\rm pl})^{-1} $, as a function of $\alpha = \sqrt{\xi_\phi} \> \phi_0 /M_{\rm pl}$ for $\xi=10,10^2,10^3, 10^4$ (from top to bottom). The solid black line shows the approximate analytic result of Eq. \eqref{Tlargexi}, which is derived under the assumption that $6 \xi_\phi \alpha^2 \gg 1$.}
\label{fig:Txiintegral}
\end{figure}

Starting from Eq. \eqref{T} and inserting the values of ${\cal G}_{\phi\phi}$ and $V(\phi)$ the period becomes
\beq
T  ={4 \sqrt{2\xi_\phi} }  \int_0 ^\alpha du    {   \sqrt{1+ 6\xi_\phi u^2 } \over   (1+u^2) } { 1\over \sqrt{ { \alpha^4  \over (1+\alpha^2)^2}  -  { u^4  \over (1+u^2)^2}}}
\eeq
where we made a change of variables $u = \sqrt {\xi_\phi} \> \phi$, and parameterized the maximum field amplitude as $\phi_{\rm max}= \alpha M_{\rm pl}/ \sqrt{\xi_\phi}$.
By assuming a maximum field amplitude such that $1< 6 \xi_\phi \alpha^2$ and approximating $1+ 6\xi_\phi u^2 \approx 6\xi_\phi u^2$, the integral can be performed analytically and the resulting Eq. \eqref{Tlargexi} shows the linear scaling of the period with $\xi_\phi$. The limit of this approximation is shown in Fig. \ref {fig:Txiintegral}, where it can be seen that the agreement between Eq. \eqref{Tlargexi} and the exact result is excellent in the large-$\xi_I$ limit for $\alpha$ not very small. The region of validity in terms of $\alpha$ increases for larger values of $\xi_\phi$, as expected from the condition $\alpha > 1/ \sqrt{6 \xi_\phi} $ used in the derivation of Eq. \eqref{Tlargexi}. Fig. \ref {fig:Txiintegral} shows the period of oscillation for different values of $\xi_I$ and $\alpha$.

\acknowledgements{It is a pleasure to thank Mustafa Amin, Bruce Bassett, Jolyon Bloomfield, Peter Fisher, Tom Giblin, Alan Guth, Mark Hertzberg, Johanna Karouby, Evan McDonough, and an anonymous referee for helpful comments. We would like to acknowledge support from the Center for Theoretical Physics at MIT. This work is supported by the U.S. Department of Energy under grant Contract Number DE-SC0012567. MPD and AP were also supported in part by MIT's Undergraduate Research Opportunities Program (UROP).  CPW thanks the University of Washington College of Arts \& Sciences for financial support. She also gratefully acknowledges support from the MIT Dr. Martin Luther King, Jr. Visiting Professors and Scholars program and its director Edmund Bertschinger. EIS gratefully acknowledges support from a Fortner Fellowship at the University of Illinois at Urbana-Champaign. }


\begin{thebibliography}{999}
\bibitem{Rsquared} As is well known, models with gravitational actions of the general form $f (R)$, where $R$ is the spacetime Ricci scalar, may be rewritten in the form of an Einstein-Hilbert action plus a scalar field with nonminimal coupling, $\xi \phi^2 R$ \cite{SotiriouFaraoni}. Hence the formalism we develop here may help illuminate processes in still broader classes of models, though such explorations remain beyond the scope of this paper.
\end{thebibliography}
\end{document}